\documentclass[twocolumn,showpacs,preprintnumbers,amsmath,amssymb,nofootinbib]{revtex4}
\usepackage{graphicx}% Include figure files
\usepackage{dcolumn}% Align table columns on decimal point
\usepackage{bm}% bold math

\usepackage{epstopdf}
\def\bea{\begin{eqnarray}}
\def\eea{\end{eqnarray}}

\def\pp{\mbox{$p$-$p$}}
\def\auau{\mbox{Au-Au}}
\def\dau{\mbox{d-Au}}

\def\pbpb{\mbox{Pb-Pb}}
\def\aa{\mbox{$A$-$A$}}
\def\pa{\mbox{$p$-$A$}}

\def\nn{\mbox{$N$-$N$}}
\def\ss{\mbox{S-S}}
\def\ee{\mbox{$e^+$-$e^-$}}
\def\ppbar{\mbox{$p$-$\bar p$}}

\def\pt{$p_t$}

\def\mmpt{$\bar p_t$}
\def\v2{$v_2$}
\def\yt{$y_t$}
\def\mt{$m_t$}

\def\ytyt{$y_t \times y_t$}

\def\ppb{\mbox{$p$-Pb}}
\def\ep{\mbox{$e$-$p$}}

\def\nch{$n_{ch}$}

\begin{document} 

\preprint{Version 1.2}

\title{Manifestations of minimum-bias dijets in high-energy nuclear collisions
}

\author{Thomas A.\ Trainor}\affiliation{CENPA 354290, University of Washington, Seattle, Washington 98195}

%%%%%%%%%%%%%%%%%%%%%%%%%%%%%%%

\date{\today}

\begin{abstract}
Dijets observed near midrapidity in high-energy nuclear collisions result from large-angle scattering of low-$x$ partons (gluons) within projectile hadrons as a signature manifestation of QCD. Within the same collisions it has been claimed that hydrodynamic flows (radial, elliptic and ``higher harmonic'' flows) carried by a dense QCD medium or quark-gluon plasma (QGP) dominate the observed hadronic final state. The flow-QGP narrative is imposed  {\em a priori} on primary particle data, and of all possible analysis methods  a subset A that seems to support that narrative is preferred. The present study explores an alternative minimum-bias (MB) jet narrative -- quantitative correspondence of MB dijet manifestations in the hadronic final state with measured {\em isolated jet} properties. The latter incorporates a different set of methods B that emerge from inductive study of primary particle data without {\em a priori} assumptions. The  resulting system of methods and data manifestations is represented by a two-component (soft + hard) model (TCM) of hadron production. A survey of methods reveals that type A tends to discard substantial information carried by primary particle data whereas type B retains almost all information in both primary particle data from nuclear collisions and from isolated jets. The main goal of the present study is a review of MB dijet  contributions to high-energy collisions in small and large systems relative to measured isolated-jet properties. Representative analysis methods from types A and B are compared in the context of MB jet manifestations. This study suggests that at least some data features commonly attributed to flows actually result from MB dijets and thereby challenges the flow-QGP narrative.
\end{abstract}

\pacs{12.38.Qk, 13.87.Fh, 25.75.Ag, 25.75.Bh, 25.75.Ld, 25.75.Nq}

\maketitle

%%%%%%%%%
 \section{Introduction} \label{intro}

This study reviews manifestations of minimum-bias (MB) dijets in the hadronic final state of high-energy nuclear collisions in the context of claimed collectivity (flows) in the same systems. There is widespread belief that hydrodynamic flows play a dominant role in high-energy collisions wherein a dense medium (quark-gluon plasma or QGP) is formed that supports a collective velocity field manifested by features of hadron distributions~\cite{keystone}. However, it is possible that at least some data features attributed to flows may relate to MB dijets~\cite{multipoles}. In order to clarify such ambiguities MB dijets should be understood in \pp, \pa\ and \aa\ collisions in relation to eventwise-reconstructed (isolated) jets derived independently from \ee\ and \ppbar\ collisions over a broad range of collision energies. That is the main goal of this study.

Flows and QGP  are intimately related by the assumption that a dense, strongly-interacting medium developed during nucleus-nucleus (\aa) collisions should respond to initial-state energy- and matter-density gradients by developing a velocity field (various flows) whose consequences may be observed in the hadronic final state~\cite{hydro}. Observation of flow manifestations and a causal relation to \aa\ initial-state geometry is sought {\em a priori} and interpreted to imply that a QGP has been established~\cite{perfect,qgp1,qgp2}.
Flow-QGP claims then rely on assignment of certain data features to flows and may refer to others as ``nonflow'' without further elaboration~\cite{nonflow}. The flow-QGP narrative forms the basis for other data analysis and interpretation, and MB jet contributions may be minimized by preferred analysis methods and interpretation strategies. 

The flow narrative is related to data by an assumption that most hadrons emerge from ``freezeout'' of a locally-thermalized, flowing dense medium~\cite{freezeout}.  Hadron \pt\ is divided into several intervals with specific physical interpretations: thermalization and flows for $p_t < 2$ GeV/c~\cite{starpidspec}, high-\pt\ jet phenomena for $p_t > 5$ GeV/c~\cite{starprl} and an intermediate region where production mechanisms are debated~\cite{recombo1,recombo2}. Such assumptions provide a preferred context for analysis and interpretation of high-energy data.

For example, transverse-momentum \pt\ spectra may be fitted with a monolithic model function interpreted to reflect a thermodynamic context and to measure radial flow~\cite{starpidspec}. Two-dimensional (2D) angular correlations are projected onto periodic 1D azimuth $\phi$ and represented by Fourier series wherein each Fourier term is interpreted to represent a type of transverse flow~\cite{luzum}. Charge and \pt\ fluctuations are addressed with statistical measures motivated by thermodynamic assumptions including some degree of local thermalization of a bulk medium. Intensive ratios or {\em ratios of ratios} are preferred over extensive measures of collision observables such as integrated charge multiplicity \nch\ or integrated $P_t$ within some angular domain~\cite{alicempt,aliceptfluct}. Jet contributions are acknowledged only within restricted \pt\ intervals including a small fraction of all hadrons, and  assumed jet properties are based on conjecture rather than actual jet measurements~\cite{raav21,starprl}.

In contrast, the properties of eventwise-reconstructed dijets, their fragment momentum distributions (fragmentation functions or FFs)~\cite{tasso,opal,eeprd} and jet (leading-parton) energy spectra~\cite{isrfirstjets,ua1jets,ua2jets} measured over thirty years predict quantitatively certain manifestations of MB dijets that appear in high-energy collision data~\cite{fragevo}. Predictions from {\em isolated-jet} measurements [as opposed to perturbative QCD (pQCD) theory with its limitations] are inconsistent with much of the flow narrative and its presumed basis in measurement as discussed below.

In  this study I examine the process of measure design in several critical areas and compare competing analysis methods. I demonstrate how measured properties of isolated jets predict certain data features from nuclear collisions corresponding to MB dijets and how analysis methods motivated by the flow narrative may lead to attribution of the same features to flows. I show how a two-component (soft + hard) model (TCM) of hadron production in high-energy collisions emerges naturally from inductive analysis of \pp\ spectrum and correlation data, is not imposed {\em a priori}, and how the TCM is manifested in other contexts. A number of examples are presented. I conclude that when data features are reexamined in the context of isolated-jet measurements little substantial evidence remains to support the flow narrative.

%%%%%%%%%%%%%%%%%%%%%%
This article is arranged as follows:
Section~\ref{meth} discusses preferred analysis methods.
Section~\ref{tcm1} introduces a two-component spectrum model for \pp\ collisions.
Section~\ref{jets} reviews the measured properties of reconstructed jets.
Section~\ref{jetppspec} describes MB jet contributions to \pp\ spectra.
Section~\ref{jetaaspec} presents MB jet contributions to \aa\ spectra.
Section~\ref{jetangcorr} reviews MB jet contributions to two-particle charge correlations.
Section~\ref{jetptfluct} presents MB jet contributions to \pt\ fluctuations and \pt\ angular correlations.
Sections~\ref{disc} and~\ref{summ}  present discussion and summary.

%%%%%%%%%
 \section{Preferred analysis methods} \label{meth}

For some aspects of data analysis alternative methods may give significantly different results and support different physical interpretations. The overall interpretation of high-energy nuclear collisions then depends on a sequence of method choices. How should such choices be made to establish an overall result that best reflects reality? One possible criterion is the fraction of information carried by primary particle data that is retained by a method for hypothesis testing. MB dijet manifestations in nuclear collisions compared to measured properties of isolated jets may provide a basis for evaluation.

\subsection{Primary and secondary observables}

All analysis methods are based on primary particle data. A charged-particle detector (e.g.\ time-projection chamber) determines the primary hadron single-particle (SP) observables in high-energy nuclear collisions: transverse momentum \pt, pseudorapidity $\eta$, azimuth angle $\phi$, charge sign and possibly hadron species via particle ID, in which case $\eta \rightarrow y_z$ (longitudinal rapidity) and  $p_t \rightarrow m_t = \sqrt{p_t^2 + m_h^2}$ (transverse mass) with $m_h$ a hadron mass. Transverse rapidity $y_t \equiv \ln[(p_t + m_t)/m_h]$ provides superior visual access to SP spectrum structure at lower \pt. For unidentified hadrons the pion mass may be assumed. SP densities are defined on \pt\ as \pt\ spectra and on $(\eta,\phi)$ as angular densities. 

Two-particle (pair) densities defined on 6D momentum space $(p_{t1},\eta_1,\phi_1,p_{t2},\eta_2,\phi_2)$ or a subspace may reveal certain pair correlations identified with physical mechanisms. 2D pair densities on $(x_1,x_2)$ may be {\em projected by averaging} onto difference variables $x_\Delta = x_1 - x_2$ to obtain a {\em joint angular autocorrelation} on the reduced space $(p_{t1},p_{t2},\eta_\Delta,\phi_\Delta)$~\cite{inverse} which can be further reduced by integration over \pt\ bins or the entire \pt\ acceptance.  In discussing 2D angular correlations it is convenient to separate azimuth difference $\phi_\Delta$ into two intervals: {\em same-side} (SS, $|\phi_\Delta| < \pi / 2$) and {\em away-side} (AS, $|\phi_\Delta - \pi| < \pi / 2$).

Charge multiplicity and particle \pt\ may be integrated over some angular acceptance $(\Delta \eta,\Delta \phi)$ or multiple bins within an angular acceptance to obtain \nch\ and $P_t$ as {\em extensive eventwise random variables} (RVs) whose fluctuations may be of interest~\cite{inverse,ptscale}. Uncorrected (observed) charge multiplicities denoted by $\hat n_{ch}$ are relevant to Poisson statistics, for instance in determining {\em void probabilities} defined below. \nch\ then denotes corrected values. 

Secondary observables may be defined as combinations of primary observables, for instance eventwise mean $\langle p_t \rangle = P_t / n_{ch}$ as an {\em intensive} RV~\cite{aliceptfluct}, event-ensemble-mean $\bar p_t = \bar P_t / \bar n_{ch}$ characterizing a collision system~\cite{alicempt}, spectrum ratio $R_{AA}$ as the ratio of a central \aa\ \pt\ spectrum to a \pp\ spectrum~\cite{starraa}, and $v_2(p_t)$ also as a ratio of distinct hadron spectra~\cite{quadspec2}. Fluctuation and pair-correlation measures (variances and covariances) may be combined with other statistics in sums, differences or ratios to define secondary statistics. SP and pair momentum spaces may be partitioned (possibly based on {\em a priori} assumptions), for instance defining certain \pt\ intervals within which specific physical mechanisms are expected to dominate collision data according to some narrative. 

\subsection{Competing analysis methods} \label{compete}

Analysis methods are generally not unique. A specific combination of methods contributing to a published analysis may comprise a subset of available methods determined by a sequence of choices among alternatives, possibly guided by a preferred narrative.  In a given context (e.g.\ SP spectra or pair angular correlations) alternative selections may lead to significantly different physical interpretations of collision data. 
For instance, each of extensive measures $P_t$ and $n_{ch}$ or their ensemble means $\bar P_t$ and $\bar n_{ch}$ may reveal certain data trends inconsistent with a temperature hypothesis but supporting an alternative hypothesis whereas fluctuations of intensive eventwise $\langle p_t \rangle$ or systematics of ensemble $\bar p_t$ may be seen as reflecting local temperature variations of a conjectured bulk medium. Critical extensive trends may be suppressed by cancellations within intensive ratios.

Hadron pair correlations from high-energy nuclear collisions projected onto 1D azimuth exhibit strong nonuniformities (literally ``azimuthal anisotropy'') that may originate from several physical mechanisms including MB dijets. However, the term ``anisotropic flow'' is commonly interpreted as synonymous with azimuthal anisotropy~\cite{ollitrault}. Any distribution on periodic azimuth, no matter what its physical origins, can be described exactly by a Fourier series (FS). The assertion ``The second Fourier coefficient of the azimuthal asymmetry [i.e.\ anisotropy] is called elliptic flow''~\cite{snellings} reflects a common assumption. Alternative modeling of azimuth distributions may favor a different physical interpretation.

This study emphasizes manifestations of MB dijets from high-energy nuclear collisions within several contexts (e.g.\ yields, spectra, correlations, statistical fluctuations) and their relation to selection of specific analysis methods: How are MB dijets, consistent with measured isolated-jet properties, revealed or concealed by method choices, and what criteria would insure conscious and unbiased choices that lead to meaningful interpretations?

%So jets are thereby defined to contribute to ``elliptic flow.'' Then there is ``nonflow.''

%Dramatic disconnect between advocates of 1D azimuth alone (the great majority) and those who analyze 2D angular correlations intact (a small minority). Another example of select abandonment of information in particle data.

%Getting rid of nonflow~\cite{nonflow}.

%``The characteristics of nonflow correlations are not well known and they cannot be calculated analytically as is done for flow correlations. These nonflow correlations are expected to exist mostly between few particles that are close to each other in pseudorapidity. In that respect they can be distinguished from flow correlations which extent over all particles independent of pseudorapidity.'' van der Kock thesis

%%%%%%%%%
 \section{Two-component spectrum model} \label{tcm1}

%Inductive inference of p-p spectrum structure

In Ref.~\cite{ppprd} a detailed analysis of \pt\ spectra was applied to ten multiplicity classes of 200 GeV \pp\ collisions. No {\em a priori} assumptions about spectrum structure were imposed. The main goal was to understand systematic variation of spectrum shape with \nch\ in terms of algebraic models {\em inferred from data alone}: given available spectrum data what is the {\em most efficient} algebraic description? For reasons given in Sec.~\ref{meth} this analysis is presented in terms of transverse rapidity \yt.

\subsection{Spectrum data and soft component}

\begin{figure}[h]
\includegraphics[width=1.66in,height=1.65in]{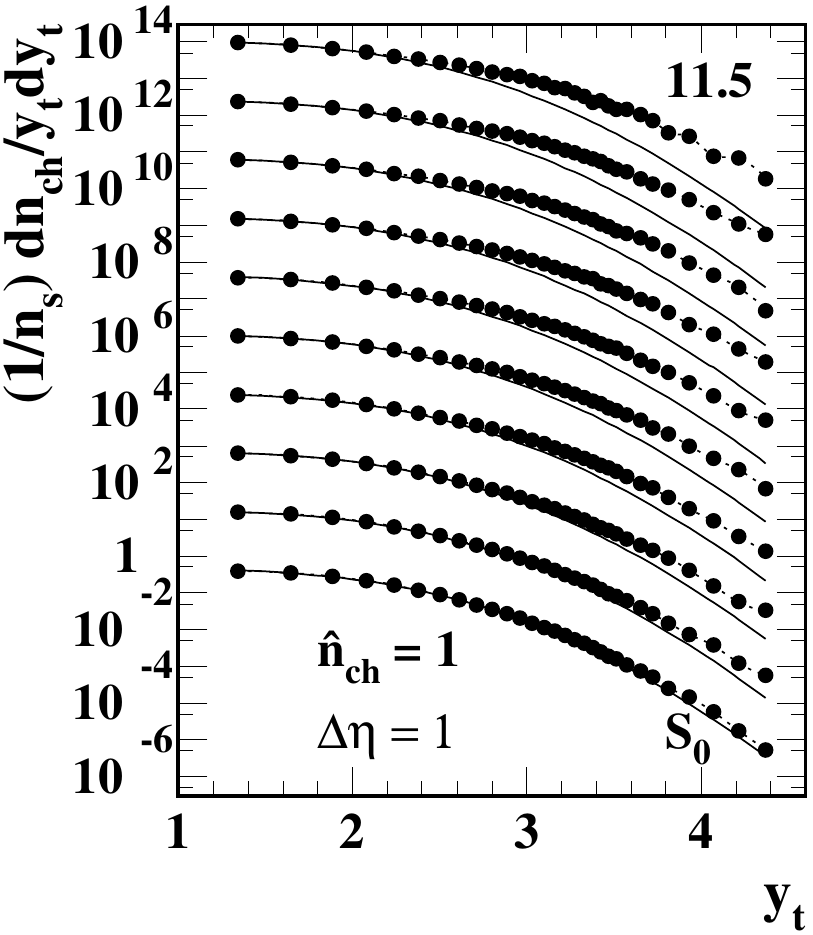}
\includegraphics[width=1.64in]{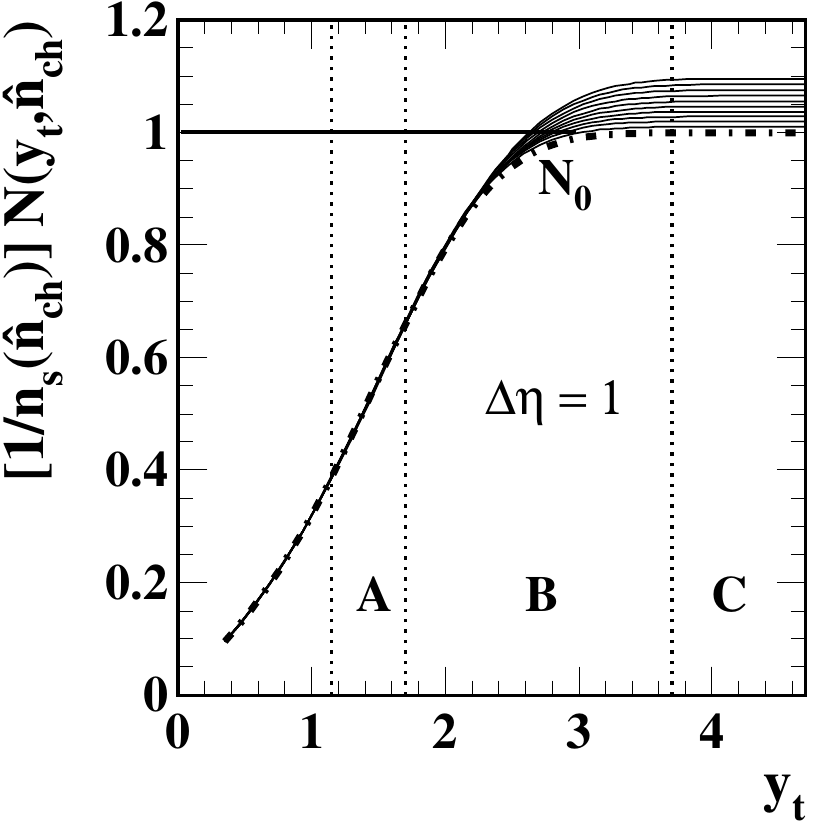}
\caption{\label{pp1}
Left: \yt\ spectra from ten charge-multiplicity classes of 200 GeV \pp\ collisions (points) compared to fixed reference $\hat S_0(y_t)$ (thin solid curves)~\cite{ppprd}.
Right: Running integrals $N(y_t)$ of normalized ( and extrapolated) $y_t$ spectra in the left panel for ten multiplicity classes (solid curves) compared to running integral $N_0(y_t)$ of fixed reference $\hat S_0(y_t)$ (dash-dotted curve).  
} % ppcomm14bb, integrals2xxx
\end{figure}

Figure~\ref{pp1} (left) shows \yt\ spectra for ten \pp\ multiplicity classes normalized by soft-component multiplicity $n_s$ (points) and displaced upward from each other by successive factors 40 relative to the lowest spectrum (the terms ``soft'' and ``hard'' are interpreted below)~\cite{ppprd}. Empirically, all spectra are observed to coincide at lower \yt\ if normalized by $n_s \approx n_{ch} - \alpha n_{ch}^2$ for some $\alpha \approx 0.01$. The definition of $n_s$ in terms of \nch\ is refined further in Sec.~\ref{spechard}.

%%%%%%%%%%%%%%%%%%%%%%%%%%%%%%%%%%
%%%%%%%%%%%%%%%%%%%%%%%%%%%%%%%%%%

Figure~\ref{pp1} (right) shows {\em running integrals} of the ten normalized spectra in the left panel. The spectra have been extrapolated to $y_t = 0$ (note that \yt\ spectra are nearly constant at lower \yt). The running integral is a means to enhance a long-wavelength signal (spectrum shape) over short-wavelength (statistical) noise. Different behavior is observed in each of \yt\ intervals A, B and C (spanning the detector acceptance). In interval A the integrals approximately coincide. In interval B the integrals diverge substantially. In interval C the integrals are nearly constant, and those constant values increase approximately linearly with \nch. The limiting case for $n_{s} \rightarrow 0$ is described by function $N_0(y_t)$ (dash-dotted curve) defined below. Those results indicate that the main \nch\ dependence lies within interval B as the running integral of a {\em peaked spectrum component} with amplitude $\propto n_{ch}$.

The limiting case  $N_0(y_t)$ is modeled by the running integral of a unit-normal L\'evy distribution on \mt~\cite{wilk}
\bea \label{s0}
\hat S_0(m_t) &=& \frac{A(T_0,n_0)}{[1 + (m_t - m_h) / n_0 T_0]^{n_0}},
\eea
where $T_0 \approx 145$ MeV controls the function mainly in interval A and $n_0 \approx 12.8$ controls the function mainly in interval C. The Jacobian factor from \mt\ to \yt\ is $p_t m_t / y_t$. The resulting $\hat S_0(y_t)$ model inferred directly from data trends can be subtracted to reveal the peaked spectrum (hard) component residing mainly within interval B.

\subsection{Spectrum hard component} \label{spechard}

\begin{figure}[h]
\includegraphics[keepaspectratio,width=1.63in]{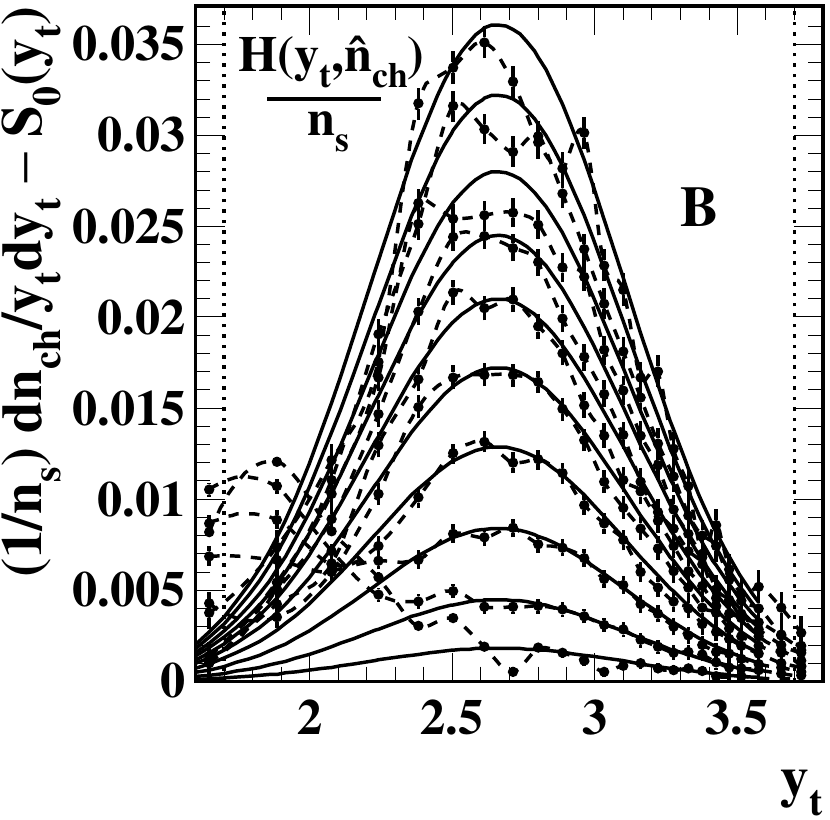}
\includegraphics[keepaspectratio,width=1.67in]{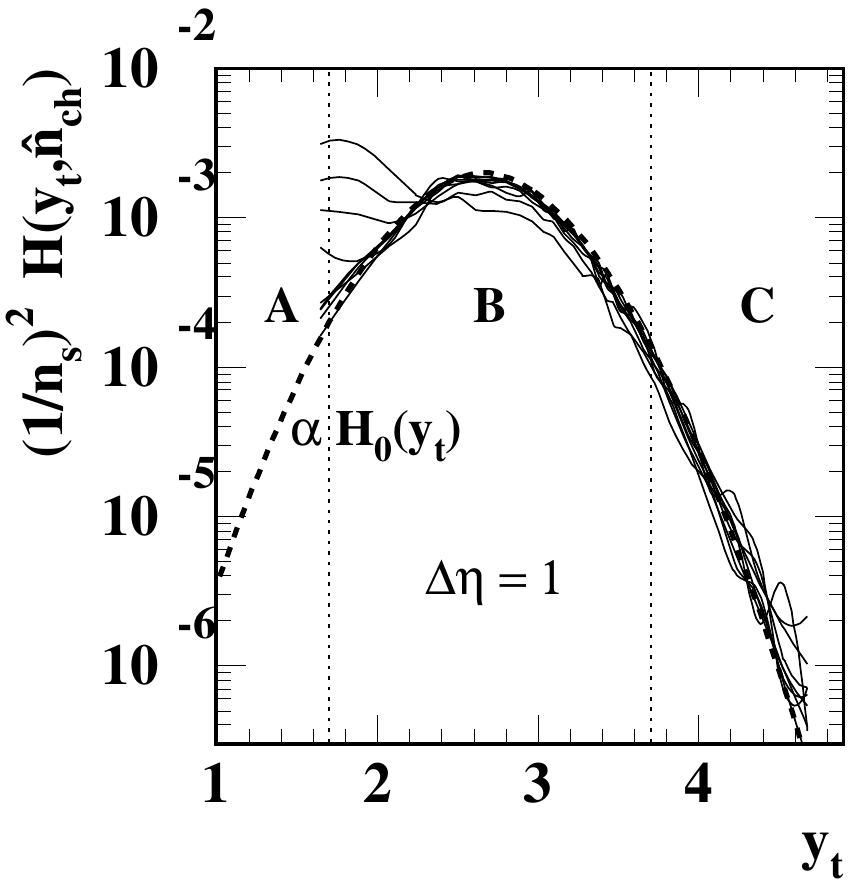}
\caption{\label{pp2}
Left: Normalized \yt\ spectra in Fig.~\ref{pp1} (left) minus unit-normal reference $\hat S_0(y_t)$ (points)~\cite{ppprd}. The vertical dotted lines enclose interval B previously defined. 
Right: Distributions $H(n_{ch},y_t)/n_s^2$ (solid curves, data in the left panel rescaled by $n_s$) compared to fixed reference $\alpha H_0(y_t)$ (dashed).  
} %  ppcomm12zzz,12bb
\end{figure}

Figure~\ref{pp2} (left) shows the normalized spectrum data in Fig.~\ref{pp1} (left) with fixed model $\hat S_0(y_t)$ inferred from Eq.~(\ref{s0}) subtracted to reveal peaked distributions (points and dashed curves) centered on interval B with amplitudes increasing approximately  $\propto n_{ch}$. With a few exceptions discussed below the distributions are well described by a two-parameter Gaussian function (solid curves). That result suggests that the peaked {\em hard component} $H(y_t,n_{ch})$ has the factorized form $H(y_t,n_{ch}) / n_s \propto n_{ch} \hat H_0(y_t)$.

%%%%%%%%%%%%%%%%%%%%%%%%%%%%%%%%%%
%%%%%%%%%%%%%%%%%%%%%%%%%%%%%%%%%%

Figure~\ref{pp2} (right) shows data in the left panel rescaled by soft-component multiplicity $n_s$ (rather than \nch) (solid) compared with a fixed Gaussian model in the form $\alpha \hat H_0(y_t)$ with centroid $\bar y_t \approx 2.7$ and width $\sigma_{y_t} \approx 0.45$ (dashed) and with coefficient $\alpha \approx 0.006$ determined by the data-model comparison. In summary, a TCM for \pp\ \yt\ spectra is inferred inductively from \pp\ spectrum data alone as
\bea \label{tcmeq}
\frac{dn_{ch}}{y_t dy_t} &=& S_{pp}(y_t,n_{ch}) + H_{pp}(y_t,n_{ch})
\\ \nonumber
&=& n_s(n_{ch}) \, \hat S_0(y_t) + n_h(n_{ch})\, \hat H_0(y_t),
\eea
where \nch\ and \yt\ trends have been factorized separately for the two spectrum components. There are two exceptions: (a) Significant systematic deviations from $\hat H_0(y_t)$ are observed for the lowest \nch\ classes. (b) Smaller systematic deviations are also observed for all \nch\ classes near the upper limit of the \yt\ acceptance. Both exceptions were reconsidered in Ref.~\cite{alicespec} and incorporated into an extended TCM describing spectrum data over the full range of collision energies (see Sec.~\ref{measurehard}).

More-detailed analysis~\cite{ppquad} shows that the relation $n_h = \alpha n_s^2$ is required by data, with $\alpha \approx 0.006$ for 200 GeV \pp\ collisions~\cite{alicespec} (and see Fig.~\ref{ppcomm}, left). Adding the condition $n_{ch} = n_s + n_h$ defines $n_s(n_{ch})$ and $n_h(n_{ch})$ in terms of {\em corrected} total multiplicity \nch\ (as opposed to {\em detected} $\hat n_{ch} \approx n_{ch}/2$). The resulting SP spectrum TCM with {\em fixed} parameters and functional forms describes \pp\ spectra accurately over an \nch\ interval corresponding to 10-fold increase in the soft component and {\em 100-fold} increase in the hard (dijet) component. There was no requirement for physical interpretation of the TCM components as the model was inferred from spectrum data.

\subsection{Alternative spectrum models} \label{altpp}

The TCM derived inductively from spectrum data requires five fixed parameters to describe \pp\ spectra for any event multiplicity within a large interval~\cite{ppquad,alicespec}. Alternative spectrum models could be proposed with free parameters determined for each event class by fits to data. One such model is the so-called ``power-law'' ($n$) or Tsallis ($q$) distribution that is similar in form to Eq.~(\ref{s0})
\bea \label{p}
P(x_t) &=& \frac{A}{[1 + x_t / n T]^{n}},
\eea
where $x_t$ is $p_t$~\cite{ua1spec,cywong} or $m_t$~\cite{tsallisblast} and $n \leftrightarrow 1/(q-1)$.

\begin{figure}[h]
\includegraphics[height=1.65in,width=3.3in]{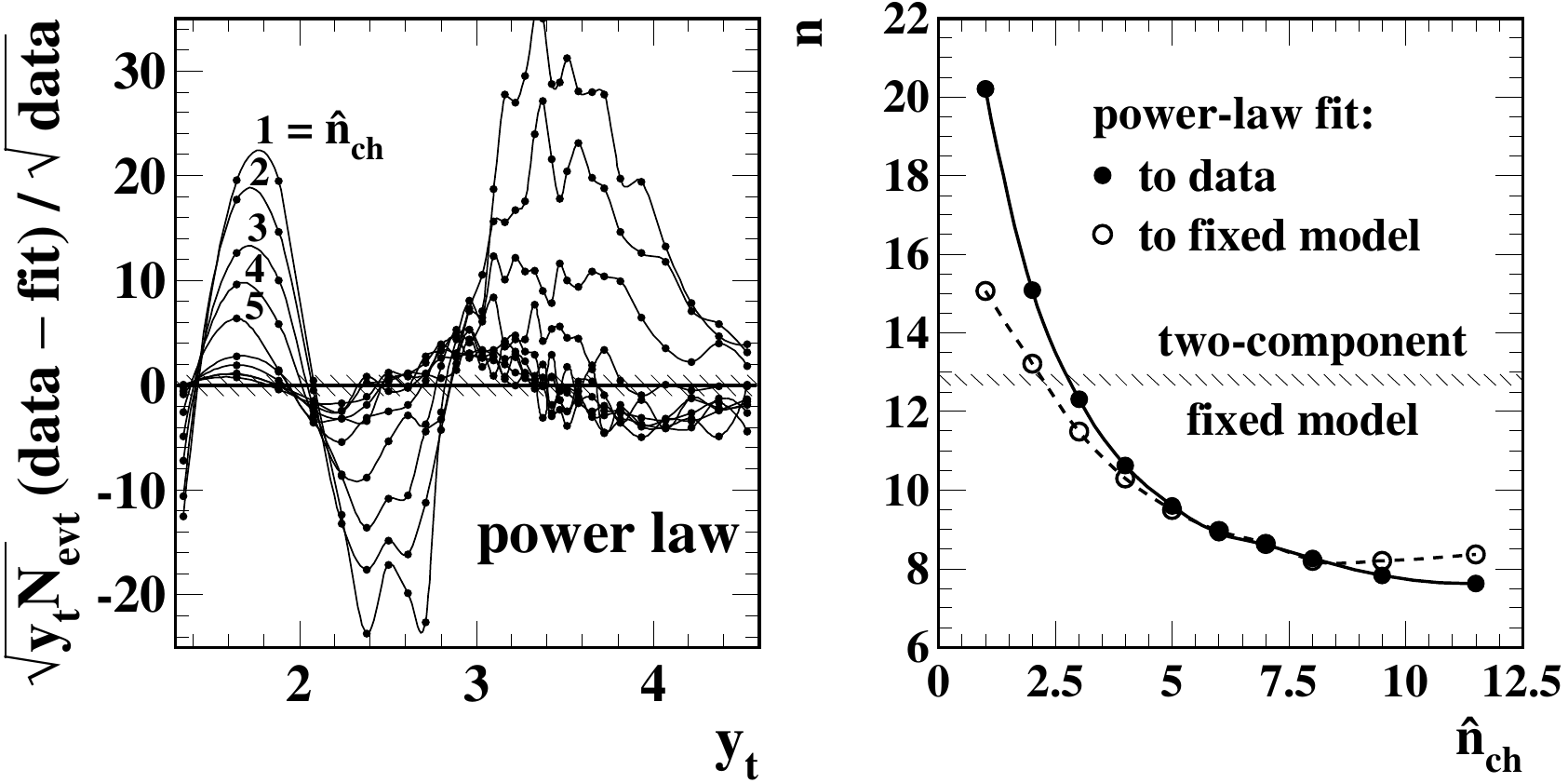}
\caption{\label{power}
Left: Relative residuals from power-law fits to $p_t$ spectra in Fig.~\ref{pp1}~\cite{ppprd}.  The hatched band represents the expected statistical errors. Right: Exponents $n$ from power-law fits to data (solid points) and to corresponding TCM fixed-model functions (open circles) compared to the fixed-model value $n_0 \approx 12.8$ (hatched band). 
} % pcomm16yyy
\end{figure}

Figure~\ref{power} shows the result of fitting the spectrum data from Fig.~\ref{pp1} (left) with Eq.~(\ref{p}). On the left are fit residuals for ten multiplicity classes {\em in units of statistical uncertainty} for each \yt\ value. That format differs from more-conventional presentations in terms of data/model ratios that strongly suppress residuals at lower \yt. Especially for lower event multiplicities those fits should be rejected given standard criteria. On the right  are values for parameter $n$ derived from fitting Eq.~(\ref{p}) to spectrum data (solid points) and to the TCM described above (open points) compared to the fixed value $n_0 = 12.8$ for the TCM itself. The variation descends from values exceeding the TCM soft-component value for lower \nch\ to values consistent with the jet-related TCM hard component for larger \nch\ as dijet production increases $\propto \hat n_{ch}^2$ per Ref.~\cite{ppquad}.

%%%%%%%%%%%%%%%%%%%%%%%%%%%%%%%%%%
%%%%%%%%%%%%%%%%%%%%%%%%%%%%%%%%%%

It could be argued that the TCM for \pt\ spectra is simply one of several competing spectrum models and that the power-law model should be preferred as requiring only two parameters compared to five for the TCM. However,  the TCM is not a simple fitting exercise applied to a single data spectrum. Ref.~\cite{ppprd} established that a spectrum TCM inferred inductively from the \nch\ dependence of \pt\ spectra without any {\em a priori} assumptions is {\em necessary} to describe consistently an ensemble of high-statistics \pt\ spectra over a large \nch\ range (e.g.\ factor 10). Separation of \pp\ spectrum {\em data} into two components does not depend on an imposed model. Specific model functions were introduced only {\em after} resolution of data spectra into two components based on  \nch\ trends.

In contrast, what emerges from power-law fits is twenty parameter values to describe ten multiplicity classes as opposed to five for the TCM. The large penalty for poor fits in Figure~\ref{power} (left)  is even more determining. The dramatic variation of power-law parameter $n$ in the right panel has no {\em a priori} explanation, whereas in a TCM context the variation occurs because an inappropriate (soft) model attempts to accommodate quadratic increase of the jet-related (hard) spectrum component with \nch.

%%%%%%%%%
 \section{Eventwise-reconstructed jets} \label{jets}

Jets may be reconstructed eventwise from final-state hadrons within the full hadron momentum space including scalar momentum and angular correlations relative to an inferred leading-parton four momentum. This section emphasizes scalar-momentum dependence of reconstructed jets and jet fragments. A joint density distribution on parton (jet) energy $E_{jet}$ and scalar fragment momentum $p_{frag}$ can be factorized as $P(E_{jet},p_{frag}) = P(E_{jet}) P(p_{frag}|E_{jet})$, where conditional distribution $P(p_{frag}|E_{jet}) \rightarrow D_{p}^h(p_{frag}|E_{jet})$ is a {\em fragmentation function} (FF) for parton type $p$ fragmenting to hadron type $h$, and $P(E_{jet}) \rightarrow d\sigma_j/dE_{jet}$ is the jet energy spectrum for a given dijet source (e.g.\ \ppbar\ collisions). In this section simple and accurate parametrizations of isolated-jet data for \ee\ and \ppbar\ collision systems are based on logarithmic rapidity variables, with $E_{jet} \leftrightarrow p_t$ to accommodate some conventional notation.

\subsection{p-p jet (scattered-parton) energy spectra}

A QCD-related energy dependence is typically of the form $\log(Q/Q_0)$ where $Q_0$ represents some characteristic energy scale. A description of fragmentation functions in terms of rapidity variable $y = \ln[(p  + E) / m_h] \approx \ln(2p/m_h)$ as in  Ref.~\cite{eeprd} is presented in the next subsection.
A jet spectrum near midrapidity for \pp\ collision energy $\sqrt{s}$ can be written in terms of a jet ``rapidity'' by
 \bea \label{curious2}
p_t \frac{d^2 \sigma_j}{dp_t d\eta} &=&  \frac{d^2\sigma_j}{dy_{max} d\eta},
 \eea
where  $y_{max} \equiv \ln(2 E_{jet} / m_\pi)$ was first defined in Ref.~\cite{eeprd} in connection with fragmentation functions as summarized in the next subsection and $E_{jet} \rightarrow p_t$ as noted above.

Study of jet-related yields, spectra and angular correlations  in 200 GeV \pp\ collisions reveals that the jet-related (hard-component) density $dn_h/d\eta$ (and presumably jet production $dn_j/d\eta$) scales with the soft-component density as $dn_h/d\eta \propto (dn_s / d\eta)^2$. 
Given that relation and  $dn_s / d\eta \propto \log(s/s_0) \equiv 2\Delta y_b$ (with $\sqrt{s_0} \approx 10$ GeV) near midrapidity~\cite{alicespec} the number of MB dijets (dominated by lowest-energy jets) appearing near midrapidity in \pp\ collisions should vary with collision energy as~\cite{alicespec,jetspec2,ppquad}.
\bea
 \frac{d^2\sigma_j}{dy_{max} d\eta} &\propto& \Delta y_b^2~~~\text{near some $E_{min}$}.
\eea

Kinematic constraints impose the upper limit $2E_{jet} < \sqrt{s}$ ($y_{max} < y_{b}$) with $y_{b} = \ln(\sqrt{s} / m_\pi)$. Evidence from jet \cite{ua1jets} and SP-hadron~\cite{fragevo} spectra suggests a lower limit  $E_{jet} > E_{min}$ ($y_{max} > y_{min}$) with $E_{min} \approx 3$ GeV. A normalized {\em jet} rapidity variable can then be defined by
\bea
u &=& \frac{y_{max} - y_{min}}{y_{b} - y_{min}} =  \frac{\log(E_{jet} / E_{min})}{\log(\sqrt{s} / 2 E_{min})} \in [0,1].
\eea

\begin{figure}[h]
     \includegraphics[width=1.67in]{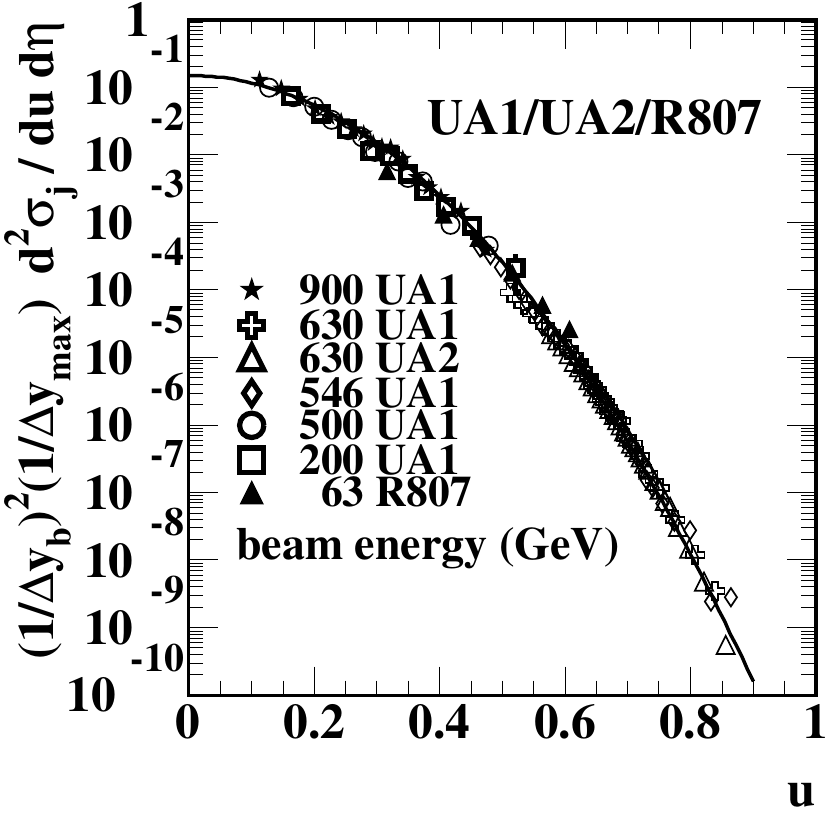}
    \includegraphics[width=1.63in]{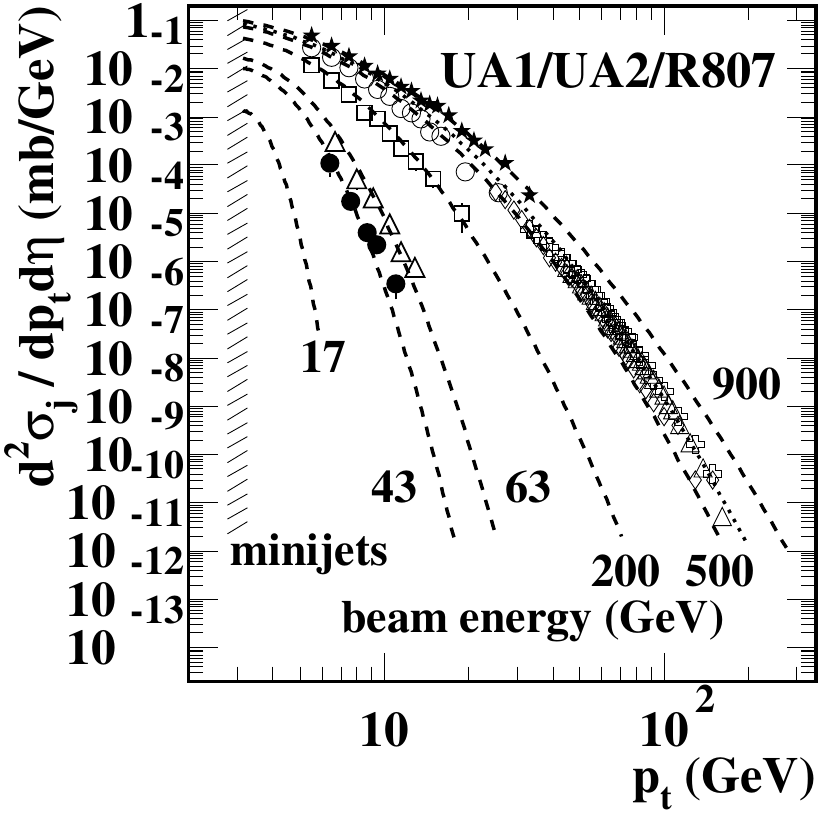}
   \caption{\label{rescale}
   Left: Jet energy spectra from \ppbar\ collisions for several energies~\cite{isrfirstjets,ua1jets,ua2jets} rescaled by factor $\Delta y_b^2$ and plotted vs normalized jet rapidity $u$. Within uncertainties the data fall on the common locus $0.15 \exp(-25 u^2)$. 
 Right: The same data plotted in a conventional log-log format. Dashed and dotted curves through the data are derived from the universal trend (solid curve) in the left panel as described in the text. 
 } %  aleph11l4, 11l5
    \end{figure}

Figure~\ref{rescale} (left) shows ISR and Sp\=pS jet spectrum data (points) with the jet spectra rescaled vertically by factor $(\Delta y_{b})^2$ and parton rapidity $y_{max}$ rescaled horizontally to $u$ assuming $E_{min} \approx 3$ GeV. 
All spectrum data for \ppbar\ collision energies below 1 TeV fall on the common locus $0.15 \exp(-u^2/2\sigma_u^2)$ (solid curve). 
The parametrized parton spectrum conditional on \pp\ beam energy is then 
 \bea \label{curious}
 \frac{d^2\sigma_j}{dy_{max} d\eta} &=& 0.026 \Delta y_b^2  \frac{1}{\sqrt{2\pi \sigma^2_u}} e^{-u^2 / 2 \sigma^2_u},
 \eea
where $0.026/\sqrt{2\pi \sigma^2_u} = 0.15$ and $\sigma_u \approx 1/7$ is determined empirically from the  data. All jet production over nine decades is then represented by parameters $\sqrt{s_0} \approx 10$ GeV, $E_{min} \approx 3$ GeV and $\sigma_u \approx 1/7$. Endpoints $\sqrt{s_0}$ and $E_{min}$ are closely related by kinematic constraints on fragmentation to charged hadrons from low-$x$ gluons~\cite{jetspec2}.

%%%%%%%%%%
   %%%%%%%%%%%%

Figure~\ref{rescale} (right) shows ISR and Sp\=pS spectrum data from the left panel plotted vs  $p_t \leftrightarrow E_{jet}$ in a conventional log-log format. The curve for each beam energy is defined by Eq.~(\ref{curious}). The dotted curve corresponds to $\sqrt{s} = 630$ GeV~\cite{ua2jets}. The model curves extend to $u = 0.9$ corresponding to partons with momentum fraction $x = 2E_{jet} / \sqrt{s} \approx 2/3$ beyond which the \pp\ collision-energy constraint should strongly influence the spectra.

%Systematic analysis of available jet production data leads to a simple parametrization based on parameters $y_{b0} \equiv \ln(Q_0 / 0.14 )$ with $Q_0 \approx 10$ GeV and $y_{m0} =  \ln(2E_{cut} / 0.14 )$.  

%We then define $\Delta y_b = y_b - y_{b0}$ and $\Delta y_{max} = y_{b} - y_{m0}$, with normalized parton rapidity $u =  (y_{max} - y_{m0}) /\Delta y_{max}$. 

% Beam rapidity $y_b \equiv \ln(\sqrt{s} / \text{0.14 GeV})$, $y_{b0} \equiv \ln(2 \times 9 / 0.14 ) = 4.86$, $y_{m0} =  \ln(2\times 3 / 0.14 ) = 3.76$

%with beam rapidity $y_{b}$ defined relative to pion mass as $y_b = \ln(\sqrt{s} / \text{0.14 GeV})$. 

\subsection{Fragmentation functions} \label{fragfunc}

Dijet formation depends on parton energy scale $Q = 2E_\text{jet}$ with rapidity $y = \ln[(E + p)/m_\pi]$ (for an unidentified hadron fragment with scalar momentum $p$) and maximum rapidity $y_{max}$ as defined below Eq.~(\ref{curious2}) to describe \ee\ FFs with $D(y|y_{max}) = 2dn_{ch,j}/dy$, the fragment rapidity density per dijet.  
The FF parametrization is $D(y|y_{max}) = 2 n_{ch,j}(y_{max}) \beta(u;p,q)/y_{max}$, where $\beta(u;p,q)$ is a unit-normal (on $u$) beta distribution, $u = (y - y_{min}) / (y_{max} - y_{min}) \in [0,1]$ is a normalized fragment rapidity, and parameters $p$ and $q$ (for each parton-hadron combination) are nearly constant over a large jet energy interval~\cite{eeprd}. The total fragment multiplicity $2 n_{ch,j}(y_{max})$ (from two jets) is inferred from the shape of $\beta(u;p,q)$ (and therefore parameters $p$ and $q$) via parton (jet) energy conservation. 
 
\begin{figure}[h]
   \includegraphics[width=1.65in,height=1.6in]{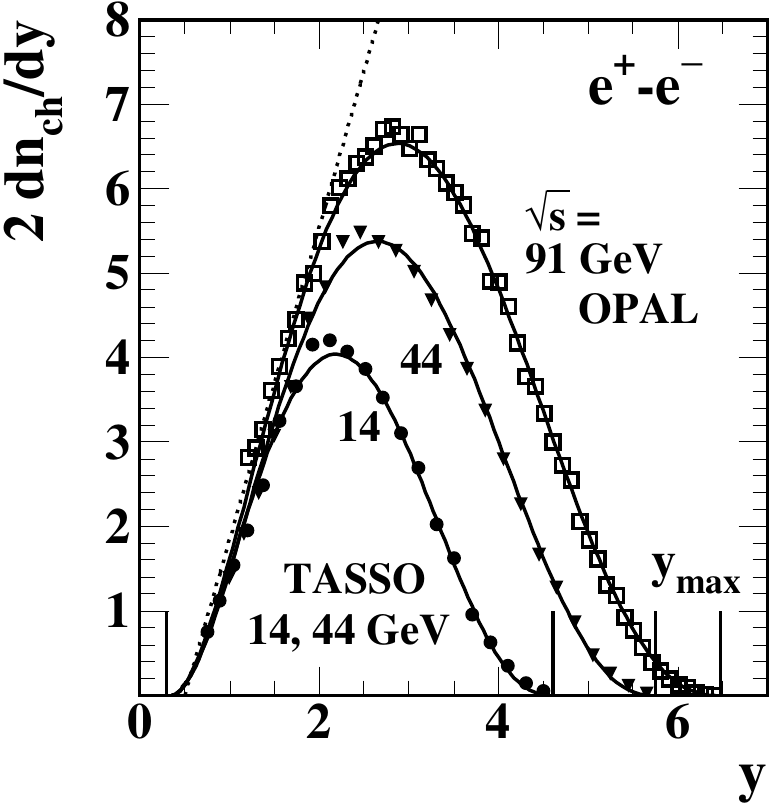}
  \includegraphics[width=1.65in,height=1.6in]{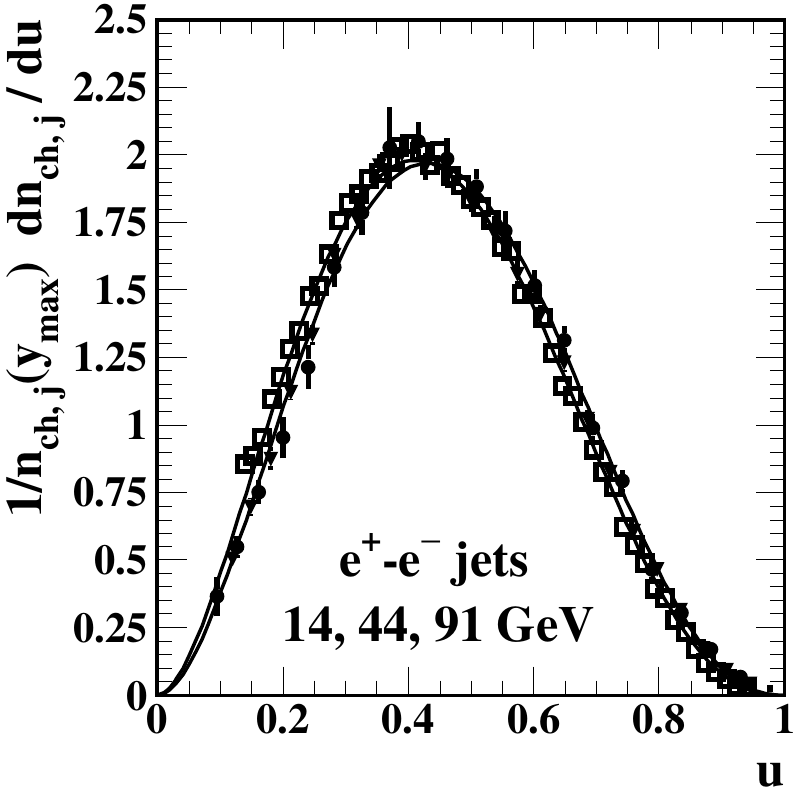}
 \caption{\label{ffs}
 Left: Fragmentation functions for three dijet energies from \ee\ collisions~\cite{tasso,opal} vs hadron fragment rapidity $y$ as in Ref. \cite{eeprd} showing self-similar evolution with parton rapidity $y_{max}$.
 Right: The same data rescaled to unit-normal distributions on normalized rapidity $u$. There is a barely significant evolution with parton energy. The rescaling result provides the basis for a simple and accurate parametrization.
  } %fragment3c,9a
  \end{figure}

 Figure~\ref{ffs} (left) shows measured FFs (points) for three dijet energies~\cite{tasso,opal} extending down to very low fragment momentum (less than 100 MeV/c). When plotted on fragment rapidity $y$ the FFs show a self-similar evolution with maximum rapidity $y_{max}$. The solid curves show the corresponding FF parametrization developed in Ref.~\cite{eeprd} based on the beta distribution as noted above.
 
 %%%%%%%%%%
 %%%%%%%%%%%%
 
 Figure~\ref{ffs} (right) shows the self-similar data in the left panel rescaled to unit integral and plotted on scaled fragment rapidity $u$ with $y_{min} \approx 0.35 $ ($p \approx 50$ MeV/c). The solid curves are corresponding beta distributions with parameters $p$ and $q$ nearly constant over a large jet energy interval. The simple two-parameter description is accurate to a few percent within the $E_{jet}$ interval 3 GeV ($y_{max} \approx 3.75$) to 100 GeV ($y_{max} \approx 7.25$)~\cite{eeprd}.  FF data for light-quark and gluon jets are parametrized separately but the parametrizations for gluon and light-quark jets converge near $E_{jet} = $ 3 GeV.
All minimum-bias jet fragment production can be described with a few universal parameters via introduction of logarithmic rapidities.

% \subsection{Comparing $\bf e^+$-$\bf e^-$ and p-\= p parton fragmentation}

FFs for isolated dijets derived from \ee\ collisions are quite different from FFs derived from \ppbar\ or \pp\ collisions as noted below. Minor differences might arise from alternative jet reconstruction algorithms, but larger observed differences suggest that the concept of universality may not apply to FFs from distinct collision systems.

\begin{figure}[h]
   \includegraphics[width=1.65in,height=1.63in]{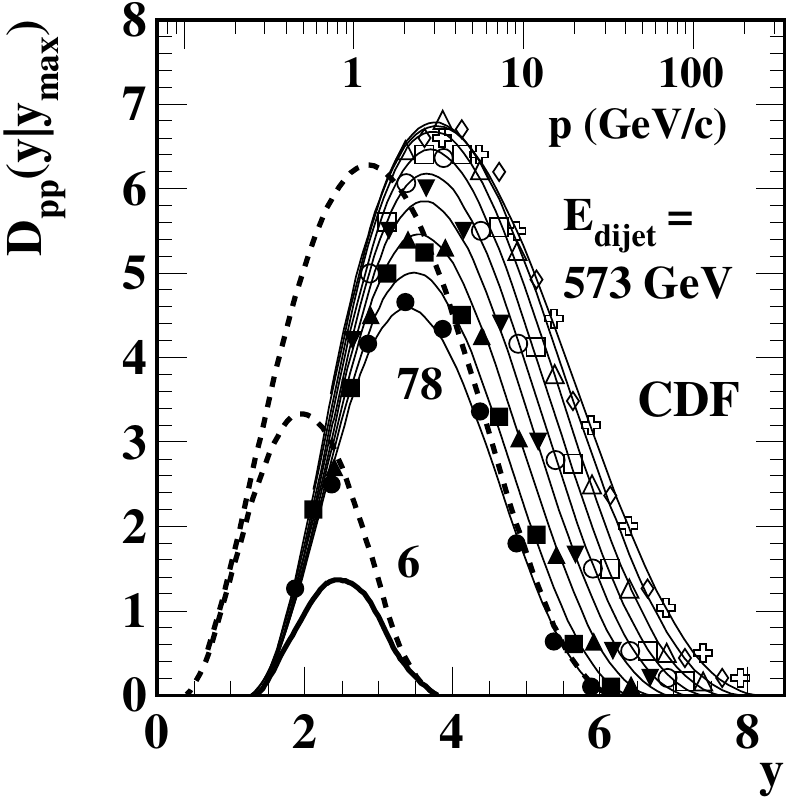}
 \includegraphics[width=1.65in,height=1.6in]{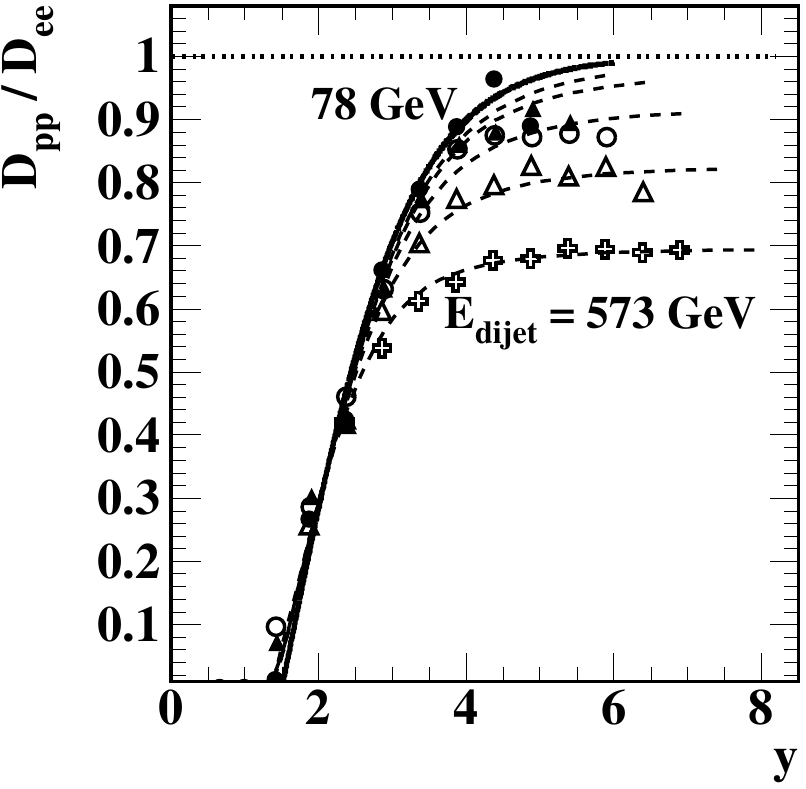}
\caption{\label{ppffs}
Left: Fragmentation functions for several dijet energies (points) from \ppbar\ collisions at 1.8 TeV~\cite{cdfff}. The solid curves represent a \ppbar\ parametrization derived from the \ee\ parametrization in~\cite{eeprd}. The dashed curves show the \ee\ parametrization itself for two energies for comparison.
Right: The ratio of \ppbar\ FFs $D_{pp}$ to corresponding \ee\ parametrizations $D_{ee}$ vs fragment rapidity showing systematic differences: a common strong suppression below $y = 4$ ($p \approx 4$ GeV/c) for all parton energies and substantial reduction at larger fragment rapidities for parton energies with $E_{dijet} > 80$ GeV.
 } %aleph17aa,aleph17b
 \end{figure}

Figure~\ref{ppffs} (left) shows FFs for ten dijet energies from 78 to 573 GeV inferred from 1.8 TeV \ppbar\ collisions (points) using eventwise jet reconstruction~\cite{cdfff}.  The solid curves are a parametrization. Comparison with the  \ee\ FF data in Fig.~\ref{ffs} (left) (e.g.\ dashed curves in this panel for $2E_{jet} =  6$ and 91 GeV) reveals that a substantial portion of \ee\ dijet FFs at lower fragment momenta may be missing from reconstructed \ppbar\ FFs. This comparison is dominated by quark jets, but quark and gluon FFs converge near $E_{jet} \approx 3$ GeV where most MB jets appear.

%%%%%%%%%%
%%%%%%%%%%%%

Figure~\ref{ppffs} (right) shows the ratio of \ppbar\ FF data in the left panel to the \ee\ FF parametrization for each jet energy (points), revealing systematic differences. The solid curve is $\tanh[(y-1.5)/1.7]$ which describes measured  \ppbar\ FFs relative to \ee\ FFs for dijet energies below 70 GeV. FF parametrizations for quark and gluon jets used for \pp\ collisions in the present study are the \ee\ parametrizations of Ref.~\cite{eeprd} and Fig.~\ref{ffs} multiplied by the same tanh factor for both quark and gluon FFs.

%%%%%%%%%
 \section{Jet contributions to $\bf p$-$\bf p$ spectra} \label{jetppspec}

The MB dijet contribution to \pt\ spectra and other momentum-space measures should consist of all hadron fragments from all dijets emerging from a given collision system and appearing within a certain angular and \pt\ acceptance. Given the jet spectrum and FF parametrizations in the previous section the resulting MB fragment distribution can be obtained from a convolution integral.

\subsection{Minimum-bias fragment distributions}

The midrapidity $\eta$ density of MB dijets from non-single-diffractive (NSD) \pp\ collision is estimated by
\bea
f_{NSD} &=&  \frac{1}{ \sigma_{NSD}} \frac{d\sigma_j}{d\eta} ~~\text{$\approx 0.028$ at 200 GeV}
\eea
given  $\sigma_{NSD} \approx 36.5$ mb~\cite{ua5} and $d\sigma_j/d\eta \approx 1$ mb~\cite{ua1jets,jetspec2} at that energy. The ensemble-mean {\em fragment distribution} for MB dijets is defined by the convolution integral
\bea \label{fold1}
\bar D_u(y) &\approx&   \frac{1}{d\sigma_j/d\eta}  \int_0^\infty \hspace{-.07in}  dy_{max}\, D_\text{pp}(y|y_{max})\, \frac{d^2\sigma_j}{dy_{max}d\eta},~~
\eea
where subscript $u$ denotes unidentified-hadron fragments.
Given that a spectrum hard component $H(y_t)$ represents hadron fragments from MB dijets it can be expressed as $y_t H(y_t) \approx \epsilon\, f_{NSD}\bar D_u(y)$, where $\epsilon(\Delta \eta,\Delta \eta_{4\pi}) \in [0.5,1]$ is the average fraction of a dijet appearing in detector acceptance $\Delta \eta$ compared to effective $4\pi$ acceptance $\Delta \eta_{4\pi}$ (which depends on collision energy). At 200 GeV $\epsilon \approx 0.6$ within detector acceptance $\Delta \eta = 2$. 
The spectrum hard component so defined represents the fragment contribution from MB  scattered parton pairs into acceptance $\Delta \eta$. It is assumed that for midrapidity jets $y_t \approx y$ (i.e.\ collinearity) except for low-momentum fragments (e.g.\ $p_t < 0.5$ GeV/c).

\begin{figure}[h]
\includegraphics[width=1.65in,height=1.66in]{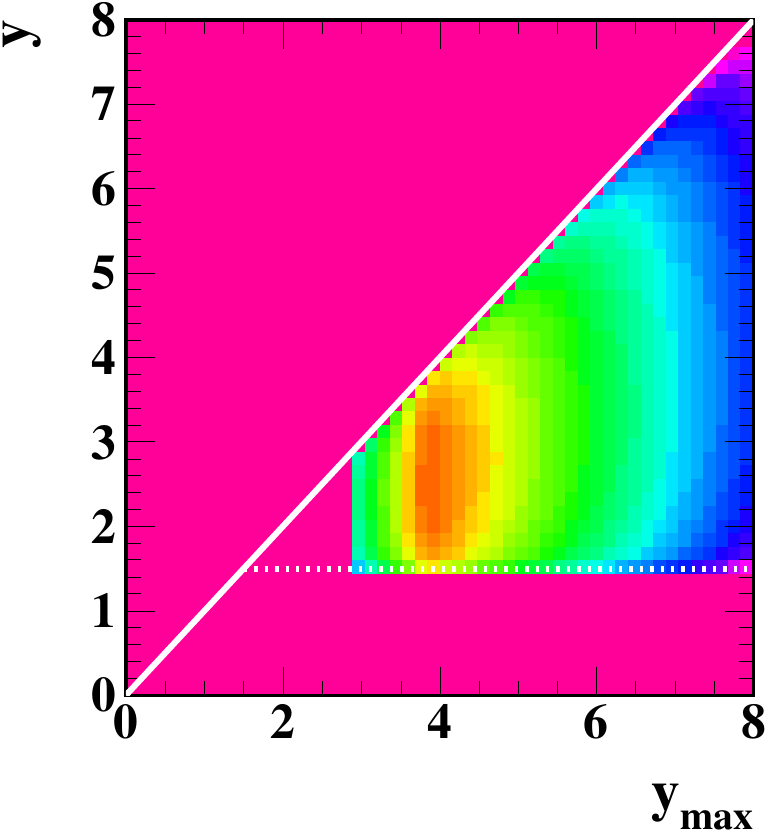}
  \includegraphics[width=1.65in,height=1.64in]{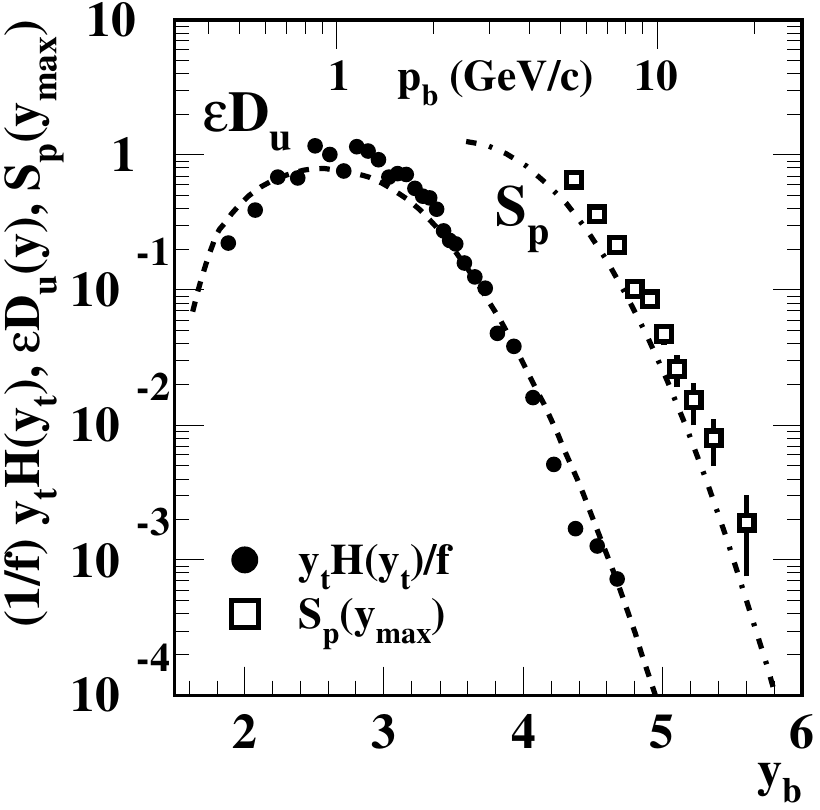}
\caption{\label{ppfd} %fig3
Left: (Color online) Argument of the pQCD convolution integral on $(y,y_{max})$ based on in-vacuum p-\=p FFs~\cite{fragevo}. The z axis is logarithmic.
Right: The spectrum hard component for 200 GeV NSD \pp\ collisions~\cite{ppprd} in the form $y_t H(y_t) / f_{NSD} \epsilon$ (solid points) compared to calculated mean fragment distribution $\bar D_u(y)$ (dashed) with $y_b = y_t$, $y$ or $y_{max}$~\cite{fragevo}.  The jet spectrum that generated $\bar D_u(y)$ is defined by Eq.~(\ref{curious}) rescaled by factor 2/3 (dash-dotted curve). The open boxes are 200 GeV \pp\ jet-spectrum data~\cite{ua1jets}.
} % aleph11cpp, 11kppqspec %11dpp
\end{figure}

Fig.~\ref{ppfd} (left) shows a surface plot of the Eq.~(\ref{fold1}) integrand---$D_\text{pp}(y|y_{max})\, d^2\sigma_j/dy_{max} d\eta$---incorporating \ppbar\ FFs from Fig.~\ref{ppffs} (left) and the 200 GeV jet \pt\ ($\rightarrow y_{max}$) spectrum from Fig.~\ref{rescale} (right).   The \ppbar\ FFs are bounded below by $y_{min} \approx 1.5$ ($p \approx 0.3$ GeV/c). The jet-spectrum effective lower bound is $E_{min} \approx 3$ GeV. The $z$ axis is logarithmic to show all distribution structure.

%%%%%%%%%%%%%%%%%%%%%%%%%%%%%%%%%%
%%%%%%%%%%%%%%%%%%%%%%%%%%%%%%%%%%

Fig.~\ref{ppfd} (right) shows the corresponding mean fragment distribution $\epsilon \bar D_{u}(y)$ as a projection (dashed) described by Eq.~(\ref{fold1}) and compared to hard-component data from 200 GeV NSD \pp\ collisions (solid points~\cite{ppprd,ppquad}) in the form $y_t H(y_t) / f_{NSD} $ corresponding to their relation in the text just below Eq.~(\ref{fold1}).  The open boxes are 200 GeV \ppbar\ jet-spectrum data from Ref.~\cite{ua1jets}. The dash-dotted curve $S_p$ is Eq.~(\ref{curious}) reduced by factor 2/3 so $\bar D_{u}(y)$ from Eq.~(\ref{fold1}) (dashed) best accommodates the \pp\ SP spectrum data (solid points). 

The apparent difference between calculated and measured jet spectra falls within the systematic uncertainties of Ref.~\cite{ua1jets}. On the other hand the FF parametrization from Fig.~\ref{ppffs} extrapolated down to $E_{jet} < 10$ GeV ($y_{max} < 5$) may overestimate the fragment yield there substantially. In any case the measured spectrum  hard component from 200 GeV NSD \pp\ collisions is quantitatively consistent with a dijet contribution derived from eventwise-reconstructed jets~\cite{ua1jets,cdfjets,jetspec2}. Given the spectrum hard-component mode near $p_t = 1$ GeV/c this comparison also establishes that a parton (jet) spectrum effective lower limit $E_{min} = 3.0 \pm 0.2$ GeV is required by 200 GeV \pp\ spectrum data. Extrapolating the ($\approx$ power-law) jet spectrum significantly below 3 GeV (e.g.\ as in some Monte Carlos) would result in a large overestimate of $H(y_t)$ for fragment momenta below 1 GeV/c.

\subsection{Measured spectrum hard components} \label{measurehard}

The comparison in Fig.~\ref{ppfd} provides convincing evidence from a specific 200 GeV NSD \pp\ SP spectrum that its TCM hard component does indeed represent a MB jet fragment distribution. A more-recent study reveals what can be learned from \nch\ and collision-energy dependence.

\begin{figure}[h]
  \includegraphics[width=1.67in]{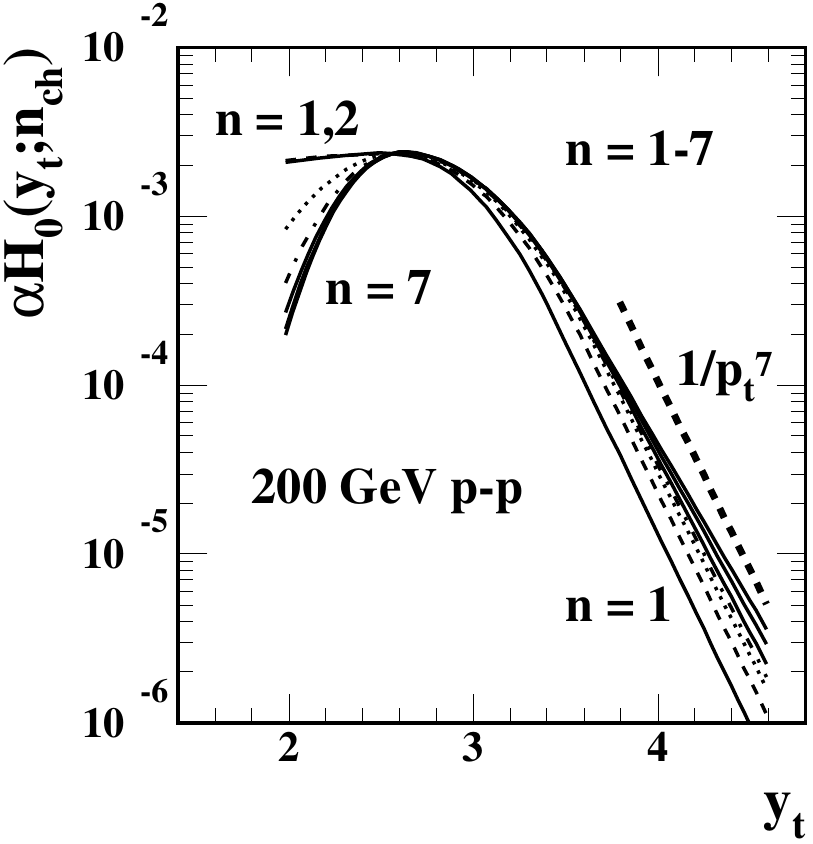}
  \includegraphics[width=1.63in]{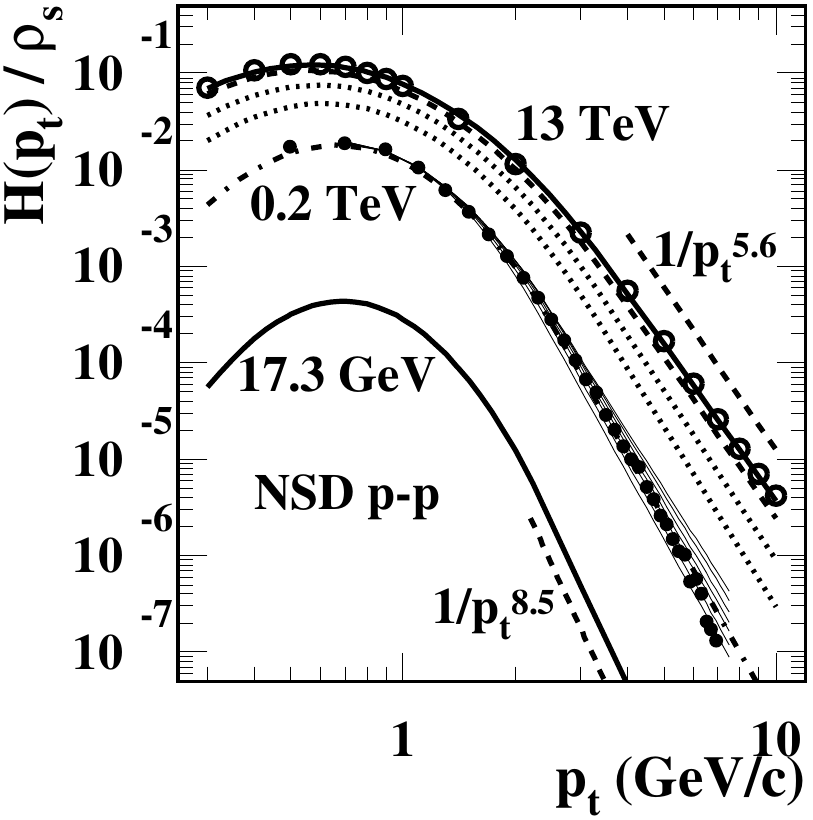}
\caption{\label{enrat3}
Left: 
Evolution of the hard-component model over seven multiplicity classes of 200 GeV \pp\ collisions~\cite{alicespec} that exhausts all information in high-statistics spectrum data from Ref.~\cite{ppquad}. Right: 
A survey of spectrum hard components for NSD \pp\ collisions over the currently accessible energy range from threshold of dijet production (10 GeV) to LHC top energy (13 TeV). The curves are defined by parameters from Table III of Ref.~\cite{alicespec} except for the 200 GeV fine solid curves obtained from the TCM curves on \yt\ in the left panel. The points are from Refs.~\cite{ppquad} (200 GeV) and \cite{alicespec} (13 TeV).
} % ppcms110g8x, alice125dnch
\end{figure}

Figure~\ref{enrat3} (left) shows the \nch\ evolution of a revised TCM spectrum hard-component model as derived in Ref.~\cite{alicespec} based on recent high-statistics spectrum data from Ref.~\cite{ppquad}. The variation below the hard-component mode near $y_t = 2.7$ is already apparent in the early data of Ref.~\cite{ppprd} shown in Fig.~\ref{pp2} (right). Whereas the model shape was initially held fixed independent of \nch\ to retain simplicity, the revised TCM of Ref.~\cite{alicespec} accommodates all significant variation of spectrum data. Shape variations below and above the mode are evidently tightly correlated.  Interpreted in a jet-related context the revised model suggests that as larger event multiplicities are required the underlying jet spectrum is biased to more-energetic jets with larger fragment multiplicities.

It is important to note  that while the \pp\  TCM hard component was initially modeled by a simple Gaussian on \yt~\cite{ppprd} subsequent detailed analysis of \pp\ and \aa\ spectrum data combined~\cite{hardspec} revealed that an exponential tail for the Gaussian on \yt\ is required for the hard-component model. The corresponding power-law trend on \pt\ required by data for 200 GeV \pp\ collisions is then $\approx 1/p_t^7$ as indicated by the dashed line in Figure~\ref{enrat3} (left). The power-law exponent evolution with collision energy in Figure~\ref{enrat3} (right) ~\cite{alicespec} is compatible with jet spectrum measurements~\cite{jetspec2}. In contrast, the ``power law'' of soft component $\hat S_0(m_t)$ corresponding to $\approx 1/p_t^{13}$ at 200 GeV appears to be unrelated to jet physics.

%%%%%%%%%%
%%%%%%%%%%%%

Figure~\ref{enrat3} (right) shows spectrum hard components for a range of \pp\ collision energies from SPS to top LHC energies based on the analysis in Ref.~\cite{alicespec}. 
%The distribution mode does not change significantly with energy suggesting that the jet (scattered-parton) spectrum cutoff $E_{min} \approx 3$ GeV does not change significantly with collision energy. 
The main variation is reduction of the power-law exponent describing the distribution high-\pt\ tail (note dashed lines at right) which can be compared with the jet-spectrum evolution shown in Fig.~\ref{rescale} (right).  The close correspondence between TCM spectrum hard components and jet spectra provides additional support for interpretation of the  hard component as a MB jet fragment distribution.

\subsection{Hadron yields and $\bf \bar p_t $ vs p-p multiplicity} \label{ppyields}

TCM analysis of differential spectrum structure as described above can be supplemented by statistical measures, e.g.\ integrated yields $n_x$ or mean angular densities $\bar \rho_x$ within some angular acceptance and ensemble-mean  \mmpt. Given TCM soft and hard components inferred from differential \pt\ spectra the integrated multiplicities $n_s$ and $n_h$ and ensemble means $\bar p_{ts}$ and $\bar p_{th}$ can be computed. Figure~\ref{pp2} (right) shows an initial comparison from Ref.~\cite{ppprd} suggesting a quadratic relation between $n_h$ and $n_{ch}$. A study of recent high-statistics \pp\ spectrum data from Ref.~\cite{ppquad} establishes more accurate relations.

\begin{figure}[h]
  \includegraphics[width=1.65in]{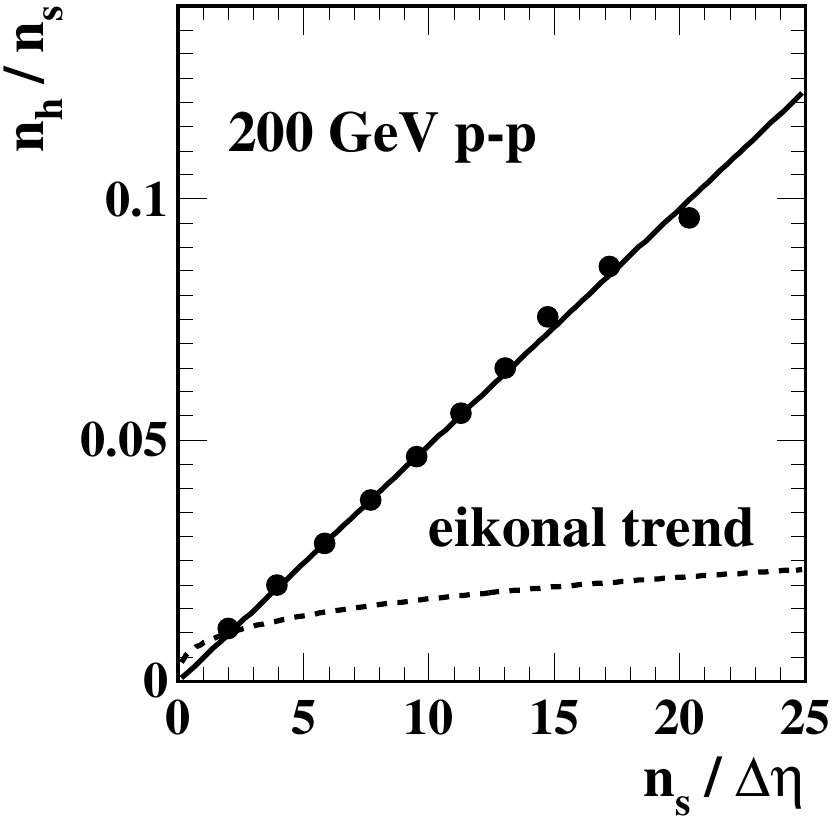}
% \put(-85,80) {\bf (a)}
  \includegraphics[width=1.65in]{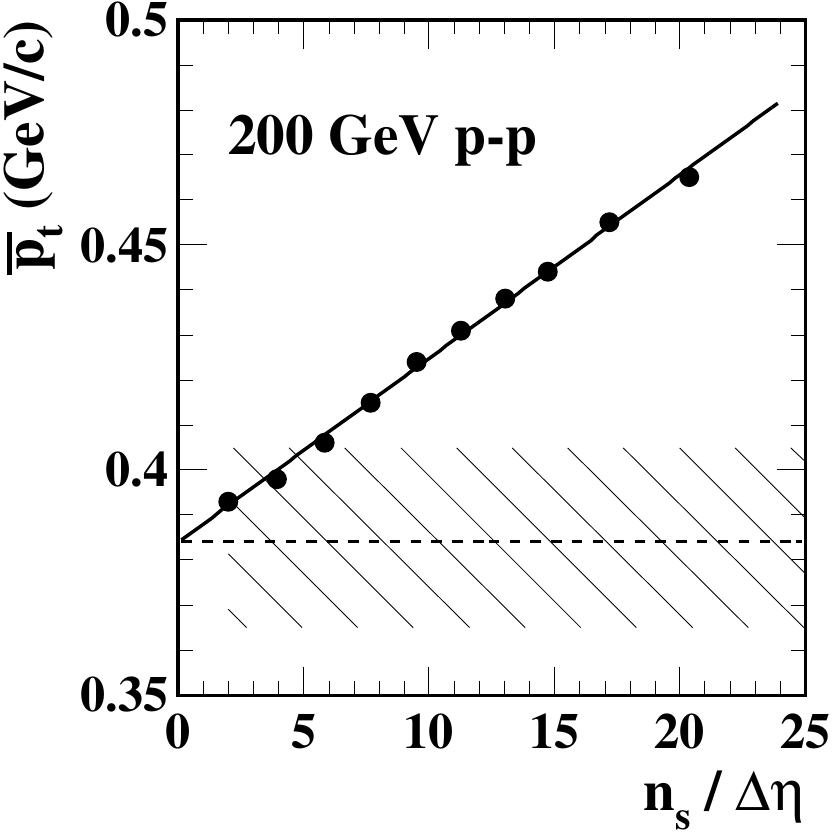}
% \put(-85,80) {\bf (b)}
\caption{\label{ppcomm} 
Left: 
Hard/soft multiplicity ratio $n_h / n_s$ (points) vs soft component $n_s$ consistent with a linear trend (line)~\cite{phenix}.  Assuming that $n_s$ represents the density of small-$x$ participant partons (gluons) and $n_h$ represents dijet production by parton scattering, an eikonal model of \pp\ collision geometry (analogous to the Glauber model of \aa\ collisions) would predict an $n_s^{1/3}$ trend for the  ratio (dashed curve).
Right: 
The \mmpt\ trend predicted by the \pp\ spectrum TCM (line) and as determined by direct spectrum integration (points). The hatched band represents the uncertainty in soft component $\bar p_{ts}$ (dashed line) from extrapolating \pt\ spectra to zero momentum.
 } % ppcomm12cc,12ddx
 \end{figure}

Figure~\ref{ppcomm} (left) shows ratio $n_h / n_s$ vs soft-component mean density $\bar \rho_s = n_s / \Delta \eta$ for ten multiplicity classes~\cite{alicetommpt}. $n_h$ is the integral of hard-component $H(y_t)$ appearing differentially in Fig.~\ref{pp2} (left) and as a running integral in Fig.~\ref{pp1} (right) (excesses above unity at right). The linear trend for $n_h/n_s$ confirms the relation $\bar \rho_h \propto \bar \rho_s^2$~\cite{ppquad}. Soft-component density $\bar \rho_s$ may be interpreted as a proxy for the density of low-$x$ gluons released from projectile nucleons in a \pp\ collision. \pp\ spectrum data then reveal that the number of midrapidity dijets $\propto \bar \rho_h$ varies quadratically with number of participant gluons. But in an eikonal collision model the number of gluon-gluon binary collisions should vary as the dashed curve representing $\bar \rho_h \propto \bar \rho_s^{4/3}$ as for the Glauber model of \aa\ collisions. The \pp\ data appear inconsistent with the eikonal model.

%%%%%%%%%
%%%%%%%%%%%%

Figure~\ref{ppcomm} (right) shows \mmpt\  vs soft-component mean density $\bar \rho_s$ well described by a constant plus linear term. The solid line is a TCM for that quantity derived from the spectrum TCM of Eq.~(\ref{tcmeq}). Given the evidence in  this section one may conclude that \mmpt\ variation with $\bar \rho_s$ is determined entirely by  the jet-related hard component.

%%%%%%%%%%%%%%%
\section{Jet contributions to $\bf A$-$\bf A$ spectra} \label{jetaaspec}

The \pp\ TCM provides an important reference for \aa\ collisions. It is reasonable to expect more-peripheral \aa\ collisions to be described by the same basic model elements modulo the Glauber model of \aa\ collision geometry -- dependence on number of participant nucleons $N_{part}$ and \nn\ binary collisions $N_{bin}$, with $\nu \equiv 2N_{bin} / N_{part}$ as the mean number of binary collisions per participant. However, it has been established that jets are strongly modified in more-central \aa\ collisions. The TCM hard-component model should then be altered to accommodate such changes.

\subsection{Identified-hadron spectrum evolution}

The TCM for \pp\ spectra can be extended to describe data from \aa\ collisions (and identified hadrons) as in Ref.~\cite{hardspec}. The TCM reference for \aa\ collisions corresponding to Glauber linear superposition (GLS) of isolated \nn\ collisions is based on the \pp\ result in Eq.~(\ref{tcmeq}) evaluated for inelastic \nn\ collisions as represented by
\bea \label{tcmpp2}
\bar \rho_0(y_t) &\approx& S_{NN}(y_t) + H_{NN}(y_t).
\eea
By hypothesis the TCM soft component in \aa\ collisions should scale with $N_{part}/2$ and the hard component should scale with $N_{bin}$ leading to the \aa\ spectrum TCM
\bea \label{tcmaa}
\bar \rho_0(y_t,b) &\approx& (N_{part} / 2) S_{NN}(y_t) + N_{bin} H_{AA}(y_t,b)
\nonumber \\
\frac{2}{N_{part}}\bar \rho_0(y_t,b) &\approx& S_{NN}(y_t) + \nu H_{AA}(y_t,b),
\eea
where the soft component is assumed to be invariant with \aa\ centrality but the jet-related hard component may vary substantially and is then a principal object of study.

%%%%%%%%%%
 \begin{figure}[h]
 \includegraphics[width=1.62in]{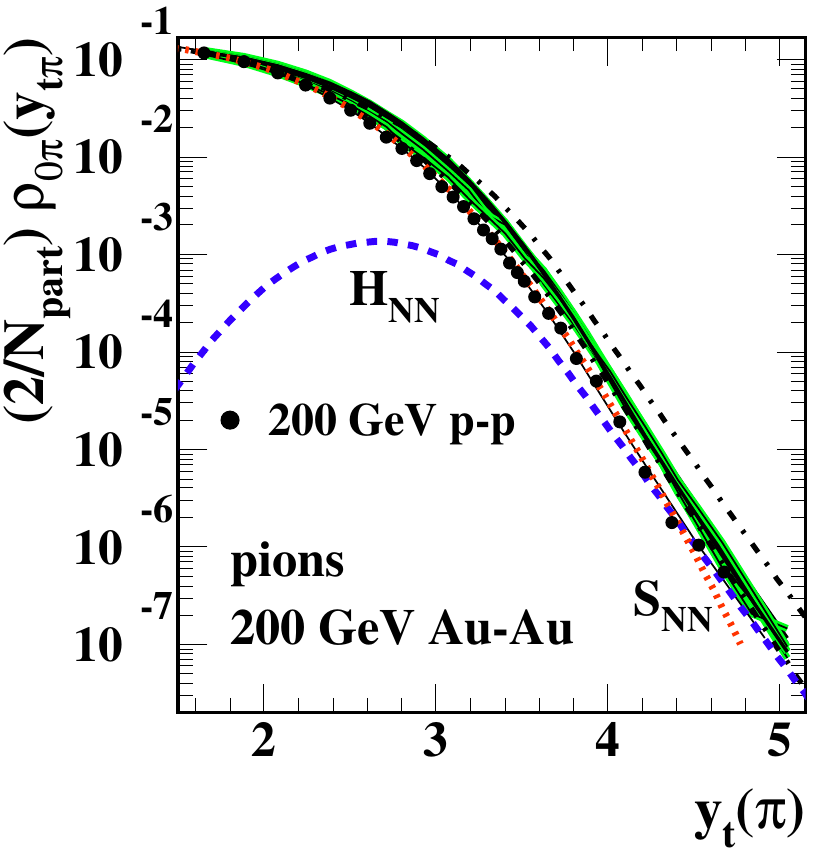}
   \includegraphics[width=1.68in]{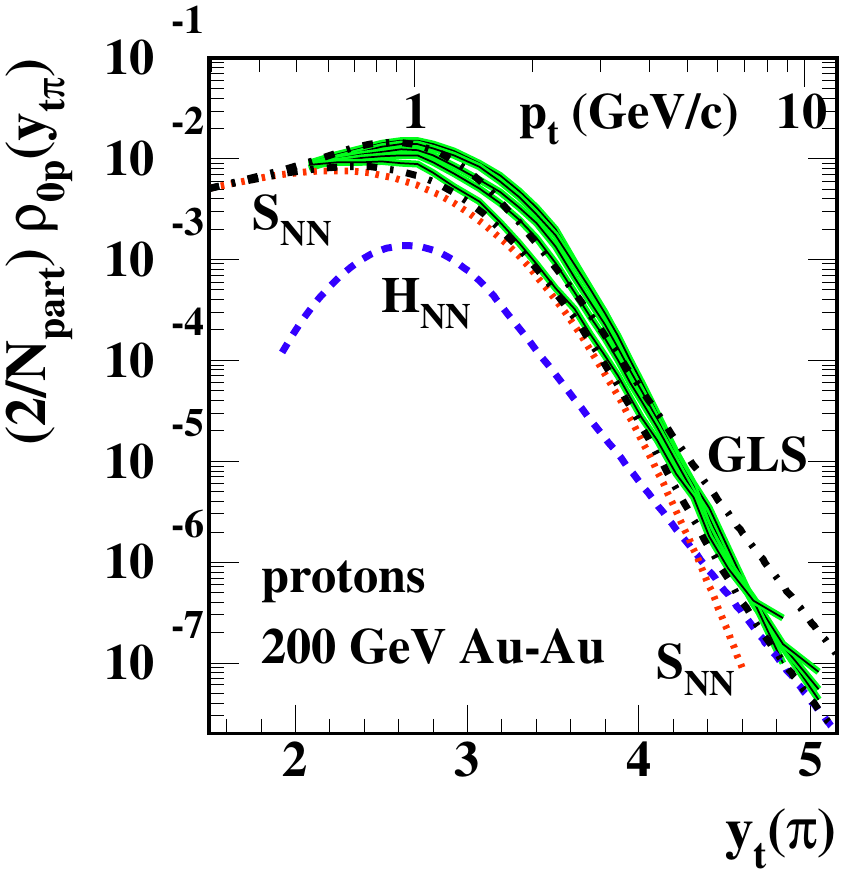}
\caption{\label{aaspec10}
(Color online) Summary of pion (left) and proton (right) per-participant-pair single-particle spectra from Au-Au collisions at 200 GeV and five centralities~\cite{hardspec}. $H_{NN}$ is the hard component (minimum-bias transverse parton fragmentation) and $S_{NN}$ is the soft component (longitudinal nucleon fragmentation), both inferred for N-N collisions. The solid points in the left panel represent the NSD p-p spectrum~\cite{ppprd}.
 } %  aaspectra10bx, 11bx 
 \end{figure}
%%%%%%%%%%

Figure~\ref{aaspec10} shows identified-pion and -proton spectra (solid curves) for five centrality classes of 200 GeV \auau\ collisions plotted vs rapidity $y_t(\pi)$\ with pion mass assumed~\cite{hardspec}. The rapidity variable is used in this case simply as a logarithmic momentum variable $y_t \approx \ln(2 p_t / m_\pi)$ but with well-defined zero. That choice is explained below. Also plotted with the pion spectra is the unidentified-hadron spectrum for 200 GeV NSD \pp\ collisions (points)~\cite{ppprd}. The TCM soft components $S_{NNx}$ (dotted) are defined as the limits of normalized spectrum data as $N_{part} \rightarrow 0$, equivalent to the definition for \pp\ collisions. The pion soft component is consistent with the \pp\ soft component inferred from unidentified hadrons. The pion hard component $H_{NN}$ is also consistent with the \pp\ analysis. The proton soft and hard components have the same algebraic structure, but model parameters are adjusted to accommodate peripheral \auau\ data as described below.  The dash-dotted curves are GLS reference spectra for $\nu = 1$ and 6 (\aa\ limiting cases).

\begin{figure}[h]
   \includegraphics[width=1.62in]{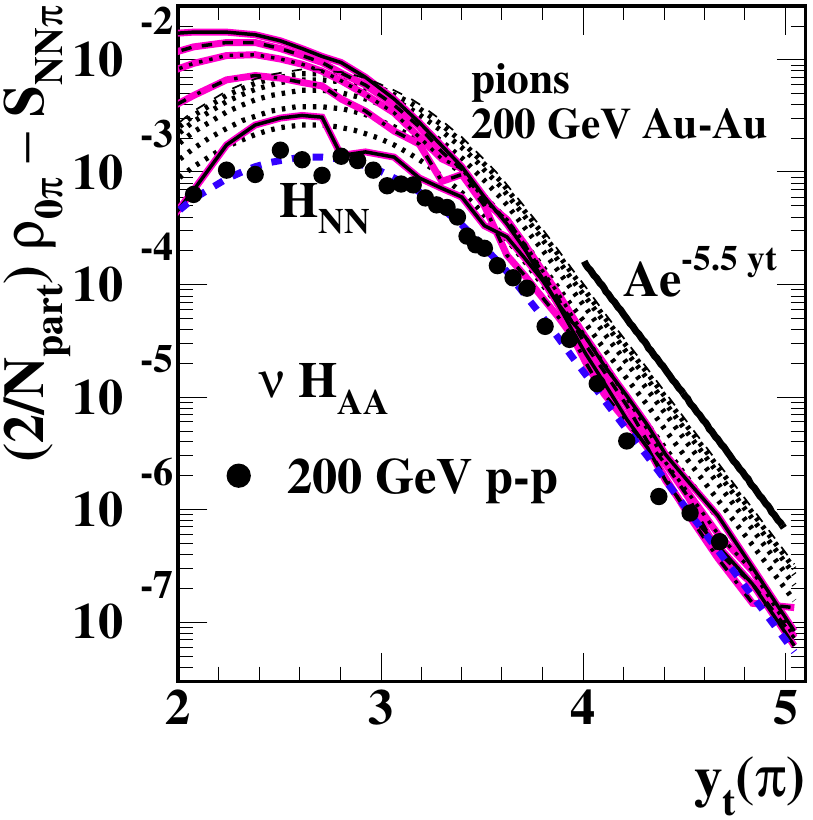}
 \includegraphics[width=1.68in]{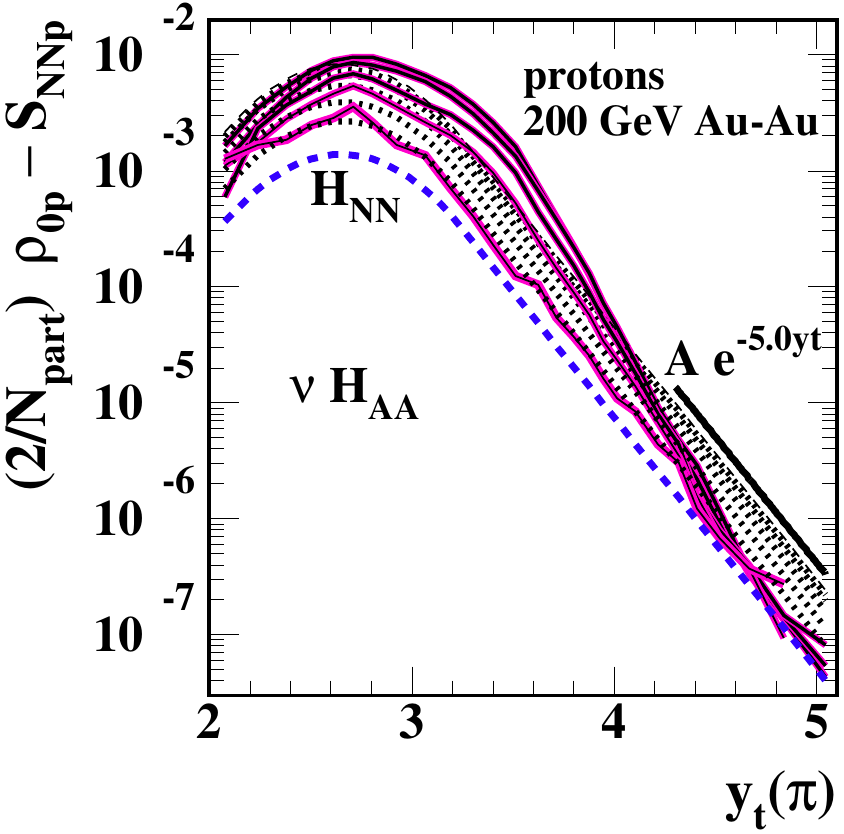}
 \caption{\label{difpi} (Color online)
Left:
The hard components of pion $y_t$ spectra in the form $\nu\, H_{AA}$ (solid) compared to two-component reference $\nu\, H_{NN}$ (dotted). The spectrum hard component for unidentified hadrons (80\% pions) from NSD \pp\ collisions (points) is included for reference.
Right:
  The hard components of proton $y_t$ spectra in the form $\nu\, H_{AA}$ (solid) compared to two-component reference $\nu \, H_{NN}$ (dotted). In either case $H_{NN}$ (dashed) is the reference \nn\ ($\approx$ \pp) hard component, and dotted curves $\nu \, H_{NN}$ represent a GLS reference.
 } %  aaspectra10ee,11ee
 \end{figure}

Figure~\ref{difpi} shows data hard components for identified pions and protons (solid) from five centrality classes of 200 GeV \auau\ collisions in the form $\nu H_{AA}$ per Eq.~(\ref{tcmaa}). Also plotted are the hard component for unidentified hadrons from 200 GeV NSD \pp\ collisions (points) and hard-component models $H_{NNx}$ (dashed). The dotted curves are GLS references corresponding to the five centrality classes assuming that $H_{AA} = H_{NN}$ (i.e.\ linear scaling with factor $\nu$). Relative to those reference curves the data exhibit substantial suppression at higher \pt\ for more-central collisions as inferred from conventional ratio measure $R_{AA}$~\cite{starraa}. However, substantial {\em enhancement} at lower \pt\ is a new feature not revealed by $R_{AA}$ data.

%%%%%%%%%%
%%%%%%%%%%

The structure of the proton hard component is a surprise. The mode for \nn\ ($\approx$ \pp) collisions appears near $p_t = 1$ GeV/c as for pions but the peak width is substantially less, and there is a significant difference in the ``power-law'' slope at high \yt\ as described below. However, it is most notable that hard-component maximum values are {\em essentially the same for protons and pions} in contrast to the soft components. The similarity of two hard components on \pt\ is the main motivation for adopting $y_t(\pi) \approx \ln(2 p_t / m_\pi)$ as the independent variable~\cite{hardspec}.

\subsection{Spectrum ratios for identified hadrons}

\begin{figure}[h]
  \includegraphics[width=1.65in]{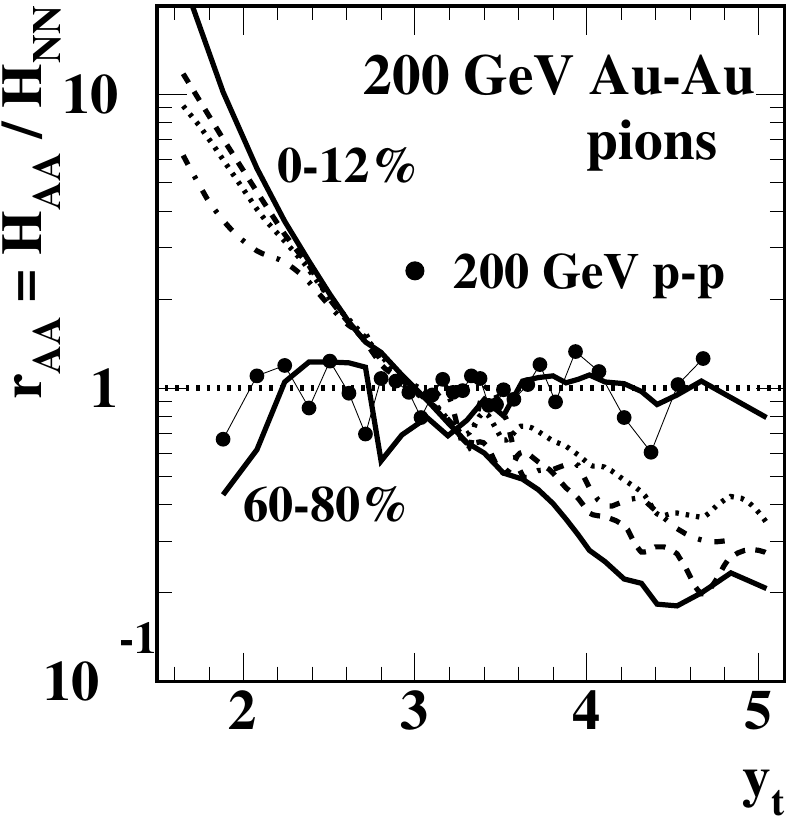}
  \includegraphics[width=1.65in]{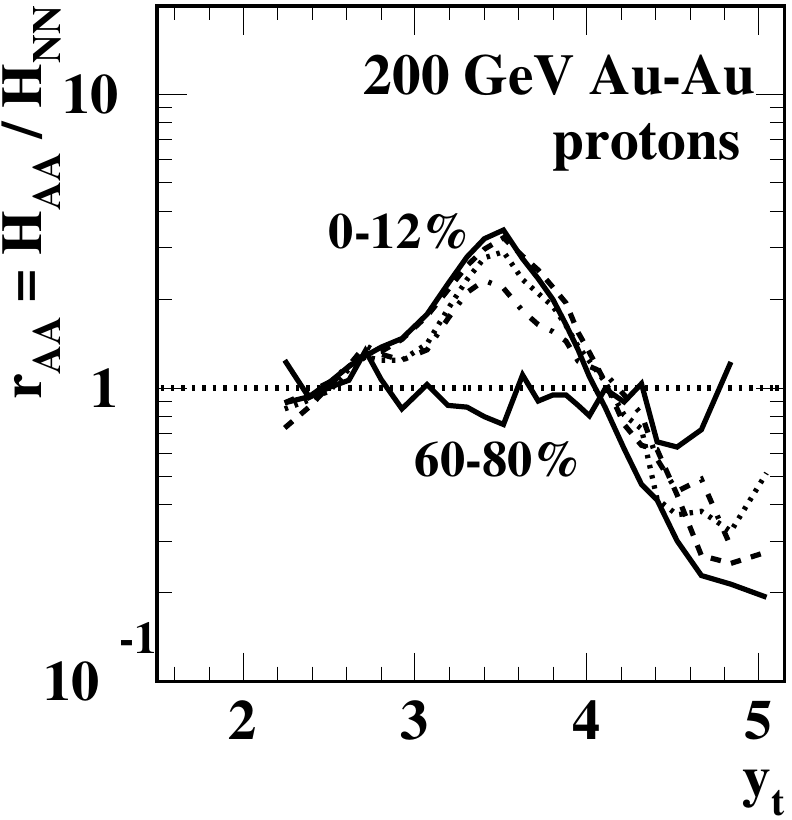}
\caption{\label{pidjet}
Hard-component ratios $r_{AA}(y_t)$ for five centralities of 200 GeV \auau\ collisions (curves) for pions (left) and protons (right)~\cite{hardspec}. Also shown are NSD \pp\ data (points, unidentified hadrons) compared to hard component $H_{NN}$.
 } % aaspectra10fy, 11fy
 \end{figure}

Figure~\ref{pidjet} shows ratios $r_{AA} = H_{AA} / H_{NN}$ for pions and protons from five centrality classes of 200 GeV \auau\ collisions (curves of several line styles). Also plotted are \pp\ hard-component data in ratio to the pion $H_{NN}$ reference (points). Several features are notable. The peripheral pion and proton data for 60-80\% central collisions indicate no jet modification ($r_{AA} = 1$), a result consistent with the observation from jet-related 2D angular correlations that below a {\em sharp transition} (ST) in jet characteristics near 50\% of the total cross section jets remain unmodified in 200 GeV \auau\ collisions~\cite{anomalous}.

%%%%%%%%%%
%%%%%%%%%%%%

Above the ST there is increasing suppression at higher \pt, with a saturation value $\approx 0.2$ for both pions and protons in central \auau\ collisions as observed with conventional ratio parameter $R_{AA}$. However, because $R_{AA}$ as defined is a ratio of entire spectra {\em including soft components} the evolution of jet-related $H_{AA}$ below $p_t = 3$ GeV/c ($y_t \approx 3.75$) is visually inaccessible. In contrast, ratio $r_{AA}$ including only hard components reveals large enhancements of jet-related hadron yields at lower \pt\ tightly correlated on centrality with suppressions at higher \pt. But whereas enhancement for pions extends below 0.5 GeV/c ($y_t = 2$) enhancement for protons peaks near 2.5 GeV/c ($y_t \approx 3.6$), and below $p_t = 1$ GeV/c the proton data for all centralities remain consistent with the \nn\ reference ($r_{AA}$ = 1) within the detector acceptance. 

These jet-related hard-component trends have important consequences for other (e.g.\ ratio) measures and for the flow narrative. In particular, whereas conjectured radial flow should boost all hadron species to higher \pt\ proportional to hadron mass the trends in Fig.~\ref{pidjet} show that while protons appear boosted to higher \pt\ relative to the \pp\ hard-component mode pions move to {\em lower} \pt.

\begin{figure}[h]
  \includegraphics[width=1.65in]{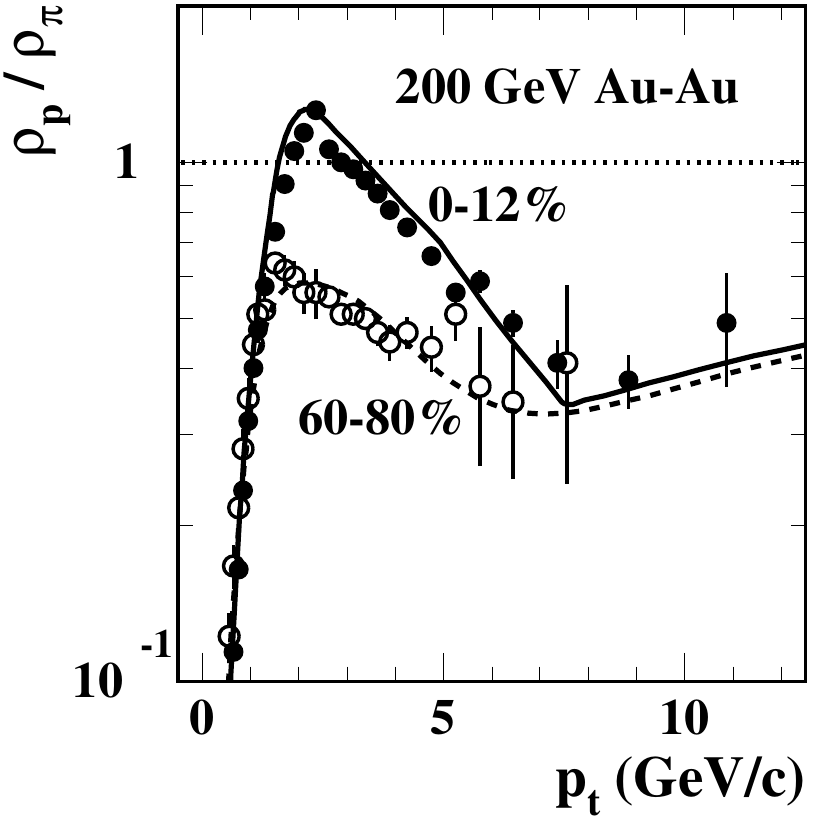} 
  \includegraphics[width=1.65in]{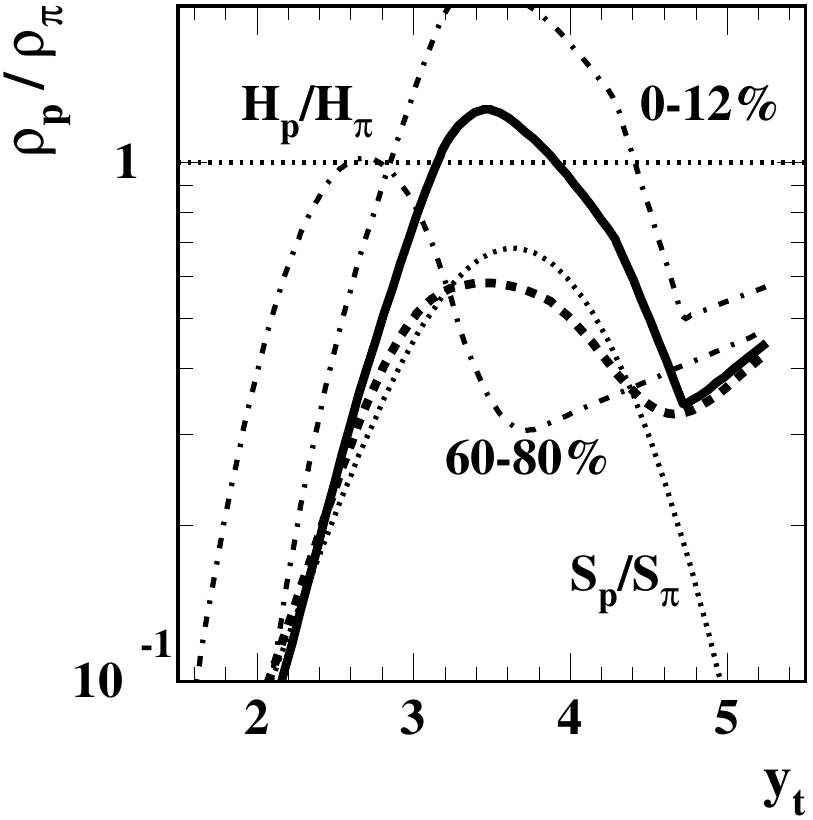} 
\caption{\label{baryonmeson}
Left:
Proton/pion spectrum-ratio data vs \pt\ for 0-12\% central (solid points) and 60-80\% central (open points) 200 GeV \auau\ collisions~\cite{barmeson} compared to curves (solid and dashed respectively) generated by the corresponding spectrum TCMs for identified hadrons described above and in Ref.~\cite{hardspec}. The curves were not fitted to those data. Right: Curves from the left panel plotted vs \yt\ (bold solid and dashed) compared to ratios determined separately from soft components (bold dotted) common to both centralities and hard components (dash-dotted) for the individual centralities.
 }  % aaspectra30a
 \end{figure}

Figure~\ref{baryonmeson} (left) shows a conventional ratio comparison of proton and pion spectra -- the proton-to-pion ratio -- for two centralities of 200 GeV \auau\ collisions (points)~\cite{barmeson}. The  curves are derived from a TCM for \auau\ spectra including hard-component modification in more-central collisions~\cite{fragevo} that describes the spectrum evolution in Fig.~\ref{difpi}. Those curves were not fitted to the ratio data.

%%%%%%%%%%
%%%%%%%%%%%%

Figure~\ref{baryonmeson} (right) shows the curves in the left panel replotted on \yt\ to improve visibility of the low-\pt\ region. Since the TCM soft and hard components contributing to those spectrum ratios are known their ratios can be plotted separately (dotted and dash-dotted curves respectively). The soft component by hypothesis does not vary with \aa\ centrality. The hard-component ratio for peripheral \aa\ (and therefore \nn) collisions is unity at the mode (\yt\ = 2.7, $p_t \approx$ 1 GeV/c). It falls off on either side of the mode because of the peak width difference but increases for larger \yt\ because the proton power-law exponent is smaller (the spectrum high-\pt\ tail is harder), all consistent with the discussion of Fig.~\ref{difpi} above.

The hard-component ratio maximum for central collisions has a substantially larger value and the mode on \pt\ shifts up to 3 GeV/c. Protons are strongly enhanced relative to pions above the \nn\ mode but pions are strongly enhanced relative to protons below the mode, a feature concealed by the conventional $R_{AA}$ ratio and plotting format. Comparison with the solid curve matching the ratio data reveals that the data peak is dominated by the jet-related hard component. In contrast, the ratio peak for peripheral collisions is dominated by the soft component; the hard component only influences the peripheral ratio above 7 GeV/c.  Note that the ratio of hard to soft hadron production in \auau\ collisions increases with centrality according to parameter $\nu$ with no jet modification (5-fold increase) and by an additional factor 3 due to jet modification~\cite{jetspec}. Thus, from peripheral to central \auau\ collisions the hard/soft ratio increases by about $3 \times 5 = 15$-fold. Although peaks in the left panel appear similar and suggest a common mechanism this TCM analysis reveals that they represent distinct soft and hard hadron production mechanisms. Comparisons of spectrum ratio data with in-vacuum \ee\ FFs and nonjet (soft) recombination or coalescence hadronization models as in Ref.~\cite{barmeson} are likely misleading.

\subsection{Hadron yields and $\bf \bar p_t $ vs A-A centrality} \label{aayields}

Just as for \pp\ collisions differential \pt\ spectrum structure for \aa\ collisions can be supplemented by trends for integrated spectrum yields and \pt\ mean values. For instance, hadron yields vs \pp\ multiplicity were discussed in Sec.~\ref{ppyields} where observed dijet production appears inconsistent with the eikonal approximation. The centrality  dependence of \pt-integral hadron yields (or mean angular densities) in \aa\ collisions is similarly of interest.

\begin{figure}[h]
%  \includegraphics[width=1.65in]{ptjets5d}
% \put(-90,70) {\bf (a)} 
 \includegraphics[width=1.65in]{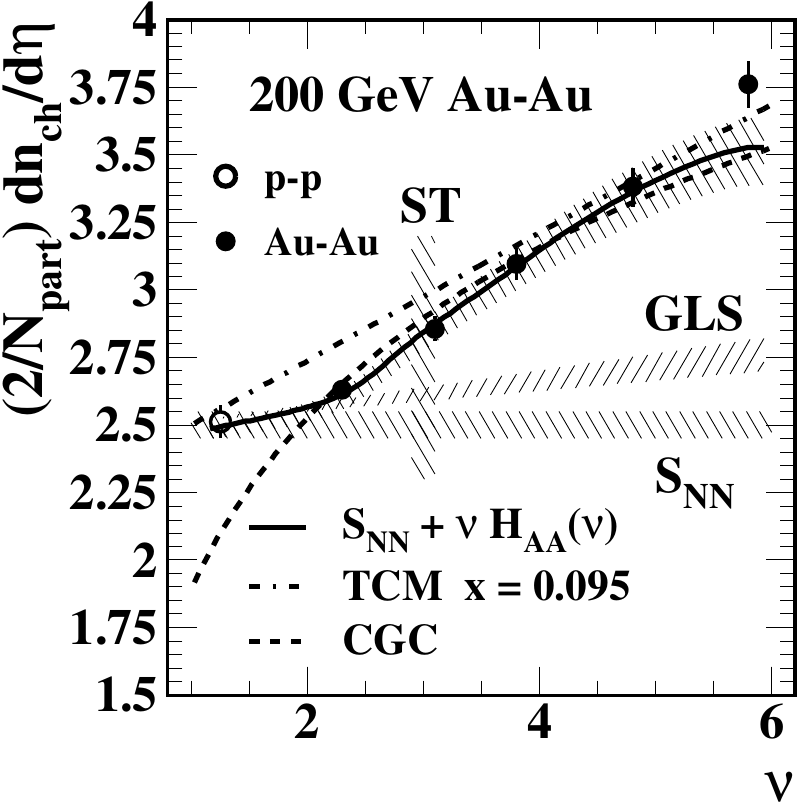}
   \includegraphics[width=1.65in]{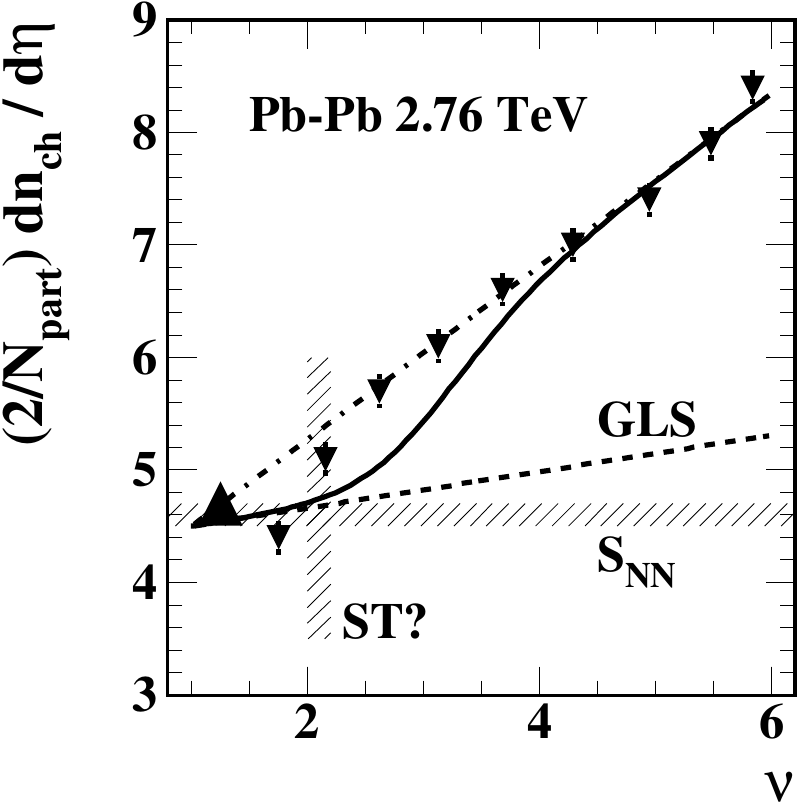}
 \put(-90,70) {\bf (b)} 
\caption{\label{corresp}
Left:
Per-participant hadron production measured by $(2/N_{part}) dn_{ch}/d\eta$ vs $\nu$ for 200 GeV \auau\ collisions (solid points) inferred from analysis of identified-hadron spectra~\cite{hardspec}. The dash-dotted line is the conventional TCM with fixed $x = 0.095$~\cite{kn}. The solid curve is a prediction obtained by analysis of jet-related angular correlations~\cite{jetspec}. The dashed curve is a prediction derived from color-glass condensate theory~\cite{glasmatom}.
Right:
 Hadron production vs centrality for 2.76 TeV \pbpb\ collisions (inverted solid triangles~\cite{alicemult}) compared to TCM trends (dash-dotted and dashed lines) extrapolated from the 200 GeV TCM based on measured RHIC energy trends as described in the text. The upright solid triangle is a 2.76 TeV \pp\ reference.
 } % ptjets6b-cgc, phenix5a
 \end{figure}

\begin{figure}[h]
  \includegraphics[width=1.63in]{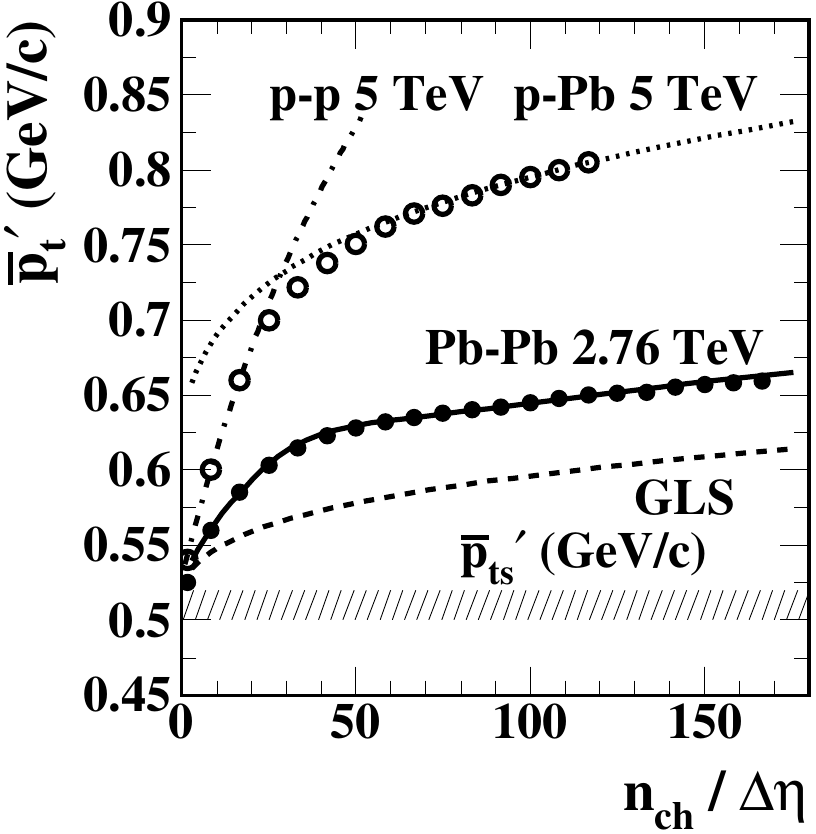}
% \put(-20,79) {\bf (a)} 
  \includegraphics[width=1.67in]{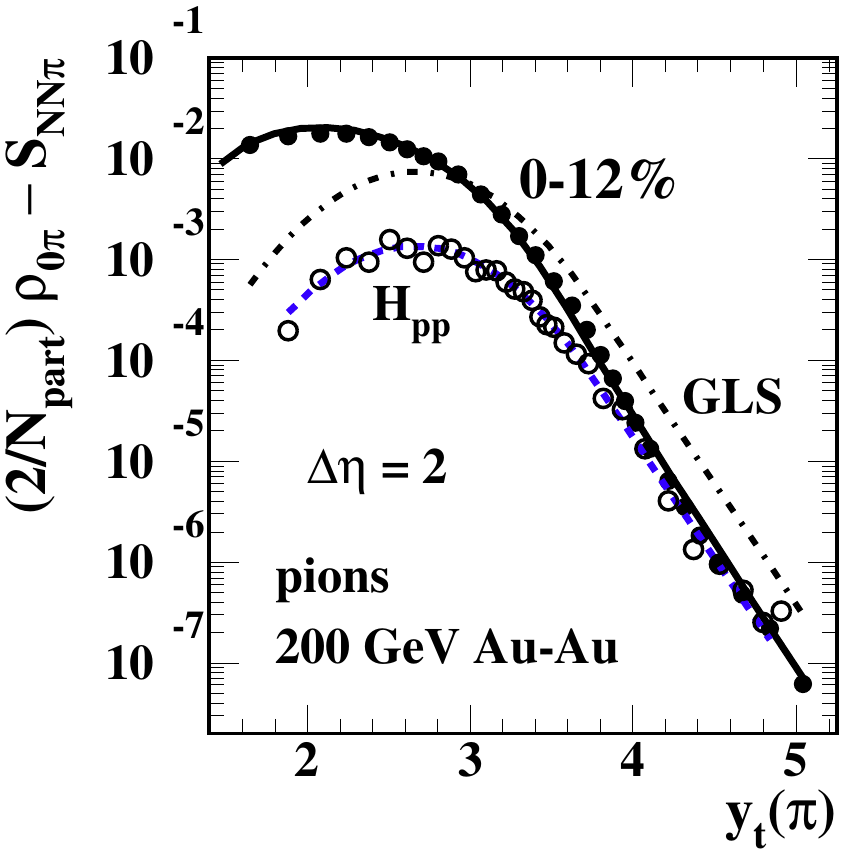}
% \put(-20,79) {\bf (b)} 
\caption{\label{mpt2}
Left:
ppb\ and \pbpb\ uncorrected \mmpt\ data from the LHC~\cite{alicempt} (points)  compared to a corresponding \mmpt\ TCM (solid and dotted)~\cite{alicetommpt}. The \ppb\ data for larger \nch\ are described by a Glauber linear superposition (GLS) form of the TCM consistent with transparent \ppb\ collisions (no rescattering). For smaller \nch\ the \ppb\ data correspond quantitatively to a TCM for \pp\ collisions at the same energy (dash-dotted).
Right: TCM for peripheral (dashed) and central (solid) 200 GeV \auau\ collisions~\cite{fragevo} compared to \pp\ (open points~\cite{ppprd}) and 0-12\% \auau\ (solid points~\cite{hardspec}) data. The dash-dotted curve is a TCM GLS reference for central \auau\ collisions.
 }  % alice3a,aaspectra10eex
 \end{figure}

Figure~\ref{corresp} (left) shows integrated unidentified-hadron densities in the form $(2/N_{part}) dn_{ch}/d\eta$ (solid points) reconstructed from spectrum data in Fig.~\ref{aaspec10}~\cite{hardspec}. Due to multiplicity fluctuations the most-central point of such a trend is typically high by an amount controlled by the detector angular acceptance: the excess is less for a larger acceptance~\cite{phenix}.  The open point is a reference value from NSD \pp\ collisions. A simple TCM with fixed constant $x = 0.095$ from Ref.~\cite{kn} is represented by the dash-dotted line. A color-glass condensate (CGC) trend approximated by $0.9 \ln(8\nu)$ is shown as the dashed curve~\cite{glasmatom}. The hatched region labeled GLS is a \auau\ reference TCM extrapolation from the \pp\ TCM of Sec.~\ref{tcm1} assuming no jet modification in \auau\ collisions. The solid curve is a TCM with varying $x$ based on dijet angular correlations~\cite{jetspec} combined with the spectrum TCM from Ref.~\cite{fragevo} that models jet modification in terms of spectrum hard components as in Fig.~\ref{mpt2} (right) below. The hatched band labeled ST marks the ``sharp transition'' in jet-related angular-correlation characteristics reported in Ref.~\cite{anomalous} (see Sec.~\ref{taanalysis}). Quantitative correspondence of those results with yield data from spectra provides additional compelling evidence that the TCM spectrum hard component remains jet-related in all \aa\ collisions~\cite{glasmatom}.

%%%%%%%%%
%%%%%%%%%%%%

Figure~\ref{corresp} (right) shows  hadron production vs centrality for 2.76 TeV \pbpb\ collisions (inverted triangles)  and a reference value for NSD \pp\ collisions (upright triangle)~\cite{alicemult}. The solid curve is the solid curve for 200 GeV in the left panel extrapolated to higher energy by two modifications. The soft component $S_{NN}$ is multiplied by 1.87 = $\log(2760/10) / \log(200/10)$ motivated by the observed soft-component energy trend $\propto \log(\sqrt{s_{NN}} / \text{10 GeV})$~\cite{alicespec}. The hard component is then multiplied by another factor 1.87 corresponding to $n_h \propto n_s^2$ for \nn\ collisions~\cite{ppquad}.  Whereas the centrality evolution of jet correlation structure for 200 GeV \auau\ collisions is consistent with a ST near $\nu = 3$ (50\% fractional cross section)~\cite{anomalous} the \pbpb\ data suggest that the ST at higher energy may have shifted down to $\nu \approx 2$. Otherwise the same TCM describes data well at two widely-spaced energies depending only on a $\log(s/s_0)$ QCD energy trend.

Figure~\ref{mpt2} (left) shows ensemble-mean \mmpt\ data from 5 TeV \ppb\ and 2.76 TeV \pbpb\ collisions (points) vs charge density $\bar \rho_0 = n_{ch} / \Delta \eta$~\cite{alicempt}. The solid and dotted curves are TCMs for the respective collision systems~\cite{alicetommpt}.  $\bar p_{ts}' \approx 0.51$ GeV/c (hatched band) is the soft-component value corresponding to a \pt-acceptance lower limit near 0.2 GeV/c, whereas $\bar p_{ts} \approx 0.38$ GeV/c is a universal feature of any \pt\ spectrum extrapolated down to $p_t = 0$ as in Fig.~\ref{ppcomm} (right).  The \pbpb\ GLS reference (dashed curve) reflects the eikonal approximation assumed for the \aa\ Glauber model and no jet modification.  The dash-dotted curve is the TCM for 5 TeV \pp\ data consistent with Fig.~\ref{ppcomm} (right) for 200 GeV data and inconsistent with the eikonal approximation. The \ppb\ data (open points) suggest a smooth transition from a non-eikonal \pp\ trend to eikonal \aa\ trend with increasing \nch\ (centrality).

%%%%%%%%%
%%%%%%%%%%%%

Figure~\ref{mpt2} (right) shows evolution of the measured TCM spectrum hard component for 200 GeV \auau\ collisions from peripheral (open points) to central (solid points) collisions. The corresponding dashed and solid curves are derived in Ref.~\cite{fragevo} from jet measurements. The dash-dotted curve is a GLS reference (no jet modification) for central \auau.  It seems likely that spectrum hard-component evolution in 2.76 TeV \pbpb\ collisions is similar but with some quantitative differences.

According to the TCM for \mmpt\ the GLS trend  for 2.76 TeV \pbpb\ in the left panel should be a weighted average of soft-component  value $\bar p'_{ts} \approx 0.5$ GeV/c (for $p_t > 0.2$ GeV/c) and hard-component value $\bar p_{th} \approx 1.7$ GeV/c (for 2.76 TeV \pp\ collisions~\cite{alicetommpt}), the average increasing with increasing fraction of jet-related hard-component hadrons. The \mmpt\ data trend differs from GLS by two consequences of jet modification: (a) the hard-component yield increases by factor 3 and (b) the hard-component mode decreases substantially from 1.7 GeV/c, as illustrated in Figure~\ref{mpt2} (right) and Fig.~\ref{difpi} (right). The combination results in a net increase of data $\bar p_t$ over GLS.

\subsection{Alternative spectrum models} \label{altaa}

%HYDRO FIG.~12

Section~\ref{altpp} presents an alternative to the \pp\ spectrum TCM in the form of a ``power-law'' or Tsallis distribution approximating the TCM soft component alone. The fit results indicate poor fit quality, and the large parameter variations are not physically interpretable. The blast-wave model is a popular alternative for \aa\ spectra in which deviations from some reference spectrum are assumed to result from radial flow of a bulk medium.
The assumed reference is usually a Maxwell-Boltzmann (M-B) exponential~\cite{uliss,starblast}, but a power-law or Tsallis distribution has been adopted in recent studies~\cite{tsallisblast}. The model parameters include slope parameter $T$ and mean radial speed $\langle \beta_t \rangle$.
Blast-wave fits may be restricted to smaller \pt\ intervals lying below 2 GeV/c under the assumption that such intervals exclude jet contributions and that any spectrum variation is then determined by flows: ``In central Au+Au collisions the flattening of the spectra [below 2 GeV/c] is likely dominated by collective transverse radial flow, developed due to the large pressure buildup in the early stage of heavy-ion collisions''~\cite{starpidspec}.

\begin{figure}[h]
  \includegraphics[width=1.65in,height=1.65in]{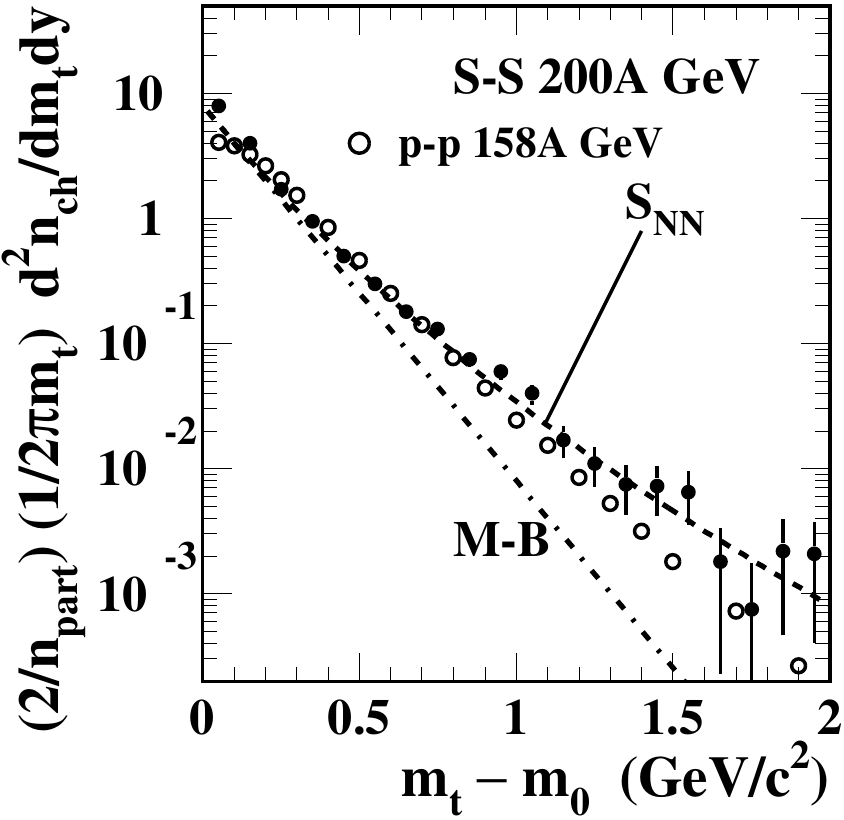}
  \includegraphics[width=1.65in,height=1.65in]{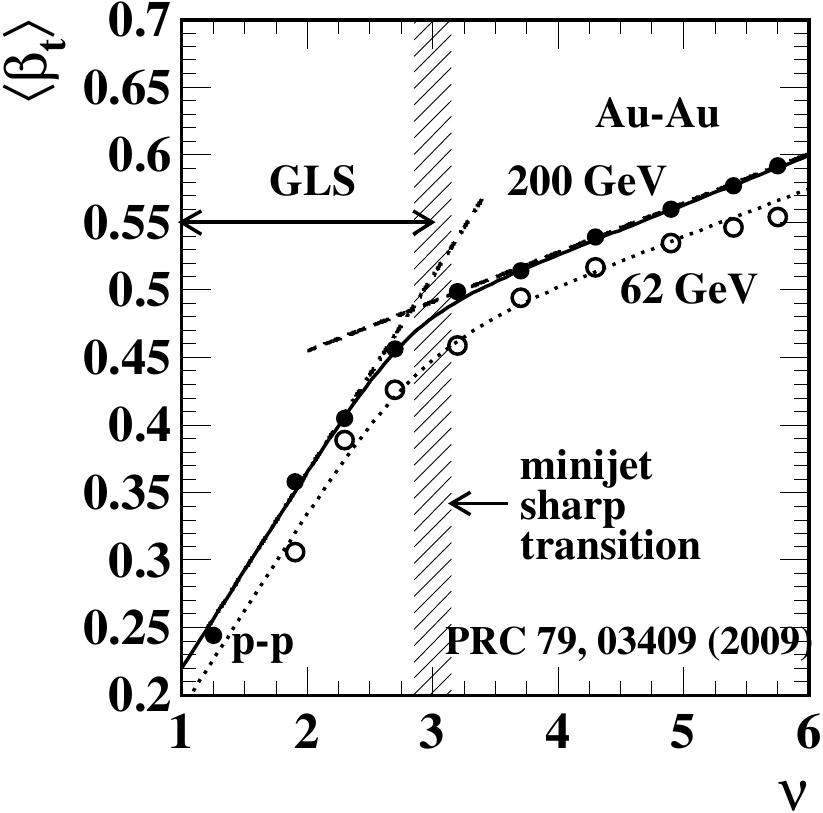}
\caption{\label{specstuff}
Left: 17 GeV \pp\ (open points) and 19 GeV \ss\ (solid points) \mt\ spectra compared to the soft component $S_{NN}$ of a TCM describing \ss\ collisions (dashed curve)~\cite{nohydro}.
Right: Radial speed $\langle \beta_t \rangle$ inferred from blast-wave fits to 62 and 200 GeV \auau\ collisions~\cite{starblast}. When plotted on participant pathlength $\nu$ the relation of the $\langle \beta_t \rangle$ trends to the minijet sharp transition (hatched band) is notable.
 }  % alephx10, hydro3bnew
 \end{figure}

Figure~\ref{specstuff} (left) illustrates an early radial-flow analysis of spectrum data from 19 GeV \ss\ collisions (solid points)~\cite{uliss}. The corresponding M-B distribution is shown by the dash-dotted line. A \pp\ spectrum for 17 GeV (open points) and a TCM soft-component L\'evy distribution (dashed) with universal $T_0 = 145$ GeV and with exponent $n_0 = 17$ adjusted to accommodate the S-S data are shown for comparison ($n_0 = 27$ describes the \pp\ spectrum~\cite{alicespec}). Deviation of the \ss\ data from the M-B reference is interpreted to represent radial flow with $\langle \beta_t \rangle \approx 0.25$ ($\beta_s \approx 0.5$ is the maximum for a radial $\beta_t$ distribution). The same fit model applied to a 200 GeV \pp\ spectrum returns a similar $\langle \beta_t \rangle$ value as noted below. 

%%%%%%%%%%
%%%%%%%%%%%%

Figure~\ref{specstuff} (right) shows published $\langle \beta_t \rangle$ values derived from fits to 62 GeV  (open points) and 200 GeV (solid points) \auau\ spectra for several collision centralities~\cite{starblast} plotted vs Glauber centrality parameter $\nu$. The location of the ST inferred from jet-related angular correlations~\cite{anomalous} is indicated by the hatched band. To the left of that point \auau\ collisions are effectively transparent linear superpositions (GLS) of \nn\ collisions~\cite{anomalous}, but to the right of that point jet structure shows substantial modification (``jet quenching''). It is notable that the inferred $\langle \beta_t \rangle$ data are not zero in the \aa\ transparency interval (or for \pp\ collisions). The  $\langle \beta_t \rangle$ data instead increase {\em more} rapidly in a centrality interval where rescattering is less likely based on jet data but increase {\em less} rapidly in an interval where jet modification is substantial suggesting copious rescattering is more likely. Results in Fig.~\ref{specstuff} (right) interpreted to indicate radial flow are in direct conflict with the results in Fig.~\ref{difpi} consistent with measured jet properties. In the blast-wave model jet-related spectrum structure described by the TCM hard component is in effect reassigned to radial flow.

%%%%%%%%%
 \section{Jets and angular correlations} \label{jetangcorr}

Angular correlation methods are introduced briefly in Sec.~\ref{meth}. A  measured pair density $\rho(x_1,x_2)$ can be compared with some reference $\rho_{ref}(x_1,x_2) = \bar \rho_0(x_1)\bar \rho_0(x_2) \approx \rho_{mix}(x_1,x_2)$ to define a {\em correlated-pair} density
\bea
\Delta \rho(p_{t1},p_{t2},\eta_\Delta,\phi_\Delta) \equiv \rho_{ref} \,(\rho' / \rho_{mix} - 1),
\eea
where $\rho'$ is an {\em uncorrected} pair density and $\rho_{mix}$ is a mixed-pair reference also uncorrected.  The correlated-pair density can be normalized to form a {\em per-particle} measure $\Delta \rho / \sqrt{\rho_{ref}} \equiv \sqrt{\rho_{ref}}(\rho' / \rho_{mix} - 1) \rightarrow \bar \rho_0(\rho' / \rho_{mix} - 1)$ or a {\em per-pair} measure $\Delta \rho / \rho_{ref} =(\rho' / \rho_{mix} - 1)$~\cite{anomalous,davidhq,davidhq2}.

\subsection{A TCM for two-particle correlations}

In addition to angular correlations on $(\eta_\Delta,\phi_\Delta)$ two-particle correlations can be studied on transverse rapidity as $(y_{t1},y_{t2}) \rightarrow y_t \times y_t$ where they are directly comparable with SP spectra on \yt\ and the \yt-spectrum TCM.

\begin{figure}[h]
  \includegraphics[width=1.65in,height=1.4in]{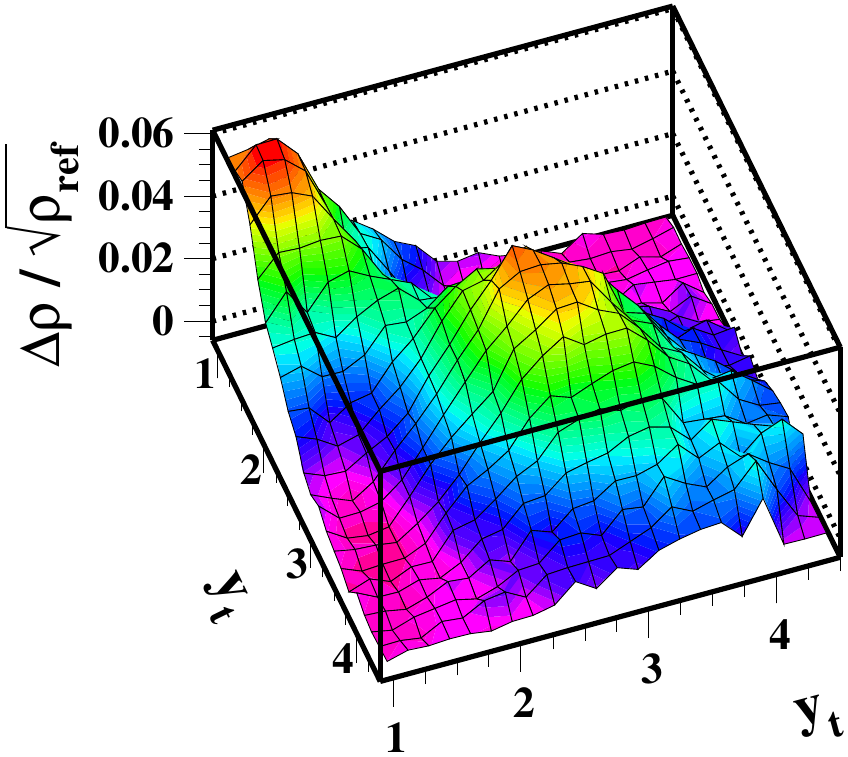}
% \put(-85,92) {\bf (a)}  
 \includegraphics[width=1.65in,height=1.4in]{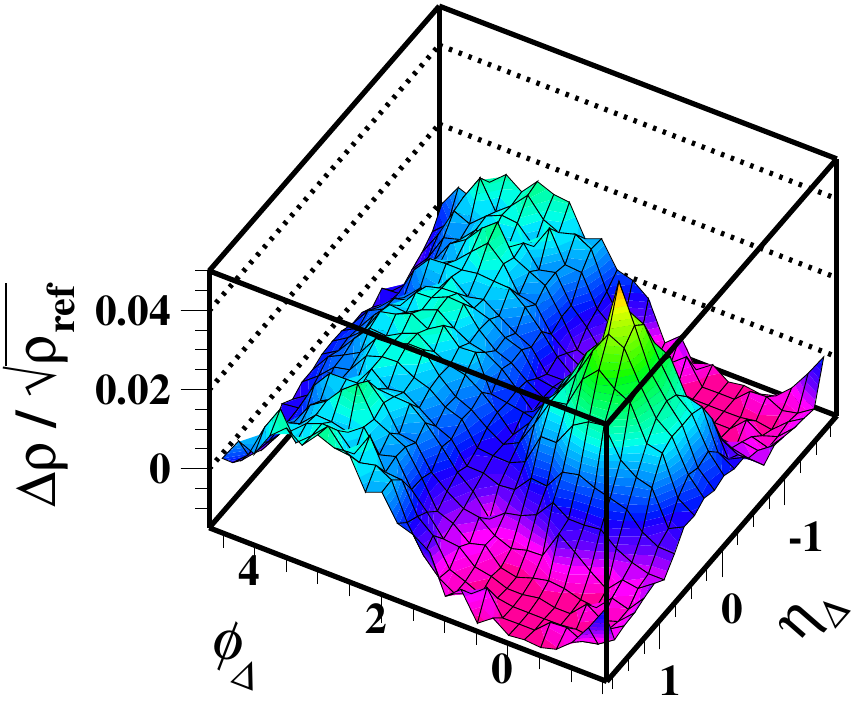}
% \put(-85,92) {\bf (b)}
 \caption{\label{ppcorr} (Color online)
(a) Minimum-bias correlated-pair density on 2D transverse-rapidity space $y_t \times y_t$ from 200 GeV \pp\ collisions showing soft (smaller \yt) and hard (larger \yt) components as peak structures.
(b)  Correlated-pair density on 2D angular difference space $(\eta_\Delta,\phi_\Delta)$. Although hadrons are selected with $p_t \approx 0.6$ GeV/c ($y_t \approx 2$) features expected for dijets are still observed: (i) same-side 2D peak representing intrajet correlations and (ii) away-side 1D peak on azimuth representing interjet (back-to-back jet) correlations~\cite{porter2,porter3}. 
 }   % ytdata1x, etaphidata01x 
 \end{figure}

Figure~\ref{ppcorr} (left) shows \ytyt\ correlations from 200 GeV NSD \pp\ collisions for $p_t \in [0.15,6]$ GeV/c~\cite{porter2,porter3}. The two peaked features are identified as TCM soft and hard components as follows. The lower-\pt\ peak falls mainly below 0.5 GeV/c ($y_t < 2$) and consists exclusively of unlike-sign (US) pairs. Corresponding angular correlations consist of a narrow 1D peak on $\eta_\Delta$ centered at the origin. That combination suggests longitudinal fragmentation of low-$x$ gluons to charge-neutral hadron pairs nearby on $\eta$, consistent with TCM spectrum soft component $S_{pp}(y_t)$. The higher-\pt\ peak falls mainly above 0.5 GeV/c with mode near $p_t = 1$ GeV/c ($y_t \approx 2.7$) corresponding to SP spectrum hard component $H_{pp}(y_t)$ and with charge structure (LS or US) depending on angular constraints.

%%%%%%%%%
%%%%%%%%%%%%

Figure~\ref{ppcorr} (right) shows angular correlations for the same collision system with the condition $p_t \approx 0.6$ GeV/c, i.e.\ near the lower boundary of the \ytyt\ hard component in the left panel. Despite the low hadron momentum the observed angular correlations exhibit structure expected for jets: a SS 2D peak representing {\em intra}\,jet  correlations and an AS 1D peak representing  {\em inter}\,jet (back-to-back jet) correlations. The SS peak is dominated by US pairs while the AS peak shows no preference, consistent with fragmentation of back-to-back {\em charge-neutral} gluons.

\subsection{2D angular correlations and model fits} \label{2dfits}

The general structure of 2D angular correlations was established for \pp\ collisions in Refs.~\cite{porter2,porter3} and for \auau\ collisions in Ref.~\cite{axialci}. Quantitative discrimination was achieved among jets, a {\em nonjet} (NJ) azimuth quadrupole and Bose-Einstein correlations (BEC). 2D angular correlations on $(\eta_\Delta,\phi_\Delta)$ have a simple structure  modeled by a few 1D and 2D functions~\cite{anomalous,v2ptb}. 
The six-element fit model of Ref.~\cite{anomalous} includes eleven model parameters but describes more than 150 data degrees of freedom for typical $25\times 25$-bin data histograms on $(\eta_\Delta,\phi_\Delta)$. Model parameters are thus strongly constrained. The NJ quadrupole component of angular correlations can be extracted accurately via such  fits. For per-particle 2D angular correlations as $\Delta \rho / \sqrt{\rho_{ref}}$ the NJ quadrupole is represented by $A_Q\{2D\} \equiv \bar \rho_0 v_2^2\{2D\}$~\cite{davidhq}.

\begin{figure}[t]
 \includegraphics[width=1.6in]{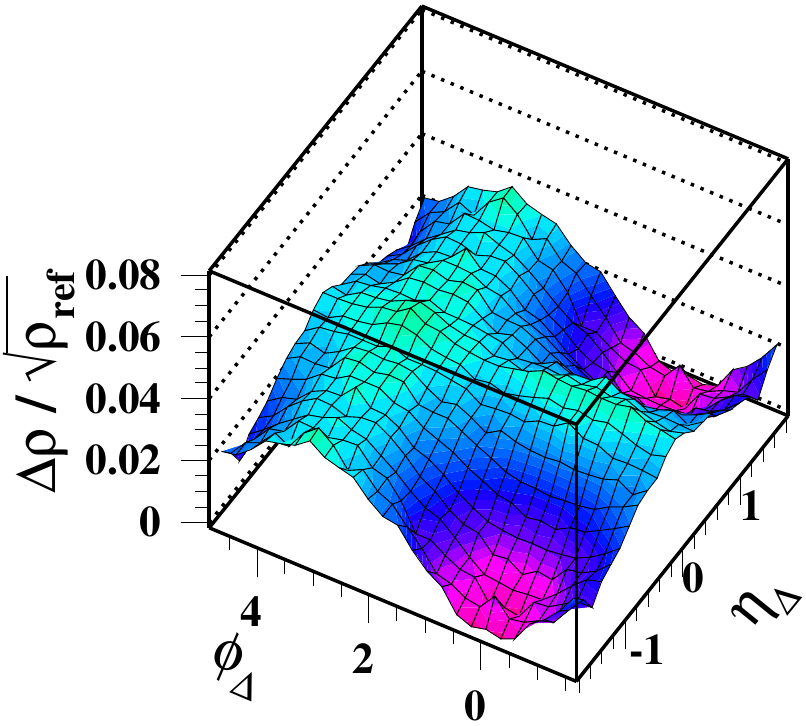}
\includegraphics[width=1.6in]{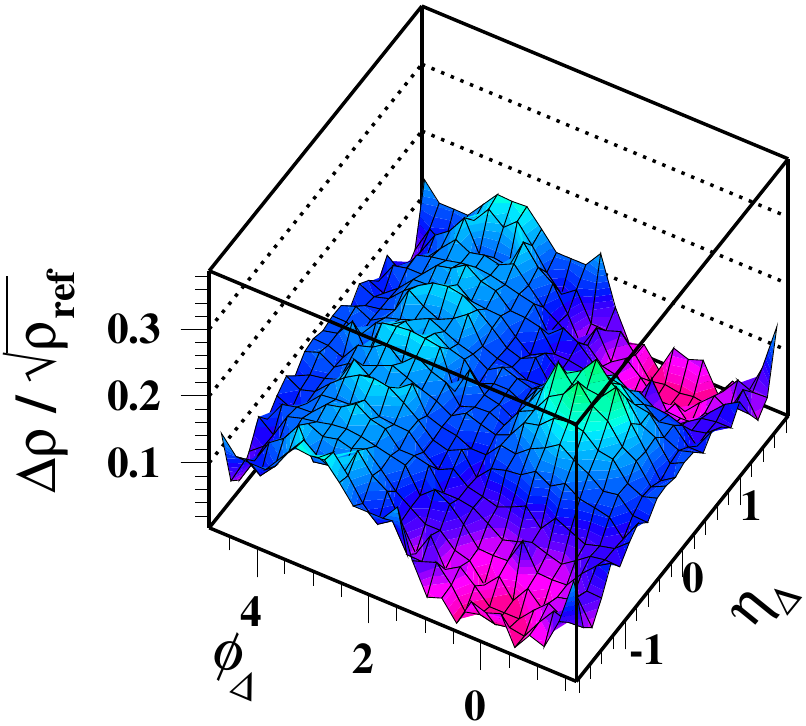}
\caption{\label{quadcomp}
2D angular correlations for  n = 1 (left) and 6 (right) multiplicity classes from 200 GeV \pp\ collisions~\cite{ppquad}. Fitted model elements for soft component, BE + conversion electrons and constant offset have been subtracted from the data leaving jet-related and NJ quadrupole data components.
}   % ppcms23-0dx, 5dx
 \end{figure}

Figure~\ref{quadcomp} shows 2D angular correlations from two of seven multiplicity classes of 200 GeV \pp\ collisions (index values $n = 1,\,6$, $dn_{ch}/d\eta \approx 1.8,$ 15, for left and right panels respectively)~\cite{ppquad}. Based on 2D model fits contributions from a soft component (1D peak on $\eta_\Delta$) and BEC (narrow 2D peak at the origin) have been subtracted. What remains is a jet contribution (broad SS 2D peak at the origin and AS 1D peak on azimuth) and a NJ quadrupole contribution manifested in the right panel by increased curvature of the AS 1D peak, reduced curvature of the SS background for $|\eta_\Delta| > 1$ and {\em apparent} narrowing of the SS 2D peak on azimuth. Systematic variations of several correlation components are presented in Ref.~\cite{ppquad}. The main message of such studies is that MB dijets dominate \pp\ correlation structure.

%%%%%%%%%%
%%%%%%%%%%%%

%%%%%%%%%%%%%
\begin{figure}[h]
\includegraphics[width=1.6in]{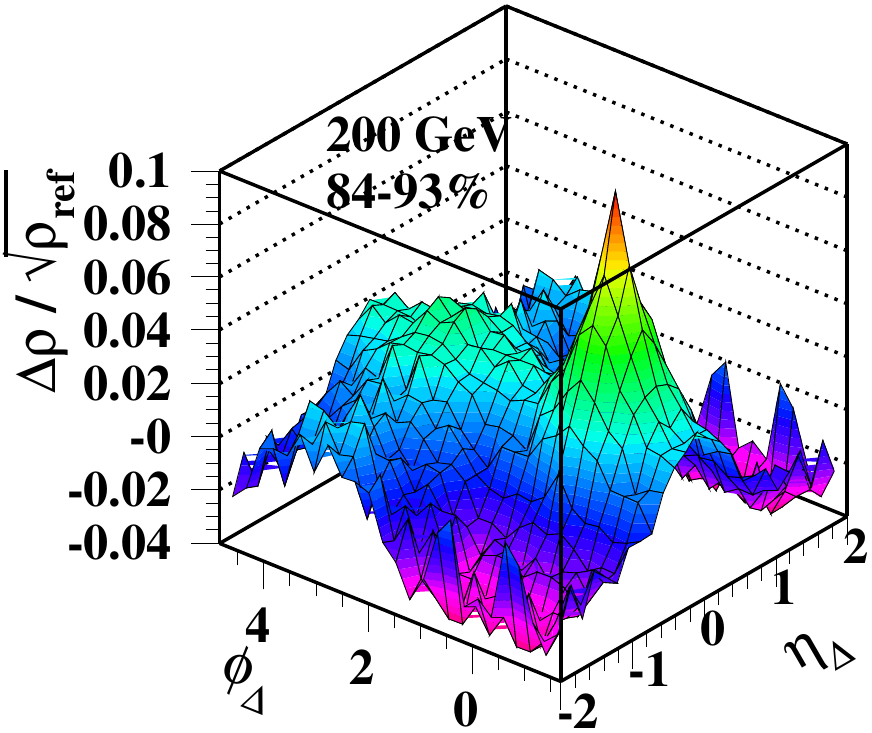}
\includegraphics[width=1.6in]{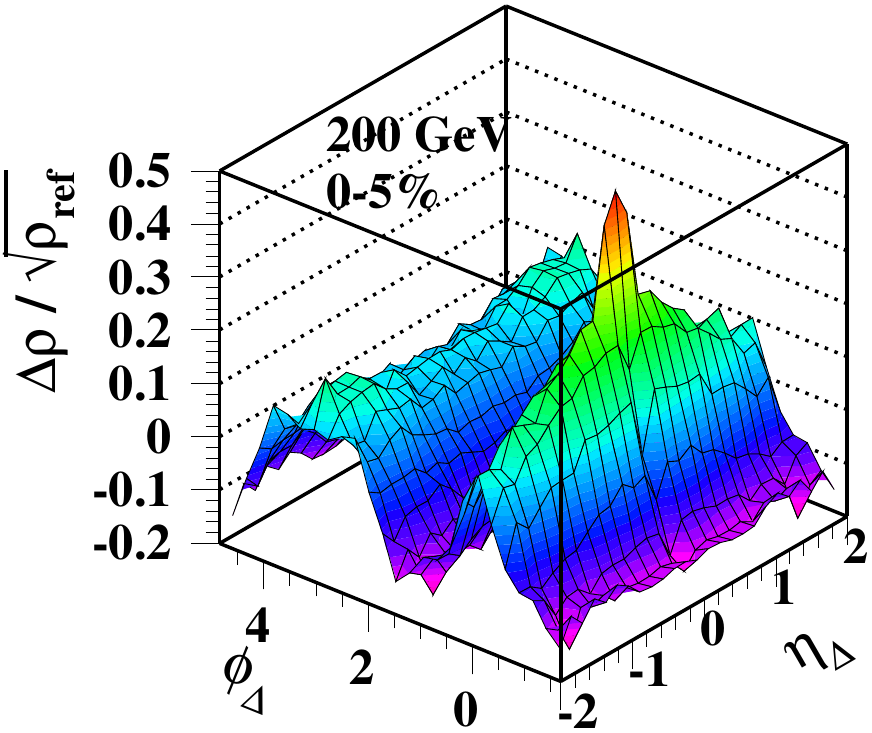}
 \caption{ \label{auau}  
(Color online)
2D angular correlations from most-peripheral (left) and most-central (right) 200 GeV \auau\ collisions~\cite{anomalous}. The prominent structures are jet-related (same-side 2D peak dominated by unlike-sign pairs and away-side ridge), soft (narrow 1D peak on pseudorapidity difference $\eta_\Delta$) and Bose-Einstein correlations (same-side 2D peak dominated by like-sign pairs). The NJ quadrupole is not visible for these cases. The same-side 2D peak is strongly elongated on $\eta_\Delta$ in central collisions.
} % mike-final200-0, 10
\end{figure}
%%%%%%%%%%

%%%%%%%%%%%%%%%%%%%%%%%%%%%%%%%%%%
\begin{figure*}[t]  
\includegraphics[width=.3\textwidth,height=.293\textwidth]{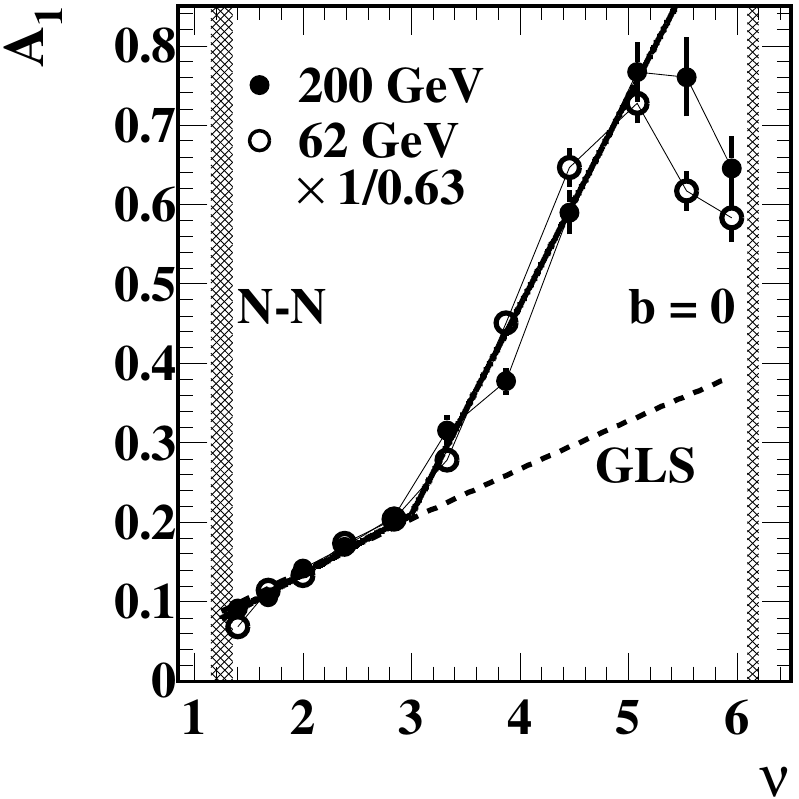}   
\put(-30,35) {\bf (a)}
\includegraphics[width=.3\textwidth,height=.3\textwidth]{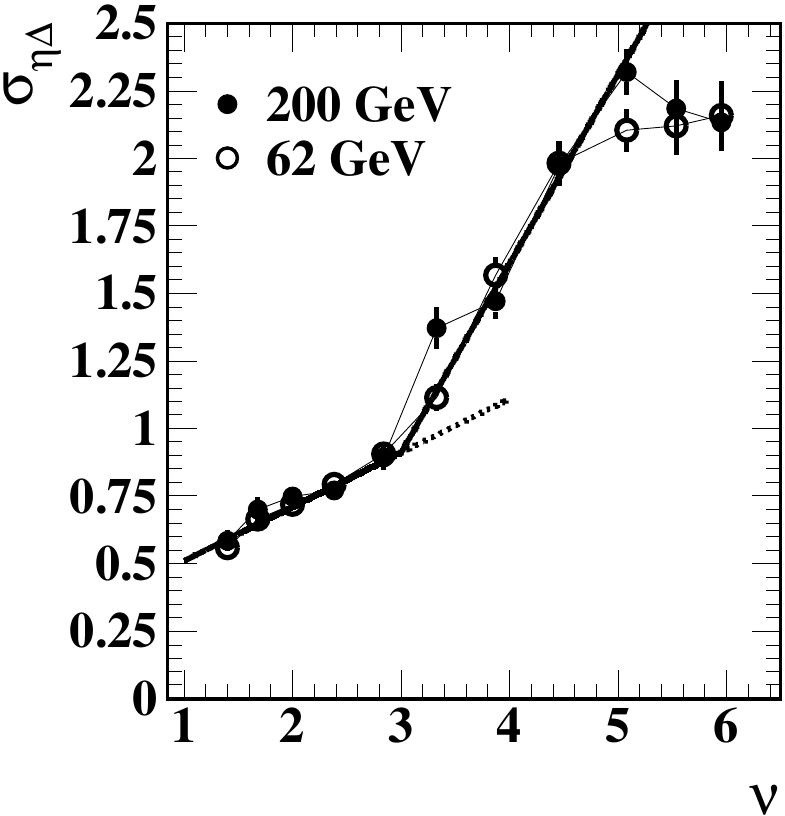} 
\put(-30,35) {\bf (b)}
\includegraphics[width=.3\textwidth,height=.3\textwidth]{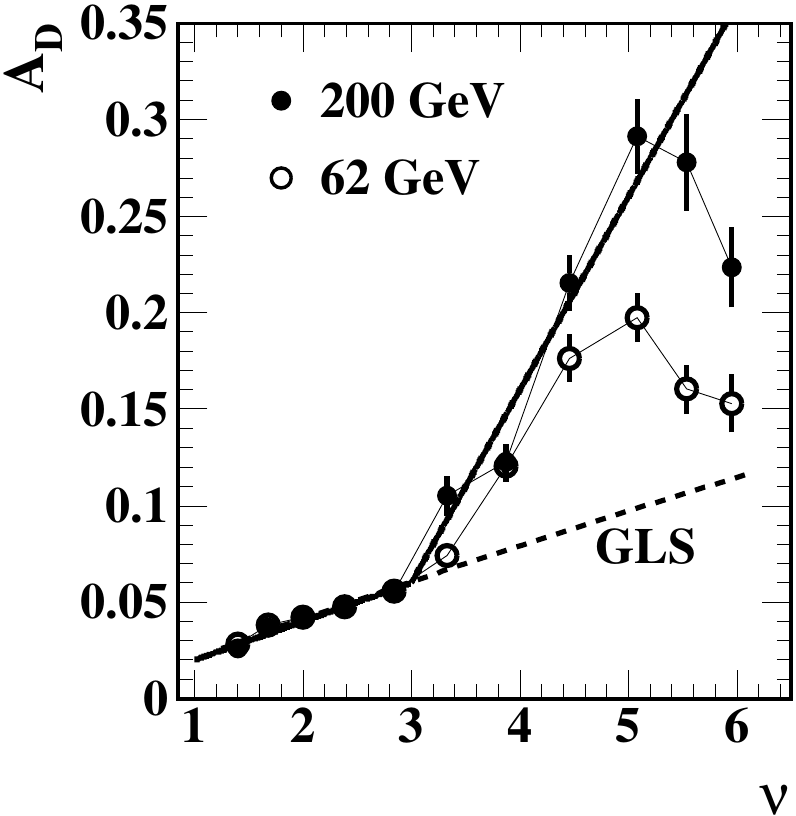}
\put(-30,35) {\bf (c)}
\caption{\label{slopes}
 Fit parameters for $(\eta_{\Delta},\phi_{\Delta})$ correlation data from Au-Au collisions at $\sqrt{s_{NN}}= 62$ (open symbols) and 200~GeV (solid symbols) versus centrality measure $\nu$ illustrating a {\em sharp transition} in centrality trends at $\nu_{trans} \approx 3$~\cite{anomalous}. The dashed curves labeled GLS are calculated predictions corresponding to linear superposition of N-N collisions (binary collision scaling). The bold solid lines illustrate slope changes by factors 3.5 in panels (a) and (b) and 5 in panel (c) within one centrality bin on $\nu$. The dotted line in panel (b) simply continues the slope trend from below the transition. Panel (a) shows that 62 and 200 GeV same-side 2D peak amplitudes $A_1$ are related by a $\log(\sqrt{s_{NN}})$ factor $\approx 0.6$~\cite{davidhq}, whereas panel (c) demonstrates that the away-side 1D peak amplitude $A_D$ is approximately independent of collision energy.
} % mike-final200-30xslope x = a,b,c
\end{figure*}
%%%%%%%%%%%%%%%%%%%%%%%%%%%%%%%%%%

Figure~\ref{auau} shows 2D angular correlations from the most peripheral (left) and most central (right) 200 GeV \auau\ collisions. The statistical errors for those histograms are about 4.5 times larger than for the high-statistics \pp\ data in Fig.~\ref{quadcomp}. The same six-element 2D fit model was applied to the \auau\ data~\cite{anomalous}. The peripheral \auau\ data are approximately equivalent to NSD \pp\ data similar to Figure~\ref{quadcomp} (left) (but before subtraction of two model elements). The NJ quadrupole  in those panels is negligible compared to both the jet-related structure (in both panels) and the soft component (1D peak on $\eta_\Delta$ at the origin in the left panel). Note the narrower BEC 2D peak atop the broader jet-related SS 2D peak in each panel.

Figure~\ref{slopes} shows fitted parameter values vs Glauber centrality parameter $\nu$ for 2D model fits to 200 GeV \auau\ data as in Fig.~\ref{auau} including (a) the SS 2D peak amplitude $A_1$, (b) the SS 2D peak $\eta_\Delta$ width $\sigma_{\eta_\Delta}$ and (c) the AS 1D peak amplitude $A_D$~\cite{anomalous}. The data systematics reveal two intervals on $\nu$ with markedly different behavior: (i) variation of three parameters consistent with Glauber linear superposition (GLS) for $\nu < 3$ equivalent to \aa\ {\em transparency} and (ii) large increases in the rate of variation for $\nu > 3$ where $\nu \approx 3$ corresponds to a factional cross section $\sigma / \sigma_0 \approx 0.5$. The rapid change from one trend to the other is characterized as a ``sharp transition'' or ST in Ref.~\cite{anomalous}. Note that the SS peak amplitudes in panel (a) for 62 GeV coincide for those for 200 GeV when rescaled by factor 1/0.63 representing a $\ln(\sqrt{s} / \text{10 GeV})$ energy trend for jet-related structure that {\em persists above the ST}~\cite{anomalous}. The data in panels (b) and (c) are not rescaled. The close correspondence among three correlation parameters strongly suggests that MB dijets remain the dominant source of correlation structure in \auau\ collisions even for the most-central collisions, although jets are substantially modified there as demonstrated by spectrum data (e.g.\ Fig.~\ref{pidjet}).

\subsection{Trigger-associated 1D correlation analysis} \label{taanalysis}

Data and analysis methods presented above correspond to MB dijets with no conditions imposed on jet structure. Model fits to 2D angular correlations integrated over the entire \pt\ acceptance arguably extract almost all angular-correlation information carried by primary particle data. Alternative methods (a) impose {\em trigger-associated} \pt\ conditions on data, (b) are typically confined to 1D projections onto azimuth and (c) subtract model-dependent backgrounds to arrive at {\em nominal} jet-related structure. Two examples are considered below.

%%%%%%%%%%
 \begin{figure}[h]
  \includegraphics[width=3.2in]{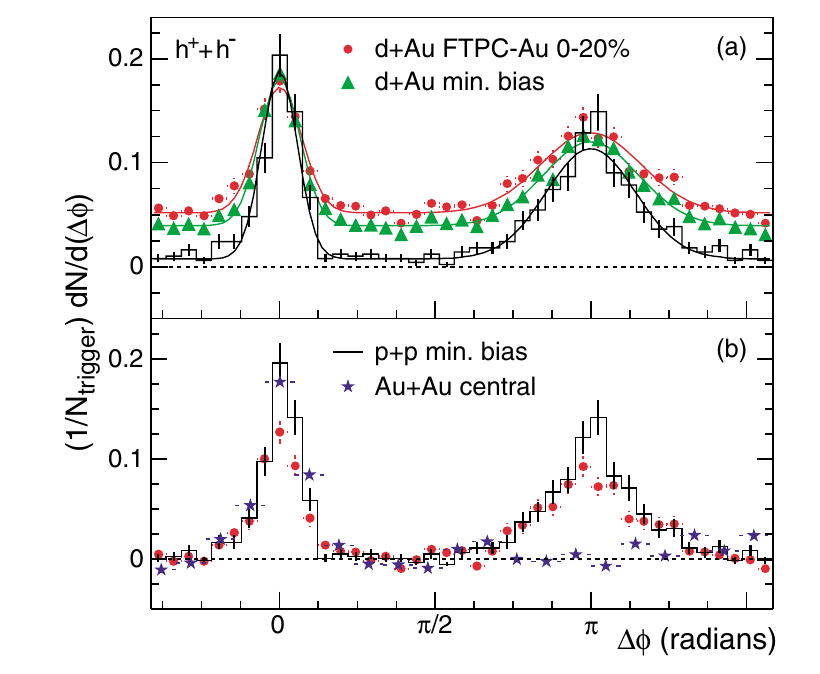} \hfill
\caption{\label{jets2} (Color online)
Jet-related dihadron azimuth correlations demonstrating ``disappearance'' of the away-side jet ($\Delta \phi = \pi$) in 200 GeV \auau\ central collisions (blue stars) compared to its presence in \pp\ collisions (histogram)~\cite{raav21}. The away-side jet remains present in more-central d-Au collisions (red solid points and green triangles).
% FROM nucl-ex/0306024
 }  % Figurx4
\end{figure}
%%%%%%%%%%%%

Figure~\ref{jets2} shows a highly-cited (over 700 citations) trigger-associated analysis of 200 GeV \auau\ azimuth correlations compared to \dau\ and \pp\ data~\cite{raav21}. Trigger-associated \pt\ conditions are $p_{t,trig} \in [4,6]$ GeV/c and $p_{t,assoc} \in [2~\text{GeV/c},p_{t,trig}]$,  admitting only a tiny fraction of all jet fragments observed in spectrum hard components (as in Fig.~\ref{difpi}). In panel (b) backgrounds have been subtracted including a $v_2$ contribution (for \auau\ data)  based on published $v_2$ data that may include a jet contribution in the form of ``nonflow.'' The principal message is that the ``away-side jet'' is suppressed (disappears) in central \auau\ collisions. The suppression is attributed to absorption of jets in a dense medium.

Indications of some form of jet modification (e.g.\ reduction of the \auau\ AS peak amplitude) are clearly apparent but the full implications are not clear. The AS peak ``disappearance'' suggested by  Fig.~\ref{jets2} is consistent (within statistics) with the factor-5 high-\pt\ reduction in SP spectra indicated in  Fig.~\ref{difpi}, but there is no information about changes in jet structure (possible {\em enhancements}) at lower \pt\ as in Fig.~\ref{pidjet}. Imposition of a high-\pt\ trigger on one jet may tend to bias its partner jet to softer fragmentation -- fewer higher-\pt\ fragments but more lower-\pt\ fragments -- independent of any medium effects. One cannot then conclude from such biased data that the AS jet has ``disappeared.'' More can be learned by relaxing the associated-particle \pt\ cut.

\begin{figure}[h]
   \includegraphics[width=3.3in,height=2.6in]{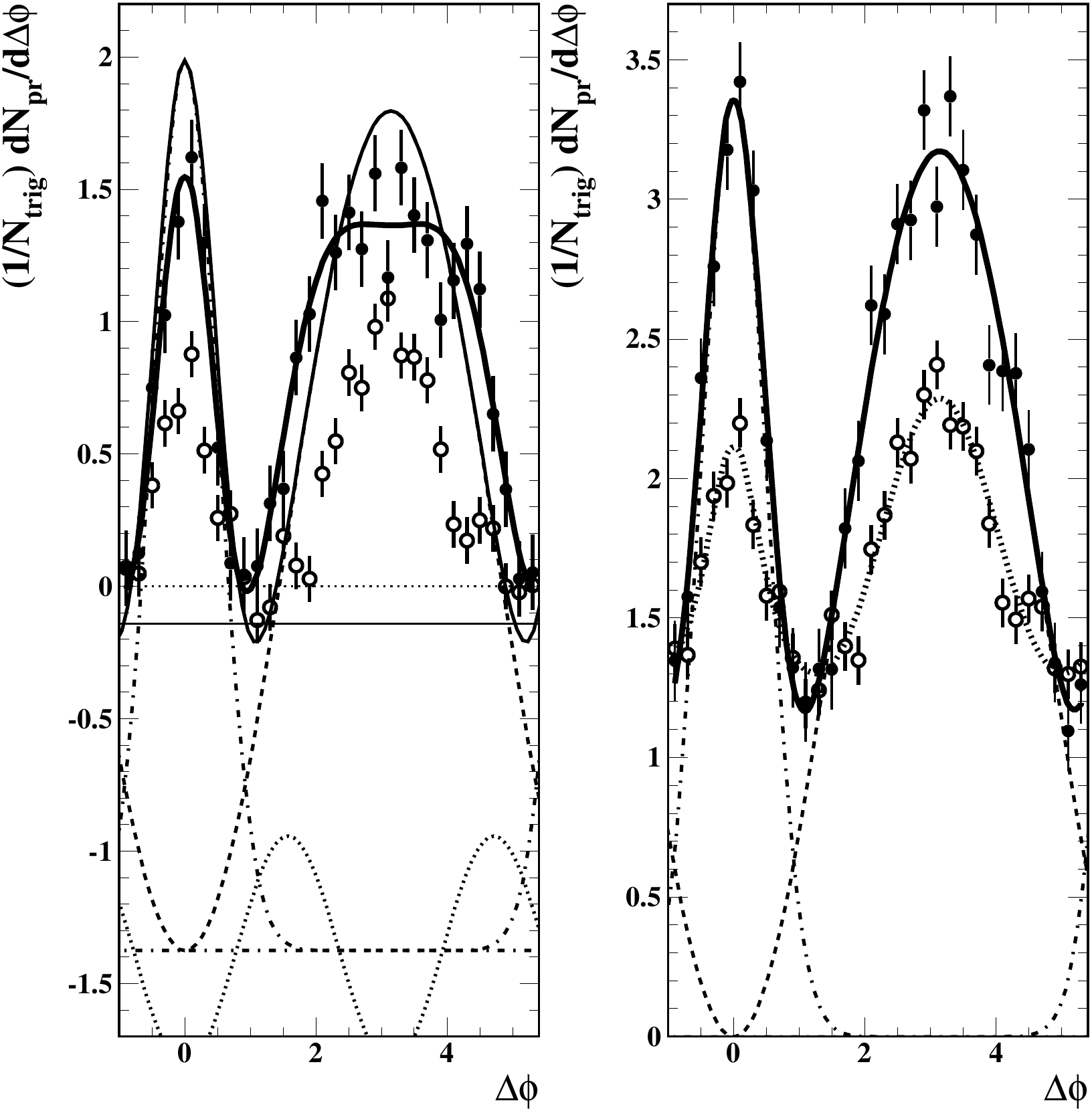}
\caption{\label{starolda}
Left: 
ZYAM-subtracted angular correlations (pairs) for 0-5\% central 200 GeV \auau\ collisions (solid points)~\cite{starprl} with free fit (bold solid) of SS Gaussian (dash-dotted), AS dipole (dashed) and quadrupole (dotted) elements. The open points are \pp\  data relative to ZYAM zero.
Right: 
The same data with ZYAM subtraction reversed. The true zero level is recovered from free fits to data (solid points) and compared to \pp\  data (open symbols) treated with the same method.
 } % newflow34abx
 \end{figure}

Figure~\ref{starolda} (left) shows a similar trigger-associated analysis of 200 GeV \auau\ azimuth correlations with the associated-particle \pt\ condition extended down to the detector-acceptance lower bound: $p_{t,assoc} \in [0.15~\text{GeV/c},p_{t,trig}]$~\cite{starprl}. The \auau\ result (solid points) is compared with \pp\ data treated similarly (open points). 
A combinatoric background to be subtracted from ``raw'' data is defined by published $v_2$ data and a ZYAM ({\em zero yield at minimum}) principle based on the {\em ad hoc} assumption that SS and AS jet peaks {\em never overlap on azimuth}. The data minimum value after background subtraction is then defined as the zero for both \pp\ and \auau\ data. 
Consequences of the updated analysis are two-fold: (a) the AS peak remains substantial, has not ``disappeared'' as reported in Ref.~\cite{raav21}, but (b) the AS peak for central \auau\ collisions appears to be softened and broadened compared to individual \pp\ (\nn) collisions -- interpreted to signal ``progressive equilibration'' of the AS jet in the medium and ``thermalization within the m
An alternative treatment of the same basic correlation data leads to different conclusions. The \auau\ data are fitted with a 1D projection of the 2D fit model from Sec.~\ref{2dfits} described in Refs.~\cite{anomalous,ppquad} including a 1D Gaussian for the SS peak (dash-dotted), a dipole term for the AS peak (dashed) and  a quadrupole term that should correspond to $v_2$ data (dotted). The inferred {\em negative} value for $2v_2^2$ is notable. The fit to \auau\ data is the bold solid curve in Figure~\ref{starolda} (left) that describes those data within statistical uncertainties. The fit parameters then estimate the actual background subtracted in Ref.~\cite{starprl}. The ZYAM offset is 1.37 (lower dash-dotted line) and the assumed background $v_2^2$ value is $|2 v_2^2| \approx 0.43$ (amplitude of dotted curve). The {\em nonjet} quadrupole amplitude for 0-5\% central 200 GeV \auau\ collisions inferred from model fits to 2D angular correlations is essentially zero (upper limit consistent with data uncertainties)~\cite{v2ptb}. The ZYAM $|2 v_2^2| \approx 0.43$ value can then be compared with the quadrupole component of the SS jet peak $2 v_2^2\{2\} \approx 0.25 \times 3.36 = 0.84$~\cite{tzyam} and an assumption that the ``true'' background value is $ v^2_2 = (v^2_2\{2\} + v^2_2\{4\})/2 \approx v^2_2\{2\}/2$ [Ref.~\cite{snellings}, Eqs.~(9) and (10)] so the {\em jet-related} value for central \auau\ is $2v_2^2 \approx 0.42$. It is then likely that the $2v_2^2$ value adopted for the ZYAM subtraction is dominated by the jet-related SS peak Fourier decomposition~\cite{tzyam}. In essence, the quadrupole component of the SS jet peak is subtracted from raw data to produce a distorted result.

%%%%%%%%%%
%%%%%%%%%%%%

Figure~\ref{starolda} (right) shows the same \auau\ data with the ZYAM background subtraction in effect reversed. The data are described within uncertainties (solid) by SS Gaussian (dash-dotted) plus AS dipole (dashed). The \pp\ data (open points) have been treated similarly except with no quadrupole term and are also described within uncertainties by the same basic model (bold dotted). Aside from the amplitude differences the \pp\ SS peak is {\em broader} ($\sigma_\phi \approx 0.7$) than the \auau\ SS peak  ($\sigma_\phi \approx 0.5$) (consistent with 2D correlation trends reported in Ref.~\cite{anomalous}). The \pp\ and \auau\ AS peaks are described by the same dipole model implying equivalent large widths. The peak amplitudes indicate that the number of jet fragments {\em per trigger} is about 70\% larger in central \auau\ than in \pp\ collisions, consistent with the lower-\pt\ enhancements evident in Fig.~\ref{pidjet}. The SS and AS jet peaks {\em strongly overlap} in all collision systems (except for very high \pt\ cuts) consistent with 2D model fits from Ref.~\cite{anomalous} and as illustrated in Figs.~\ref{quadcomp} and~\ref{auau}.  Jet-related correlation analysis on 1D azimuth based on ZYAM background subtraction is thus strongly inconsistent with data properties derived from other contexts.

\subsection{Bayesian analysis of azimuth correlations}

It could be argued that the results in the previous subsection depend on a chosen fit model and are therefore arbitrary. Bayesian analysis provides neutral criteria for comparison of data models based on competition between goodness of fit (e.g.\ the $\chi^2$ measure) and cost of model complexity (based on information parameter $I$) as measured by {\em evidence} $E$ with $-2\ln(E) = \chi^2 + 2I$~\cite{bayes}.

%%%%%%%%%%%%%%%%%%%%%%%
\begin{figure}[h]  
\includegraphics[width=.48\textwidth]{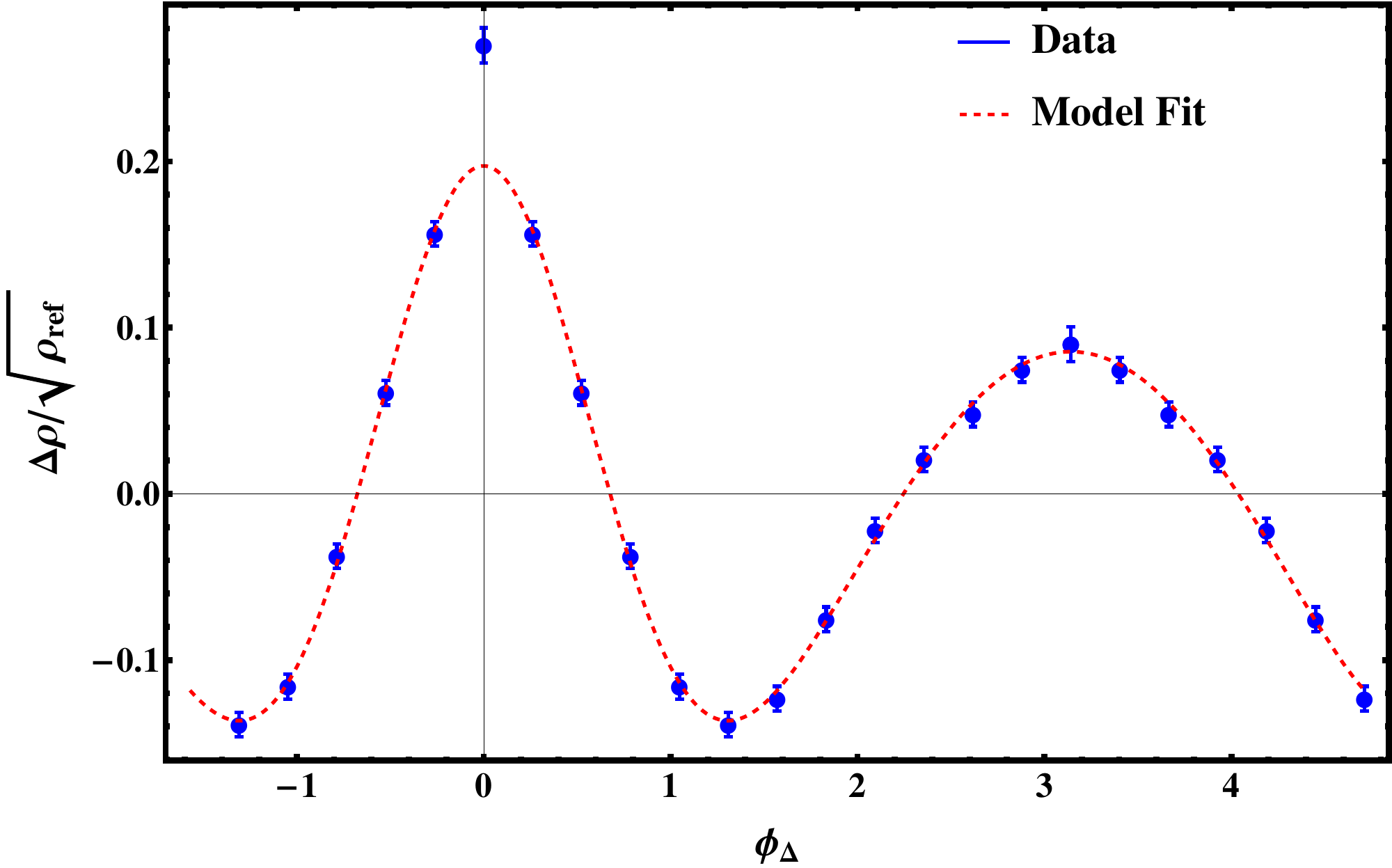}   
%\put(-28,90) {\bf (a)}
%\put(-158,120) {\bf bin 10}
\caption{\label{1ddata} (Color online)
A 1D projection onto azimuth (points) from the 2D data histogram for 0-5\% central 200 GeV \auau\ collisions from Ref.~\cite{anomalous}. The bin-wise statistical errors 0.0037 have been multiplied by 2 to make them visible outside the data points. The (red) dashed curve is obtained from a fit to data with the same 1D model used in  Figure~\ref{starolda}. A fit with a Fourier-series model including four or more terms would appear identical on the scale of this plot.
} % AuAu2002GeV-bin10-Projection
\end{figure}
%%%%%%%%%%%%%%%%%%%%%%%

Figure~\ref{1ddata} shows 2D angular correlation data from 0-5\% central 200 GeV \auau\ collisions as reported in Ref.~\cite{anomalous} projected onto 1D azimuth difference $\phi_\Delta$. A calculated distribution integral has been subtracted from the data. The dashed curve is a fit to data with the same 1D model used in  Figure~\ref{starolda}. A fit with a Fourier-series (FS) model would achieve the same apparent result given a sufficient number of terms. Which model should be preferred?

%%%%%%%%%%%%%%%%%%%%%%%
\begin{figure}[h]  
\includegraphics[width=.48\textwidth,height=.3\textwidth]{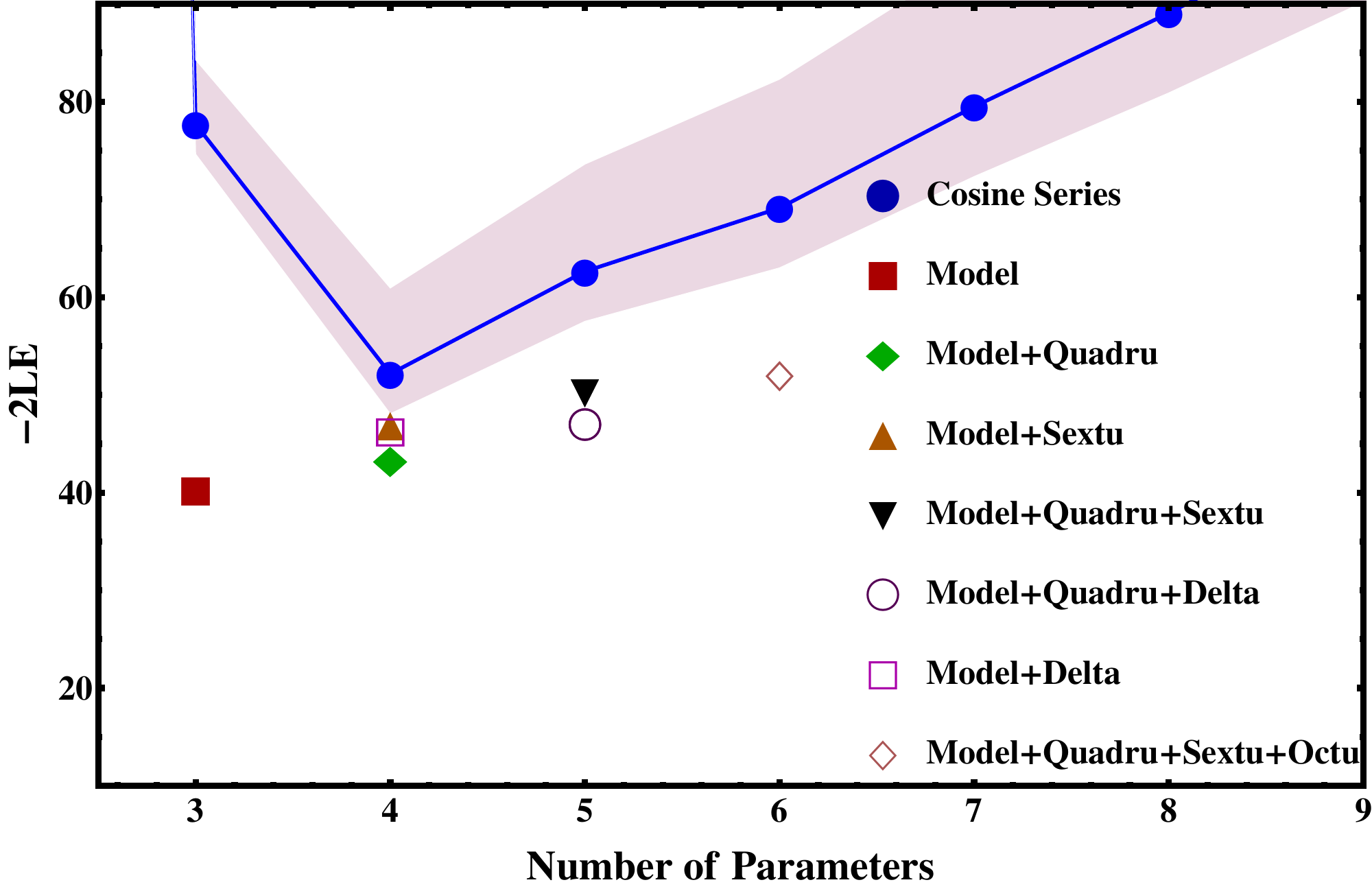}   
%\put(-28,90) {\bf (a)}
%\put(-158,120) {\bf bin 10}
\caption{\label{modelcomp} (Color online)
 Negative log evidence $-2LE$ vs number of parameters $K$ for several models~\cite{bayes}. The basic Model (solid square) is strongly favored over all others (lowest $-LE$). The hatched band indicates the common uncertainty of priors assigned to cosine terms in all models. FS-only models for all $K$ (solid dots and line) are strongly rejected by evidence $E$.
} %AuAu200GeV-bin10-EvidenceModels
\end{figure}
%%%%%%%%%%%%%%%%%%%%%%%

Figure~\ref{modelcomp} shows {\em negative log evidence} $-$2LE $\equiv -2\ln(E)$ values for several competing models, where Bayesian evidence $E$ measures the competition between goodness of fit (better fit increases the evidence) and cost of model complexity (more complexity decreases the evidence). With increasing model complexity (e.g.\ number of parameters) improved fit quality could reduce $\chi^2$ but at the cost of increasing information $I$. The entry labeled ``Model'' (solid square) consisting of SS Gaussian and AS dipole as applied to the data in Fig.~\ref{starolda} achieves the minimum log-evidence value (largest evidence). Adding an $m = 2$ quadrupole term (solid diamond) increases $-$2LE ({\em decreases} evidence $E$) because the additional model parameter does not achieve a compensating improvement in fit quality (consistent with results from Refs.~\cite{anomalous,v2ptb}). The same is true for other additions ($m = 3$ sextupole and $m = 4$ octupole terms).

The $-$2LE trend for a Fourier cosine-series (FS) model (solid dots) achieves a minimum at four elements (including an $m = 3$ sextupole term) and then increases monotonically. In all cases the FS model (commonly interpreted to represent flows) is strongly rejected by Bayesian analysis compared to the simple two-peaked model (consistent with MB dijet production as in Fig.~\ref{starolda}). In Ref.~\cite{bayes} the large difference is traced to the {\em predictivity} of a model. The term in the negative log evidence that increases with model complexity is information $I$ gained by a model upon acquisition of new data.\footnote{Information can be described as a logarithmic measure of volume reduction. In this case ``volume'' refers to some part of a fit-model parameter space. Information $I$ compares the volume allowed before data (prior) to that allowed after data (posterior)~\cite{bayes}.} A fixed model or one with few parameters may gain little or no information from the addition of new data and is thus highly predictive, might be falsified by new data. Predictivity and falsifiability are equivalent concepts. In contrast a Fourier series on periodic azimuth has no predictivity, can describe any data distribution. Thus, introduction of new data adds substantial information to the FS model in the form of parameter adjustments that lead to corresponding increase in the negative log evidence. In effect, the two-peak model predicts that any azimuth distribution from high-energy nuclear collisions should include peaks at $0$ (SS) and $\pi$ (AS),  and the SS peak should be substantially narrower than the AS peak (as expected for MB dijets). New data should modify only the SS peak width and the peak amplitudes. Any data with different features would falsify that model but would remain well described by a FS model.

\subsection{Predicting trigger-associated correlations} \label{trigpredict}

Trigger-associated (TA) analysis of jet structure in 1D azimuth distributions, such as summarized in Sec.~\ref{taanalysis},  follows precedents established at lower energies during a period when the jet concept was not well established and detector technology was quite limited~\cite{isrfirstjets}. With much-higher collision energies, established jet phenomenology, higher-statistics \pp\ and \aa\ data and much-improved particle detectors the analysis of full pair-momentum space is both possible and necessary to access all information carried by particle data. Correlation data with and without \pt\ conditions should be combined quantitatively with SP-spectrum and yield data to provide the strongest possible challenge to competing theories and better inductive understanding of underlying mechanisms.

This subsection summarizes a study in Ref.~\cite{jetcorr} where the pair density on asymmetric TA rapidity space $(y_{t,trig},y_{t,assoc})$ for 200 GeV \pp\ collisions is predicted based on jet data and the TCM for \pp\ SP spectra. When symmetrized the distribution on  $(y_{t,trig},y_{t,assoc})$ can be compared with the distribution on \ytyt\ in Fig.~\ref{ppcorr} (left). 
In a MB TA analysis all \pp\ collision events and all hadrons within a collision are accepted for analysis. A ``trigger particle'' is the single hadron in each collision with  the highest \pt; all other hadrons are ``associated.'' Some fraction of all trigger hadrons may be related to jets (the leading {\em detected} hadron in a jet serving as proxy for the leading parton), and some fraction of all associated hadrons may be fragments from a triggered jet or its back-to-back partner jet. The TA pair distribution on  $(y_{t,trig},y_{t,assoc})$ resulting from such conditions includes soft-soft (SS), soft-hard (SH) and hard-hard (HH) pair combinations. It is assumed that the HH contribution is amenable to prediction based on measured jet properties. A full analysis involves two parts: (a) predict the HH contribution to MB TA correlations~\cite{jetcorr} and (b) extract the HH contribution from measured MB TA correlations for direct comparison based on the SP spectrum TCM~\cite{tacorrexp}. The full derivations  are summarized schematically below.

(a) The HH component of TA correlations can be predicted from measured conditional FFs $D_u(y|y_{max})$ and jet (parton) spectrum $\hat S_p(y_{max}) = (1/\sigma_j) d^2\sigma_j/dy_{max}/d\eta$ introduced in Sec.~\ref{jets}~\cite{jetcorr}. FF distribution $D_u(y|y_{max})$ is first decomposed into a trigger component $\hat S_t(y_{trig}|y_{max})$ and an associated component $D_a(y_{assoc}|y_{max})$ based on void probability $G_t(y|y_{max})$. The void probability is defined as the Poisson probability that no fragment appears for $y > y_{trig}$ based on the FF integral over that interval, which is just the probability that a fragment at $y = y_{trig}$ is a trigger particle for given $y_{max}$. Trigger spectrum $\hat S_t(y_{trig})$ is then obtained by convoluting $\hat S_t(y_{trig}|y_{max})$ with the measured jet spectrum $\hat S_p(y_{max})$. An intermediate conditional spectrum is obtained from Bayes' theorem as $\hat S_p(y_{max}|y_{trig}) = \hat S_t(y_{trig}|y_{max})\hat S_p(y_{max}) / \hat S_t(y_{trig})$. The associated spectrum $D_a(y_{assoc}|y_{max})$ is the complement of $\hat S_t(y_{trig}|y_{max})$ in $D_u(y|y_{max})$. The associated fragment distribution $D_a(y_{assoc}|y_{trig})$ is obtained by convoluting $D_a(y_{assoc}|y_{max})$ with $\hat S_p(y_{max}|y_{trig})$, thus eliminating jet rapidity $y_{max}$. The TA HH distribution {\em per jet} derived from eventwise-reconstructed jet data is then
\bea
F_{at}(y_{assoc},y_{trig}) &=& \hat S_t(y_{trig}) D_a(y_{assoc}|y_{trig}).
\eea

\begin{figure}[h]
\includegraphics[width=3.3in]{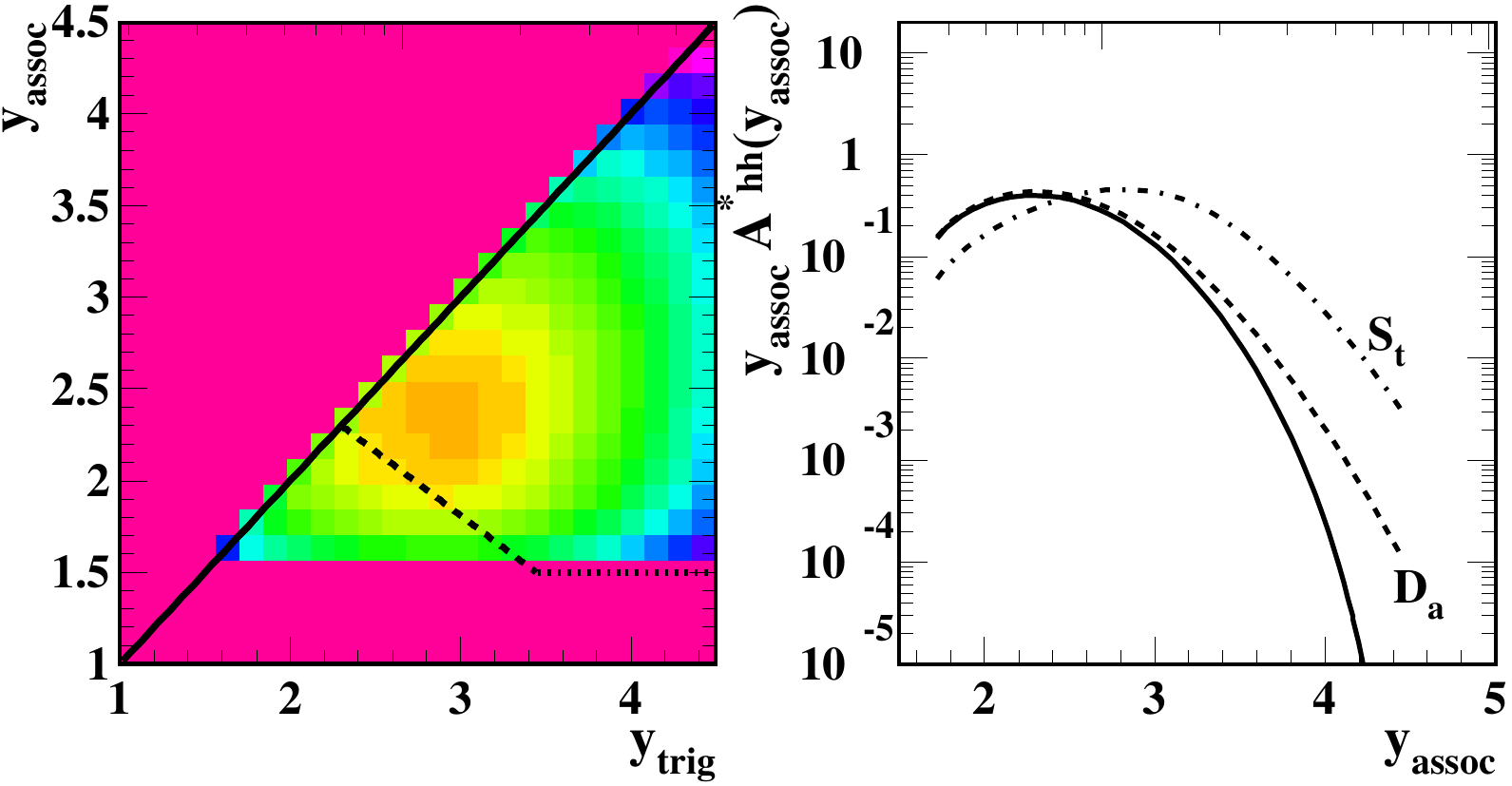}
\put(-67.2,109) {\scriptsize \bf 1}
 \put(-30,109) {\scriptsize \bf 4}
 \put(-78,103) {\scriptsize $\bf p_{assoc} (GeV/c)$}\\
\caption{\label{alephgg}
(Color online) 
Left: TA joint distribution $F_{at}(y_{assoc},y_{trig})$ as the product of associated-fragment distribution $D_a(y_{assoc}|y_{trig})$ with  trigger-fragment spectrum $\hat S_t(y_{trig})$~\cite{jetcorr}. The z axis is logarithmic.
Right: Projection of the histogram at left onto $y_{assoc}$ to obtain $D_{a}(y_{assoc})$ (dashed and solid curves) and trigger spectrum $\hat S_t(y_{trig})$ (dash-dotted), both derived from \ppbar\ FFs and jet spectrum.
 }  %aleph11ggnew
 \end{figure}

Figure~\ref{alephgg} (left) shows HH distribution $F_{at}(y_{assoc},y_{trig})$ for 200 GeV \pp\ collisions based on the measured \ppbar\ FFs and 200 GeV jet spectrum summarized in Sec.~\ref{jets}. The mode on $y_{trig}$ corresponds to $p_t \approx 1.2$ GeV, and the mode on $y_{assoc}$ corresponds to $p_t \approx 0.6$ GeV/c. Data from eventwise-reconstructed jets thus predict that the {\em great majority} of jet-related TA pairs appear near 1 GeV/c.

%%%%%%%%%%
%%%%%%%%%%%%

Figure~\ref{alephgg} (right) shows 1D projections $D_a(y_{assoc})$ (solid) and $\hat S_t(y_{trig})$ (dash-dotted). The dashed curve shows $D_a(y_{assoc})$ with $F_{at}(y_{assoc},y_{trig})$ extrapolated to large $y_{trig}$ (beyond the 2D plot boundaries at left).

(b) Isolation of hard component HH from correlation data requires a TA TCM based on the SP spectrum TCM in Sec.~\ref{tcm1}~\cite{tacorrexp}. The TA TCM assumes that hadrons from soft and hard components of the SP spectrum are {\em uncorrelated in pairs}. The TA TCM is factorized in the form $F_{at}(y_{ta},y_{tt}) = \hat T(y_{tt}) A(y_{ta}|y_{tt})$, the product of a unit-integral trigger spectrum and a conditional associated-particle distribution. The derivation distinguishes \pp\ soft events (no jet within acceptance) from hard events (at least one jet within acceptance) with probabilities $P_s$ and $P_h$ depending on event multiplicity \nch\ based on measured \pp\ dijet systematics~\cite{ppquad}. Because \pp\ dijets $\propto n_s^2$, multiple dijets may appear within acceptance $\Delta \eta$ for higher \pp\ multiplicities. The trigger spectrum is then
\bea
\hat T(y_{tt};n_{ch}) &=&  P_s(n_{ch}) \hat T_s(y_{tt}) + P_h(n_{ch}) \hat T_h(y_{tt}),
\eea
where each $\hat T_x$ is defined by a corresponding void probability based on the appropriate TCM spectrum model -- with or without a spectrum hard component. The TCM for the full TA distribution is then
\bea
F_{at}(y_{ta},y_{tt};n_{ch}) &=& P_s \hat T_s A_{ss} + P_h \hat T_h (A_{hs} + A_{hh})~~
\eea
that can be compared with a measured $F_{at}$ TA distribution. Each column of $A_{xy}(y_{ta}|y_{tt})$ on $y_{ta}$ is a SP spectrum TCM soft (+ hard where applicable) component set to zero above trigger rapidity $y_{tt}$ (comprising the void) as the condition, each normalized to the associated-particle number $\hat n_{ch} - 1$ for that event class. The object of analysis for comparison with jet correlations is the {\em per jet} hard component of hard events $A^*_{hh}$. To obtain that quantity from TA data the appropriate nonjet elements of the TA TCM are subtracted from the TA data. $A = F_{ta} / \hat T$ is first obtained from TA data using the TCM trigger spectrum. The associated hard component HH is then isolated by subtracting SS and HS TCM elements
\bea
P_h R_h A_{hh} = A - P_s R_s A_{ss} - P_h R_h A_{hs},
\eea
where $R_x \equiv \hat T_x / \hat T$. Two more steps are required to obtain $A^*_{hh}$ (correlations from trigger jet only) from $A_{hh}$ (correlations from soft and hard triggers and from both trigger jet and partner jet if present) as described in Ref.~\cite{jetcorr}.

\begin{figure}[h]
 \includegraphics[width=3.3in]{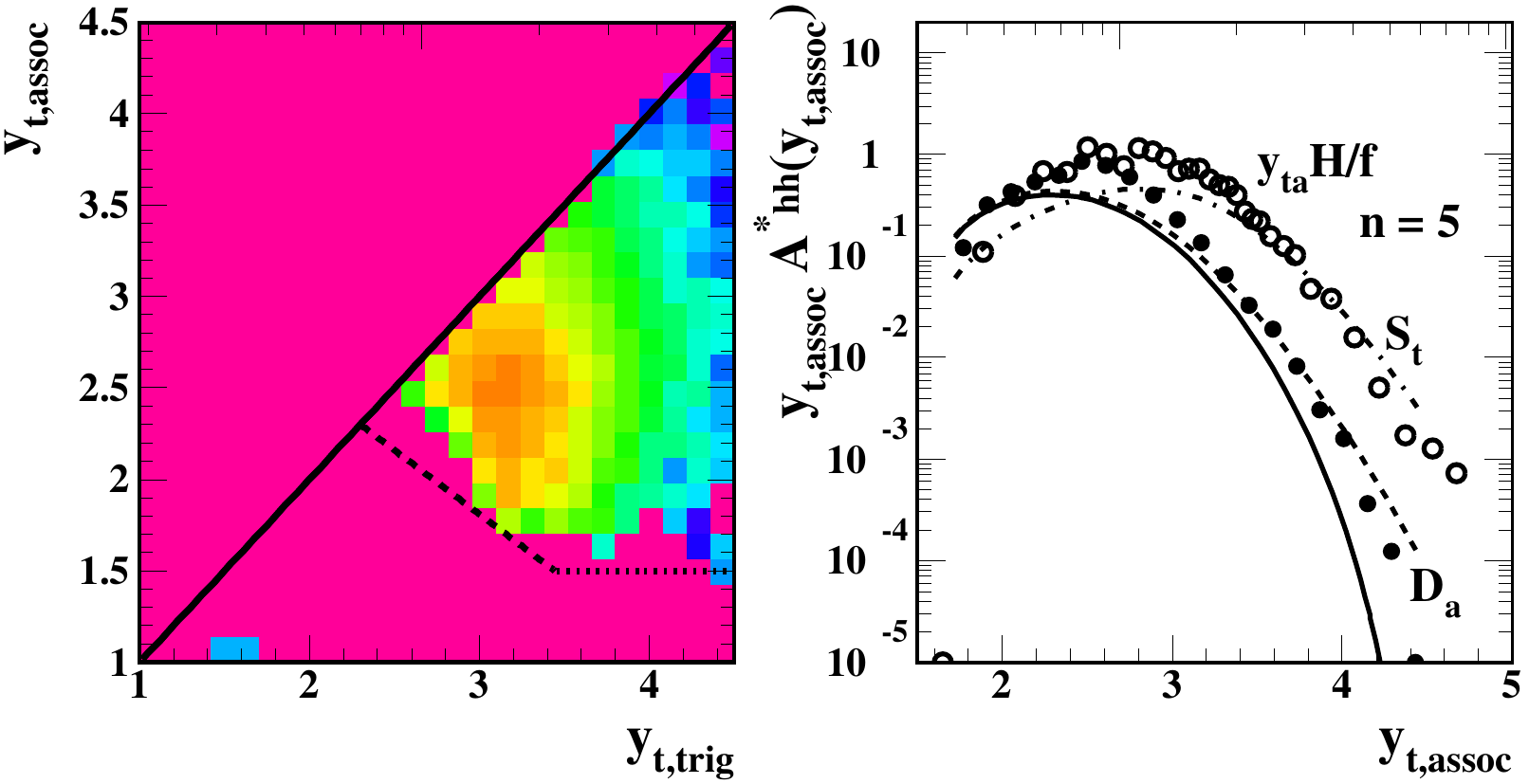}
%\put(-212,108) {\bf (a)}
% \put(-90,108) {\bf (b)}
\put(-65.6,109) {\scriptsize \bf 1}
 \put(-30,109) {\scriptsize \bf 4}
 \put(-78,103) {\scriptsize $\bf p_{t,assoc} (GeV/c)$}\\
 % \includegraphics[width=3.3in]{ppcms116d3}
%\put(-212,108) {\bf (c)}
% \put(-90,108) {\bf (d)}\\
%  \includegraphics[width=3.3in]{ppcms116d5y}
%\put(-212,108) {\bf (c)}
% \put(-90,108) {\bf (d)}
\caption{\label{compare2}
(Color online) 
Left:  
The per-dijet hard component of TA joint distribution $F_{at}$ in the form $\hat T_{h}(y_{tt},n_{ch}) y_{ta}A^*_{hh}(y_{ta}|y_{tt},n_{ch})$ from 200 GeV \pp\ collisions for multiplicity class $n = 5$~\cite{tacorrexp}. The z-axis limits (log scale) are the same as Fig.~\ref{alephgg} (left).
Right: 
Projections of the histogram in the left panel onto $y_{t,assoc}$ (solid points) compared to equivalent projections (solid and dashed curves) of $D_a(y_assoc|y_{trig})$ derived from measured FFs~\cite{fragevo} and reconstructed-jet spectra~\cite{jetspec2}. The 200 GeV \pp\ \yt\ spectrum hard component (open circles) and FF trigger spectrum $\hat S_t(y_{trig})$ are included for comparison.
 }  %ppcms116d5y
%\\ this is [k]7 = Th yta A*hh 
 \end{figure}

Figure~\ref{compare2} (left) shows the {\em per jet} hard component of TA correlations $F_{at}$ in the form $\hat T_h y_{ta} A^*_{hh}$ for multiplicity class $n = 5$ from 200 GeV \pp\ collisions. That result can be compared directly with the prediction in Fig.~\ref{alephgg} (left) derived from eventwise-reconstructed jet data. The z-axis limits are the same for the two plots. There is quantitative correspondence except near the low-\yt\ boundary.

%%%%%%%%%%
%%%%%%%%%%%%

Figure~\ref{compare2} (right) shows a projection of the left panel onto $y_{ta}$ (solid points) that can be compared directly with projection $D_a(y_{assoc})$ (mean associated fragment distribution) from Fig.~\ref{alephgg} (right) (solid and dashed curves). Also plotted is the hard component from 200 GeV NSD \pp\ collisions (open points) in the {\em per jet} form $y_{t,assoc} H / f$ [also plotted in Fig.~\ref{ppfd} (right)] where $f$ is the dijet $\eta$ density per \pp\ collision. For both SP spectra as in Fig.~\ref{ppfd} and more-complex TA correlations as in Figs.~\ref{alephgg} and \ref{compare2} there is a quantitative correspondence between eventwise-reconstructed jets and MB dijet manifestations in \pp\ spectra and correlations. The correspondence extends down to at least $p_t = 0.3$ GeV/c, and most jet fragments ($> 90$\%) appear {\em below} 2 GeV/c ($y_t \approx 3.3$).

It is notable that trigger-associated \pt\ cuts applied to data as in Fig.~\ref{jets2} from Ref.~\cite{raav21} or Fig.~\ref{starolda} from Ref.~\cite{starprl} correspond respectively to rectangles on $(y_{t,assoc},y_{t,trig})$ in Fig.~\ref{compare2} (left) bounded by $([3.3,y_{t,trig}],[4,4.5])$ and $([1,y_{t,trig}],[4,4.5])$ for the two cases. Such cuts include only a tiny fraction of correlated pairs and exclude the dominant TA peak near $y_t = 2.7$ ($p_t \approx 1$ GeV/c). It has been argued that since pQCD calculations are valid only for ``high-\pt'' fragments and energetic partons high-\pt\ cuts are required for valid jet analysis. But the validity of pQCD calculations is not relevant to comparisons between MB jet phenomena and {\em measured} jet properties (FFs and jet spectra) as above.

%%%%%%%%%
 \section{Jets and $\bf p_t$ fluctuations} \label{jetptfluct}

When the $\sqrt{s_{NN}} = 17$ GeV \pbpb\ program commenced at  the CERN super proton synchrotron it was expected that certain {\em eventwise fluctuation} measures might reflect thermodynamic properties of a quark-gluon plasma. In particular, eventwise mean-\pt\ fluctuations might serve as a proxy for temperature fluctuations within a locally-thermalized plasma~\cite{na49ptfluct}. Exceptional fluctuations (i.e.\ deviations from independent particle emission) might signal proximity to a QCD phase boundary and reveal its nature. That concept continues to provide the principal context for recent fluctuation studies~\cite{aliceptfluct} but does not include possible contributions to \pt\ fluctuations from MB jets that are considered below.

\subsection{Fluctuation measures}

Fluctuation measure definitions described below follow those presented in Ref.~\cite{aliceptflucttom}. $\bar n_{ch}$ and $\bar p_t$ are event-ensemble means averaged over a detector acceptance. To reduce notation complexity $n_{ch} \leftrightarrow n$ where there is no ambiguity. Symbols $\Delta \sigma^2_x$ denote {\em per-particle} variance differences that are negligible in case of central-limit conditions (CLT) wherein data are comprised of (a) independent samples from (b) a fixed parent distribution~\cite{clt}. Deviations from the CLT may represent significant two-particle correlations. Extensive RVs $n_{ch}$ and $P_t$ represent eventwise integrals over some angular domain: a full detector acceptance or smaller bins within an acceptance as described in Ref.~\cite{ptscale}. Overlines or bars represent ensemble means whereas angle brackets represent eventwise means. The charge-multiplicity variance is
$\sigma^2_{n} = \overline{(n_{ch} - \bar n_{ch})^2} \equiv \bar n_{ch}(1 + \Delta \sigma^2_n)$, where $\Delta \sigma^2_n$ represents a non-Poisson contribution from correlations.

Most measurements of ``mean-\pt\ fluctuations'' are based on one of two statistical measures with different interpretations. If the emphasis is on eventwise mean \pt\ represented by {\em intensive} ratio $\langle p_t \rangle = P_t / n_{ch}$ as a proxy for local temperature the conventional fluctuation measure is
$\sigma^2_{\langle p_t \rangle} = \overline{(\langle p_t \rangle - \bar p_t)^2} \approx (\sigma^2_{p_t} + \Delta \sigma^2_{\langle p_t \rangle}) / \bar n_{ch}$. Alternatively, a system based on {\em extensive} quantities is conditional measure $\sigma^2_{P_t|n} = \overline{(P_t - n_{ch} \bar p_t)^2} = \bar n_{ch} (\sigma^2_{p_t} + \Delta \sigma^2_{P_t|n})$ where in each case $\Delta \sigma^2_x$ represents a non-CLT contribution from two-particle correlations and/or dynamical fluctuations that may shed light on collision mechanisms. 

In Ref.~\cite{aliceptflucttom} several statistical measures are compared in reference to \pt\ fluctuation measurements reported in Ref.~\cite{aliceptfluct}. An extensive differential measure is denoted by $\bar B = \sigma^2_{P_t|n} - \bar n_{ch} \sigma^2_{p_t}$ with associated per-particle measure $\Delta \sigma^2_{P_t|n} = \bar B / \bar n_{ch}$, both negligible for CLT conditions. In Ref.~\cite{aliceptfluct} the preferred intensive measure is $C \equiv \bar B / \overline{n_{ch}(n_{ch}-1)} \approx \bar B / \bar n_{ch}^2$. \pt\ fluctuations are then reported in terms of the r.m.s.\ quantity $\sqrt{C} / \bar p_t \approx \sqrt{\bar B / \bar P_t^2}$. One motivation for that construction may be that $C \approx \sigma^2_{\langle p_t \rangle} - \sigma^2_{p_t} / \bar n_{ch}$, so $\sqrt{C} / \bar p_t$ may be interpreted as a {\em fractional} r.m.s.\ $\langle p_t \rangle$ fluctuation measure. However, that choice is based on several questionable assumptions, in particular that a local temperature is a relevant concept and that {\em non}\,thermodynamic sources (e.g.\ MB dijets) do not dominate collision dynamics. Per-pair measure $C$ tends to decrease strongly with increasing system size (i.e.\ \aa\ collision centrality) other things being equal, which might suggest that thermalization increases with \aa\ centrality. In contrast, fluctuations of extensive measures $n_{ch}$ and $P_t$ can (and do) exhibit informative TCM trends (strong increases with \aa\ centrality) that are concealed by the intensive ratio $\langle p_t \rangle = P_t / n_{ch}$.

\subsection{A-A fluctuation data at 200 GeV and 2.76 TeV}

\begin{figure}[h]
\includegraphics[width=1.65in]{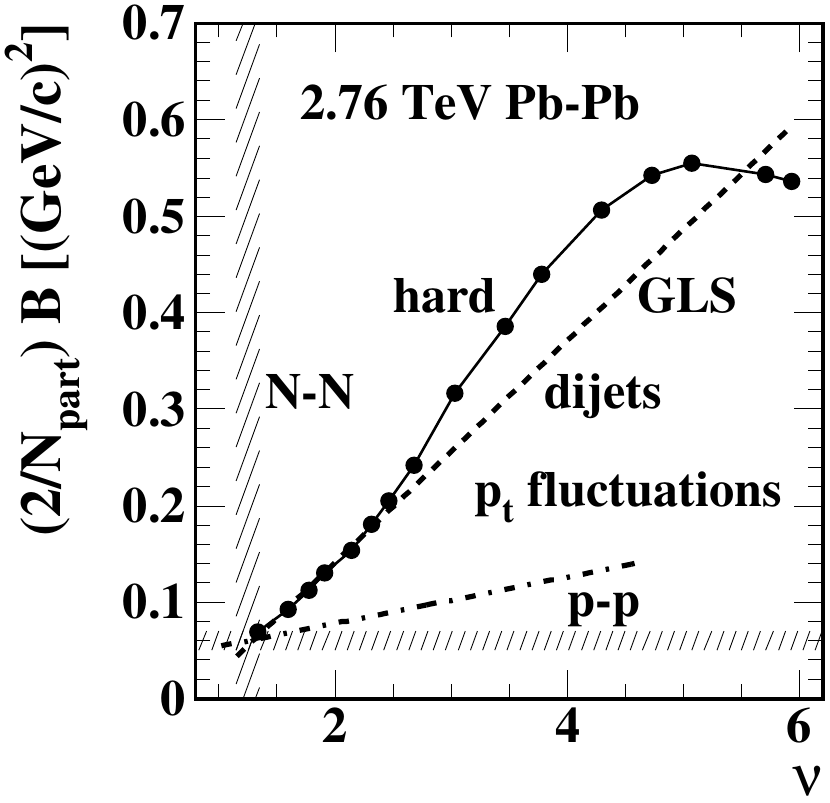}
\includegraphics[width=1.65in]{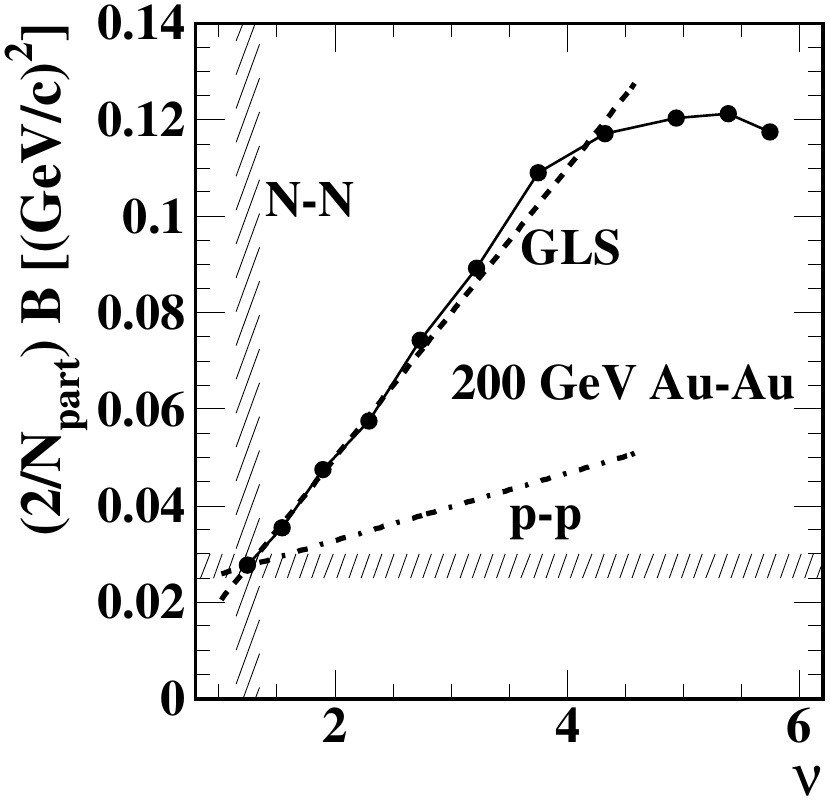}
\caption{\label{xxx}
Left:
Per-particle \pt\ fluctuation data from Ref.~\cite{aliceptflucttom} converted to $\bar B / \bar n_{ch}$ and multiplied by factor $2\bar n_{ch}/N_{part}$ to obtain per-participant trends for 2.76 TeV \pbpb\ collisions. The dashed lines represent a TCM GLS reference (transparent \aa\ collisions). The dash-dotted lines represent TCM ``first-hit'' trends extrapolated from \pp\ data. Solid lines guide the eye.
Right:
Comparable results for 200 GeV \auau\ collisions derived from results reported in Ref.~\cite{ptscale}.
} %alice11e, 11i
\end{figure}

Figure~\ref{xxx} (left) shows \pt\ fluctuation data for 2.76 TeV \pbpb\ collisions from  Ref.~\cite{aliceptfluct} in the form $\sqrt{C} / \bar p_t$ (which decreases monotonically with increasing centrality) transformed to $(2/N_{part}) \bar B$ rather than $\bar B / \bar n_{ch}$ (a per-particle measure in terms of initial-state participant nucleons rather than final-state charged hadrons) plotted vs mean participant pathlength $\nu$. The trend for more-peripheral collisions is consistent with a TCM GLS trend expected for dijet production in transparent \aa\ collisions (dashed line). In more-central collisions the data significantly exceed the GLS trend consistent with increased fragment yields from jet modification as noted for 200 GeV \auau\ collisions in Fig.~\ref{slopes} (a) and (c).

%%%%%%%%%%%%%%%%%%%%%%%%%%%%%%%
%%%%%%%%%%%%%%%%%%%%%%

Figure~\ref{xxx} (right) shows comparable TCM results for 200 GeV \auau\ collisions reported in Ref.~\cite{ptscale}. The general trend is similar but with reduced overall amplitude as expected given the $\log(s/s_0)$ collision-energy dependence of dijet production~\cite{jetspec2}. The dash-dotted curves are the GLS expectations extrapolated from isolated \pp\ collisions. The dashed lines relate to production from ``wounded'' projectile nucleons after a first \nn\ collision.

\subsection{Inversion of fluctuation scaling at 200 GeV}

Fluctuation data for a given collision system depend strongly (and possibly nonmonotonically) on the detector acceptance or bin size of a particular analysis as demonstrated in Ref.~\cite{inverse}. Results from different experiments are therefore not simply comparable. However, measurements of the scale (bin size) dependence of fluctuations are ``portable'' and may be inverted to infer underlying angular correlations as first demonstrated in Ref.~\cite{ptscale}.

\begin{figure}[h]
  \includegraphics[width=1.65in]{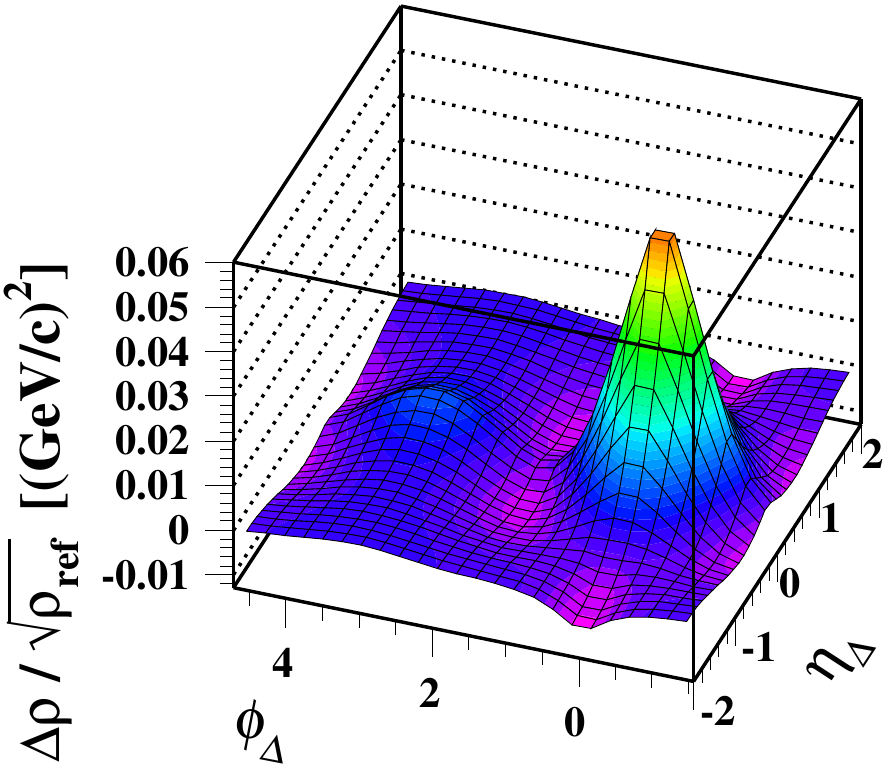}
  \put(-95,90) {\bf (a)}
  \includegraphics[width=1.65in]{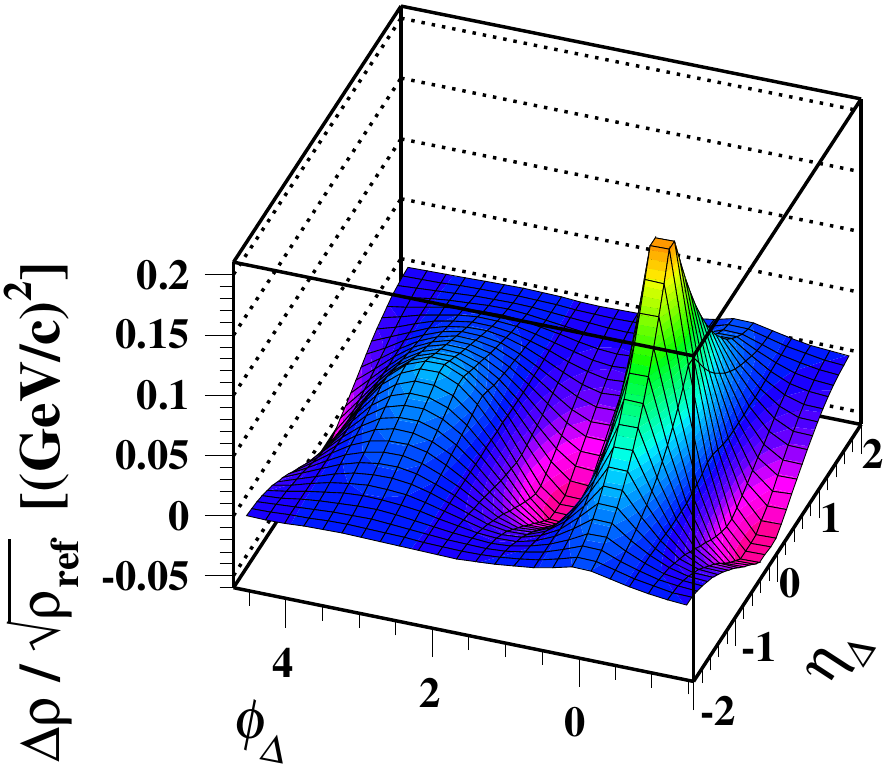}
  \put(-95,90) {\bf (b)}  \\
  \includegraphics[width=1.65in]{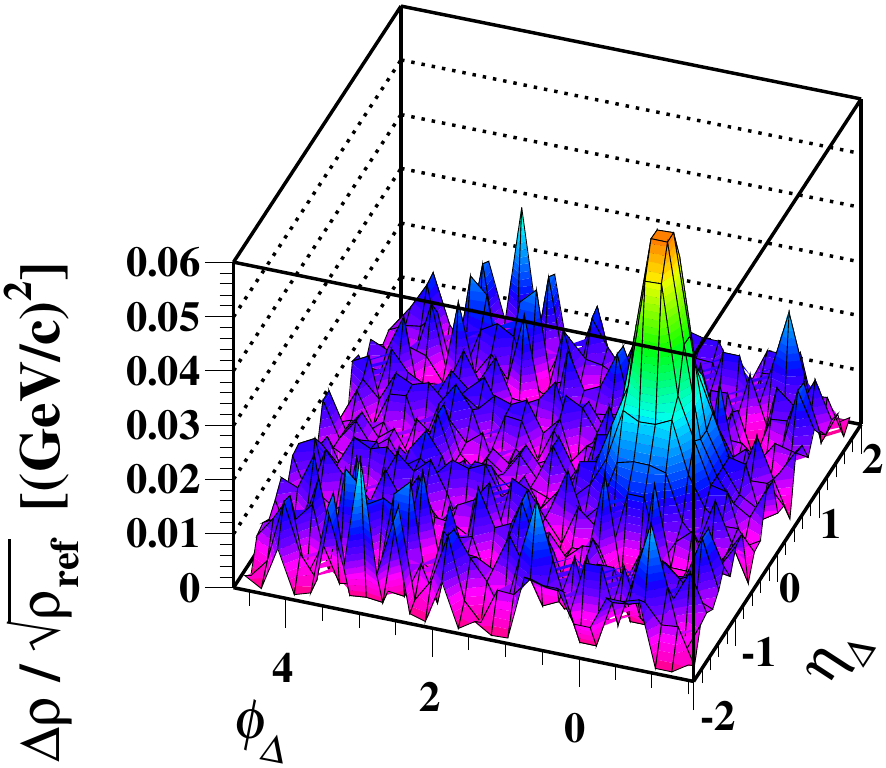}
  \put(-95,90) {\bf (c)}  
  \includegraphics[width=1.65in]{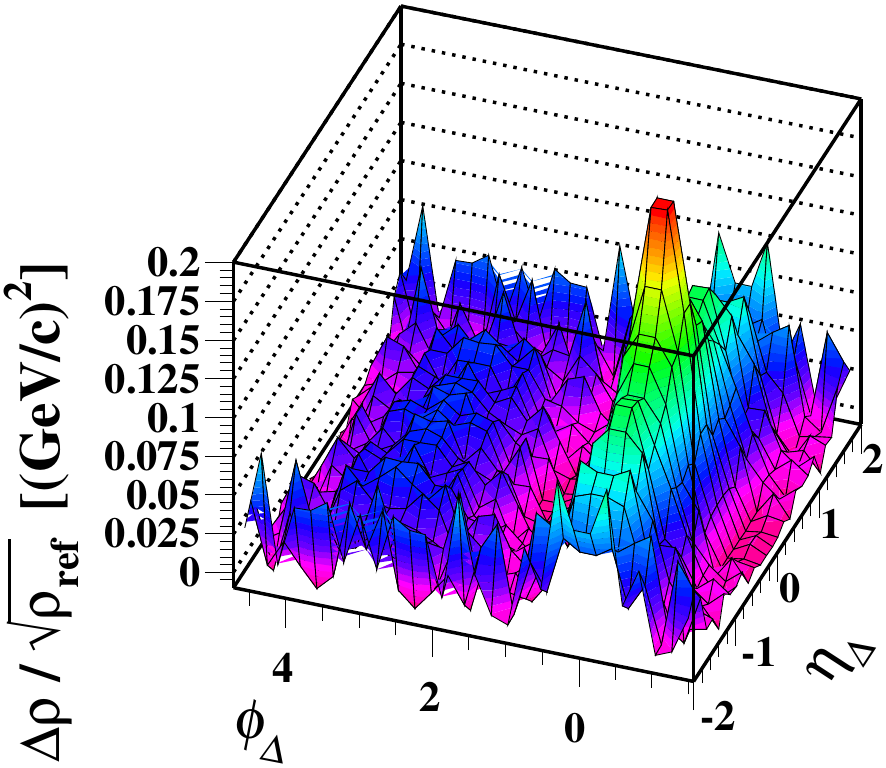}
  \put(-95,90) {\bf (d)}\caption{\label{ptscale} (Color online)
Upper: \pt\ angular correlations for  (a) 85-95\% and (b) 10-20\% central 200 GeV \auau\ collisions inferred by inverting \pt\ fluctuation scale dependence~\cite{ptscale}. AS dipole and nonjet quadrupole components of 2D model fits to the data are subtracted.
Lower:  Results for the same collision systems but \pt\ correlations are obtained by direct pair counting rather than fluctuation inversion. Improved angular resolution and unfiltered statistical fluctuations are evident. Data for all panels include an additional acceptance factor $4\pi$.
 }  %ptautono200-0,8, ptautonewmich0,8
 \end{figure}

Figure~\ref{ptscale} (a) and (b) show \pt\ angular correlations from peripheral and central 200 GeV \auau\ collisions inferred from inversion of fluctuation scale dependence~\cite{ptscale}. Model elements representing an AS 1D peak and nonjet quadrupole have been subtracted to isolate the SS 2D peak structure. The structure is equivalent to that in Figs.~\ref{quadcomp} and \ref{auau}, and those results appear to confirm that \pt\ fluctuations are dominated by MB dijets. However, the complex inversion procedure could be questioned.

%%%%%%%%%
%%%%%%%%%%%%

Figure~\ref{ptscale} (c) and (d) show \pt\ angular correlations from the same collision systems determined directly by pair counting, not by inversion of fluctuation scaling. The  results are in quantitative agreement with (a) and (b) confirming that dijets are the dominant source of \pt\ fluctuations in high-energy nuclear collisions. There are significant differences due to reduced angular resolution of the inversion process, but the equivalence is clearly apparent. \pt\ fluctuation data compel the conclusion that a local temperature is not relevant  for \pt\ fluctuation measurements and their interpretation in high-energy nuclear collisions, and direct correlation measurements by pair counting supply equivalent information more accurately.

%%%%%%%%%
 \section{Discussion} \label{disc}

As noted in the introduction, interpretation of high-energy collision data from more-central \aa\ collisions near midrapidity tends to follow one of two themes: (a) a flowing bulk medium identified as a QGP with unique properties or (b) a combination of two or three hadron production mechanisms including dijet production, albeit with jet modification increasing with centrality. Theme (a) assumes the existence of flows and local thermalization {\em a priori} and prefers methods and data interpretations that favor a flow narrative. Theme (b) assumes dijet production as the principal manifestation of QCD in high-energy nuclear collisions and uses measurements of eventwise-reconstructed (isolated) jets in elementary collisions to form expectations for other jet manifestations in \pa\ and \aa\ collisions. It is not unreasonable to pursue both, but theme (b) should be engaged at least as thoroughly as (a) to insure a balanced scientific outcome.

To pursue theme (b) properly requires direct and quantitative comparisons of all available information on isolated jets in elementary collisions with all available information on MB dijets in nuclear collisions over the largest possible range of collision systems.  Comparisons should be based on accurate isolation of jet-related hard components from other contributions to yields, spectra, correlations and fluctuations combined with extensive measures that preserve TCM trends, as demonstrated by various examples and cited references in the present study.

\subsection{Universality of the TCM}

The TCM represents a conceptually simple idea -- two complementary mechanisms dominate hadron production near midrapidity consisting of projectile-nucleon dissociation (soft) and MB dijet production (hard). The TCM concept was first related to RHIC data fifteen years ago~\cite{kn}. Elaboration of the TCM has proceeded in a number of subsequent publications (e.g.\ Refs.~\cite{ppprd,hardspec,fragevo,anomalous,ptscale,alicetommpt,aliceptflucttom,alicespec2}). The TCM hard component in hadron yields, spectra, fluctuations and correlations from \pp, \pa\ and \aa\ collisions corresponds quantitatively to eventwise-reconstructed jet properties over a large range of \pp\ multiplicity, \aa\ centrality, collision energies and hadron momenta. The general TCM pattern for {\em extensive} quantities $X$ (e.g.\ $n_{ch}$ or $P_t$) is
\bea
(1 /\bar \rho_s) X &=&  X_s + \alpha \bar \rho_s X_h~~~p\text{-}p
\eea
or
\bea
(2/N_{part}) X &=& X_s + \nu X_h~~~A\text{-}A,
\eea
where $\bar \rho_s$ is by hypothesis a proxy for the number of participant low-$x$ partons (gluons) in \pp\ collisions and $N_{part}$ is the number of participant nucleons in \aa\ collisions according to the Glauber model of that collision system. When left-hand-side quantities are plotted vs $\bar \rho_s$ (\pp) or $\nu$ (\aa) the TCM signature is a constant (soft component) plus an approximately linear increase (jet-related hard component).  Those trends are illustrated for $X =$ integrated charge \nch\ in Fig.~\ref{ppcomm} (left) (\pp) or Fig.~\ref{corresp} (\aa), for integrated $P_t$ in Fig.~\ref{ppcomm} (right) (\pp), for $P_t$ variance difference $\bar B$ in Fig.~\ref{xxx} (\aa) and for number of correlated hadron pairs $\Delta \rho$ in Fig.~\ref{slopes} (a) and (c). 
Intensive ratios  tend to conceal jet-related TCM trends by partial cancellations.

\subsection{Phenomenology of MB dijets}

For MB analysis as described in this study all collision events and all hadrons within each collision are accepted for analysis. No conditions are imposed (except those that define a detector acceptance). All hadron fragments from dijets that enter the detector acceptance are therefore retained within the primary particle data. If properties of the MB dijet population for a given collision system are known (i.e.\ measured) the detected fragment system should correspond quantitatively to predictions derived from jet data as a {\em critical test} for any jet interpretations of data. An example is provided in  Sec.~\ref{trigpredict}.

Section~\ref{jets} summarizes a comprehensive description of properties of isolated jets in terms of FFs and jet (scattered-parton) energy spectra represented by simple and accurate parametrizations for an array of collision systems~\cite{eeprd,fragevo,jetspec2,alicespec2}. Those parametrizations permit quantitative comparisons with MB fragment data from nuclear collisions. For some analysis methods (e.g.\ those associated with the TCM) the comparisons demonstrate accurate correspondence (e.g.\ Refs.~\cite{jetspec,alicespec2}). For other methods (some of those associated with the flow narrative) there is no clear pattern (e.g.\ Refs.~\cite{nohydro,noelliptic}).

Descriptions of dijet contributions in terms of pQCD theory cannot represent MB dijets. Due to its inevitable limitations pQCD can represent less than 10\% of jet fragments~\cite{eeprd,fragevo}. Next-to-leading-order (NLO) predictions of \pt\ spectrum structure are usually constrained to $p_t > 2$ GeV/c with large systematic uncertainties near the lower bound (see the next subsection), whereas MB fragment distributions peak well below that point (Figs.~\ref{pp2} and \ref{enrat3}).

Biased representations of fragment data distributions that rely on  model-dependent background subtraction, {\em ad hoc} \pt\ cuts or intensive (spectrum or statistics) ratios represent only a distorted fraction of the MB fragment population. Comparisons of  such biased data to eventwise-reconstructed jet properties is likely misleading and cannot test jet-related interpretations of data components.
 
\subsection{Comparisons with NLO $\bf p_t$ spectrum predictions}

A substantial effort has been expended to compare NLO pQCD theory predictions of inclusive hadron production with measured hadron \pt\ spectra~\cite{nlorefs}. The form of such predictions is represented schematically by
\bea \label{pqcdeq}
\frac{d^2\sigma}{dp_t d\eta} &=& \int \frac{dz}{z} \iint dx_1 dx_2 f_{p_1}^{h_1}(x_1) f_{p_2}^{h_2}(x_2) 
\\ \nonumber
&\times& \frac{d^2\hat \sigma(x_1,x_2,\hat p_t)}{d\hat p_t d\eta} D_{p_3}^{h_3}(z),
\eea
a convolution of projectile-hadron parton distribution functions (PDFs) $f_{p}^{h}(x)$, QCD parton scattering cross section $d^2\hat \sigma(x_1,x_2,\hat p_t)/d\hat p_t d\eta$ and scattered-parton FF $D_{p}^{h}(z)$. It is argued that comparisons of NLO predictions to inclusive spectra may test the role of jets in nuclear collisions, but there are problems with that argument.

Such NLO predictions (e.g. Ref.~\cite{nlospectra}) assume that almost all hadrons from \pp\ collisions are the result of binary parton-parton collisions. But that assumption conflicts with the TCM inferred from 200 GeV \pp\ collisions (Sec.~\ref{tcm1}) wherein most hadrons emerge from a (soft) process that does not scale with a number of binary collisions and is interpreted to represent dissociation of single projectile nucleons as the result of a soft \nn\ interaction. Instead, Eq.~(\ref{pqcdeq}) corresponds to Eq.~(\ref{fold1}) describing (within a constant factor) the \pt\ spectrum hard component.%
\footnote{Note that $\hat p_t$ is here the scattered-parton (jet) momentum, \pt\ is a fragment momentum, $z = p_t / \hat p_t$, $dz/z = d\ln(z) = dy_{max}$ for given \pt, $y_{max} \equiv \ln(2\hat p_t / m_\pi)$ and $\ln(1/z) = y_{max} - y = \xi$ (conventional FF parameter). The assumption that FFs $D(z)$ depend only on $z$ or $y_{max} - y$ is contradicted by \ppbar\ FF data as in Fig.~\ref{ppfd}, and \ee\ FFs have a universal form on $u \approx y/y_{max}$~\cite{eeprd}.}
Comparisons between NLO predictions and isolated spectrum hard components might provide a useful test of MB dijet production, but there are further issues.

Because of limitations on the applicability of perturbative methods NLO predictions are typically restricted to hadron $p_t > 2$ GeV/c (approximating $p_t \gg 1$ GeV/c). But examples above and cited references demonstrate that most jet fragments ($> 90$\%) appear in $p_t < 2$ GeV/c, and that fraction plays a dominant role in yields, spectra, fluctuations and correlations from \pp\ and \aa\ collisions.

An alternate (perhaps the main) purpose for comparison of NLO predictions to hadron spectra is tests of QCD factorization and FF universality and optimizing FF parametrizations from {\em global fits} to \ee, \ep\ and \pp\ or \ppbar\ data. A recent review stipulates that NLO predictions for jet spectra combining PDFs with pQCD parton-parton cross sections describe the spectrum data well~\cite{nlospectra}. The major remaining uncertainty lies with current FFs, especially gluon-to-unidentified-hadron FFs whereby hadron spectra are typically overpredicted by a factor 2. It is proposed to reduce such errors by refinement of FFs through improved global fits to data.

However, one can question such a strategy, especially the assumption of FF universality as it is conventionally invoked. Section~\ref{fragfunc} illustrates the large discrepancies between light-quark FFs derived from \ee\ collisions and those inferred from \ppbar\ collisions. The differences lie well outside systematic uncertainties. \ee\ light-quark FFs used to describe the \pp\ spectrum hard component  already produce a large overprediction~\cite{fragevo}. But low-\pt\ hadrons that dominate the spectrum hard component near midrapidity arise mainly from low-$x$ gluons. If \ee\ gluon FFs are applied instead the overprediction is worse due to the increased (and softer) hadron yield from gluon jets. The approximate agreement between jet systematics and spectrum hard components as in Fig.~\ref{ppfd} is achieved only by using \ppbar\ light-quark FFs~\cite{fragevo}. The result suggests that there are substantial differences in FFs depending on the jet environment in \pp\ collisions. Strong jet modification in \aa\ collisions is a notable result of the RHIC experimental program confirmed at the LHC. But jets appear to be modified significantly already in \pp\ collisions relative to \ee\ collisions. As a result, attempts to optimize ``universal'' FFs via global fits to multiple systems may produce a strongly-biased result.

Another issue for NLO comparisons is the structure of spectrum hard components below 5 GeV/c (e.g.\ Fig.~\ref{enrat3}). The maximum near 1 GeV for pions and protons~\cite{hardspec} (and therefore possibly kaons and other hadrons) is probably outside the scope of pQCD. The evolution of jet-related structure with \aa\ centrality as in Sec.~\ref{jetaaspec} is certainly so, consistent with the conclusion that ``...only the region above $p_\text{T} \approx 10$ GeV/c of these charged-hadron data, with theoretical scale uncertainties below $\pm20$\%, should be included in forthcoming global fits of parton-to-hadron fragmentation functions''~\cite{nlospectra}. However, the effectiveness of such global fits may still be questioned as noted above.

Three areas may then be distinguished: (a) comparison of measured isolated-jet characteristics with various MB jet manifestations in high-energy nuclear-collision data as described in the present study, (b) comparisons of PDF and pQCD parton-spectrum combinations with measured jet spectra as one test of QCD factorization and (c) improved measurement of scattered-parton FFs in elementary collisions with the understanding that FFs may depend on collision context -- e.g.\ \ee\ vs \pp\ -- as an extension of jet modification in \aa\ collisions.

\subsection{Methods that misidentify or distort MB dijets}

Section~\ref{taanalysis} provides an example of analysis methods that tend to minimize and/or distort MB dijet manifestations in nuclear-collision data. Other methods tend to reveal contributions from MB dijets consistent with isolated-jet measurements as described in this study. A principal difference is the amount of information carried by primary particle data that is retained by a method and may be utilized for direct comparison with isolated jets.
Examples are provided below for integrated \nch\ or $P_t$ vs \nch\ (\pp) or centrality (\aa), SP \pt\ spectra vs \nch\ or centrality, \ytyt\ correlations, 2D angular correlations, \pt\ fluctuations and choice of plot formats.

Integrated charge \nch\ from \aa\ collisions may be plotted as $(2/N_{part}) n_{ch}$ vs $\nu$ over the complete \aa\ centrality range to reveal a TCM trend with strong MB dijet contribution as in Fig.~\ref{corresp}, varying from a GLS trend extrapolated from \pp\ for more-peripheral \aa\ collisions to a trend reflecting strong jet modification for more-central collisions. Alternatively, the same per-participant quantity may be plotted vs $N_{part}$ over a limited centrality interval (e.g.\ top 40\% as in Ref.~\cite{global3}), and the \pp\ GLS extrapolation is then effectively concealed. Alternative nonjet hypotheses may seem to describe the more-central data but could be falsified by more-peripheral data (e.g.\ the CGC dashed trend in Fig.~\ref{corresp}, left).

A full analysis of \pt\ spectra vs \pp\ \nch\ or \aa\ centrality over a large \pt\ acceptance may reveal all available information in spectrum data through inductive study as in Refs.~\cite{ppprd,hardspec,ppquad,alicespec} leading to or consistent with the TCM. Alternatively, spectra for single collision systems fitted over limited \pt\ intervals with {\em a priori} model functions may appear to support nonjet interpretations (e.g.\ radial flow as in Ref.~\cite{starblast}). Studies that seem to emphasize jets by applying ``high-\pt'' cuts may actually discard almost all evidence of MB jets (more than 90\% of MB jet fragments).

Possible two-particle correlation studies include \ytyt\ correlations as in Fig.~\ref{ppcorr} (left) for various collision systems and several combinations of pair angular acceptance, and 2D angular correlations for various \pt\ conditions including full pair \pt\ acceptance~\cite{porter2,porter3,axialci,anomalous,ppquad}. The result in Fig.~\ref{ppcorr} (left) provides compelling evidence for the TCM, is directly comparable to the SP \yt\ spectrum TCM, and the hard-component peak centered near (2.7,2.7) corresponds directly to jet-related angular correlations as in Fig.~\ref{ppcorr} (right). It is notable that \ytyt\ correlations remain largely unexplored despite the essential information they convey.

2D angular correlations for a number of collision systems have been studied and characterized by a universal 2D fit model~\cite{anomalous,ppquad}. The results are quantitatively compatible with TCM results from other analysis including MB jet spectrum manifestations~\cite{jetspec}. An important feature of 2D angular correlations is resolution of the jet-related SS 2D peak from other structure that may or may not be jet related. In contrast, projection of 2D angular correlations onto 1D azimuth discards information relating to $\eta$ dependence and reduces the ability to distinguish jet-related structures from others. 
In particular, the NJ quadrupole cannot be uniquely distinguished from the quadrupole component of the jet-related SS 2D peak. Some fraction of the latter may then be attributed to ``elliptic flow'' $v_2$ leading to distortion of nominal jet structure following ZYAM background subtraction~\cite{starprl} or attribution of all MB jet structure to flows including ``triangular''~\cite{triangular} and ``higher harmonic''~\cite{luzum} flows~\cite{multipoles,higherharm}.%
\footnote{While some data features attributed to flows may represent MB dijet manifestations the NJ quadrupole is most likely a distinct {\em nonjet} phenomenon with a unique physical mechanism~\cite{gluequad,quadspec2}.}

While extensive quantities reveal clear TCM trends (as in this study) intensive ratios of such quantities discard essential information in primary data by canceling data trends required to interpret collision data, especially MB jet manifestations. Examples include spectrum ratio $R_{AA}$ that obscures contributions from MB jets below $p_t \approx 3$ GeV/c including more than 99\% of MB jet fragments,  eventwise mean $\langle p_t \rangle = P_t / n_{ch}$ and corresponding ensemble mean $\bar p_t$ that partially obscure TCM trends in the extensive numerators and denominators, and ratio $v_2(p_t)$ including NJ quadrupole spectrum (numerator) and SP hadron spectrum (denominator) that may represent different hadron populations confused by the $v_2$ ratio~\cite{quadspec,quadspec2}. Extensive fluctuation measures (e.g.\ variance differences such as $\Delta \sigma^2_{P_t|n}$) convey almost all information on underlying correlations including MB jet manifestations, whereas intensive ratios of statistical quantities (or {\em ratios of ratios} such as $\sqrt{\bar C} / \bar p_t$) tend to suppress jet manifestations and may favor interpretations within a thermodynamic context (see Sec.~\ref{jetptfluct} for notation).

The choice of plot format, including independent-variable choice, may further suppress essential information. The choice of participant number $N_{part}$ as \aa\ centrality measure over mean participant pathlength $\nu$ visually suppresses the more-peripheral part of centrality dependence, especially GLS trends extrapolated from \pp\ collisions that provide a valuable reference for \aa\ collisions. The choice of linear \pt\ over rapidity \yt\ [or at least $\log(p_t)$] visually suppresses the low-\pt\ region where most MB jet fragments appear. Semilog plots showing model curves passing through spectrum data points are often misleading; highly significant deviations may appear much smaller than the plotted points (as are the statistical uncertainties). Plots of spectrum data/model {\em ratios} are similarly misleading because deviations at lower \pt\ may be strongly suppressed. Fit quality is tested only by comparing [data $-$ model] {\em differences }in ratio to statistical uncertainties, as demonstrated in Fig.~\ref{power} (left).

\subsection{Consequences for physical interpretation}

Section~\ref{compete} notes that a specific analysis may rely on sequential selection from several alternative methods. A given combination of methods may then contribute to support of a preferred narrative whereas a different combination might falsify that narrative. A possible criterion for optimum selection is maximized use of the information carried by primary particle data. Another is insistence on consistent application and interpretation across a range of collision systems and measured quantities.

The present study demonstrates that one can assemble a combination of methods providing clear evidence for manifestations of MB dijets in all collision systems. Measure choices are guided by the principle that almost all available information from nuclear collision data is compared with almost all available information from isolated jets. No information is intentionally discarded. Measured MB jet manifestations then correspond quantitatively and consistently to properties of isolated jets established by separate experiments. Recognition of MB jet manifestations as such then precludes much of the evidence advanced to support flow interpretations.

In contrast, alternative combinations of methods can be assembled that minimize MB jet manifestations and appear to support a flow narrative to describe high-energy nuclear collisions (possibly even in small systems). However, as argued and demonstrated with examples in the present study, such preferred methods tend to discard major fractions of the information in primary particle data -- in the form of projections to lower dimensions, \pt\ cuts based on a preferred narrative, intensive ratios and nonoptimal plotting formats -- and discard information in isolated jet data by invoking pQCD as an imposed intermediary context (a sort of filter). Established jet physics is thereby largely excluded from descriptions of high-energy nuclear collisions in favor of a flow narrative.

%%%%%%%%%
 \section{Summary} \label{summ}

Primary particle data from high-energy nuclear collisions may be processed with alternative classes of analysis methods that seem to support one of two narratives: (a) collisions are dominated by flows carried by a dense medium (quark-gluon plasma or QGP) or (b) collisions are dominated near midrapidity by two hadron production mechanisms consisting of projectile-nucleon dissociation and minimum-bias (MB) dijet production.

Because a data analysis system leading to physical interpretations typically relies on multiple selections from among several possible methods there is no unique method combination. Results of the present study suggest that flow-QGP interpretations tend to arise from a certain class A of analysis methods  imposed {\em a priori} within a narrative context that exhibit substantial {\em information discard}. Interpretations based on MB dijets are favored by an alternative class B inferred inductively from primary particle data that retain most information.

This article presents a detailed study of measured manifestations of MB dijets in high-energy nuclear collisions and quantitative comparisons of those manifestations with measured properties of eventwise-reconstructed (isolated) jets. The study considers jet manifestations in yields, spectra, correlations and fluctuations from a range of collision systems. The MB dijet system is then used to evaluate properties of alternative analysis methods and the quality of support for the two competing narratives.

The two-component (soft + hard) model (TCM) of hadron production near midrapidity provides a context for MB dijets and narrative (b). The TCM as applied in this work was derived inductively from the \nch\ dependence of \pp\ \pt\ spectra. The soft and hard components of yields, spectra, correlations and fluctuations are observed to evolve independently with \pp\ charge multiplicity or \aa\ centrality. For each form of primary data the TCM hard component is quantitatively related to measured isolated-jet properties. For instance, evolution of the \pp\ \pt\ spectrum hard component with collision energy tracks quantitatively with measured isolated-jet spectra. Evolution of hadron yields with \aa\ centrality and collision energy follows simple TCM trends on centrality and a QCD trend $\log(s / s_0)$ on collision energy with $\sqrt{s_0} \approx 10$ GeV. The observed TCM simplicity applies to methods based on {\em extensive} variables such as \nch\ and $P_t$ integrated over some acceptance and supports narrative (b).

Alternative methods that appear to support narrative (a) rely on {\em intensive ratios} (or ratios of ratios), {\em a priori} imposed fit models, projections from higher- to lower-dimensional spaces, imposed \pt\ cuts and subtraction of {\em ad hoc} backgrounds that tend to discard essential information carried by primary particle data and may then lead to biased and distorted results. Those conclusions are based on manifestations of MB dijets using such methods in quantitative comparison to the measured properties of isolated jets. A number of examples are provided in this study. The overall results suggest that at least some data features interpreted to arise from flows actually represent MB dijet manifestations.

I conclude that the TCM for high-energy nuclear collisions emerges naturally from inductive analysis of yield, spectrum and correlation data, is not imposed {\em a priori}. The TCM hard component agrees quantitatively with the measured properties of isolated jets. Statistical measures based on extensive variables retain substantially more of the information carried by primary particle data. When various data features are reexamined in the context of isolated-jet measurements little substantial evidence remains to support a flow narrative. And any description of high-energy nuclear collisions that omits a clear, quantitative description of MB dijets consistent across all measures and collision systems may be questioned.

%%%%%%%%%%%%%%%%%%%%%%%%%%%%

\end{document}